\documentclass{aa}

\usepackage{graphicx,multirow,xcolor}
\graphicspath{{./figures/}}
\usepackage{txfonts}
\usepackage[hidelinks]{hyperref}
\begin{document} 

   \title{Selection of powerful radio galaxies with machine learning}

   \author{R. Carvajal\inst{1, 2}
          \and
          I. Matute\inst{1, 2}
          \and
          J. Afonso\inst{1, 2}
          \and
          R. P. Norris\inst{3, 4}
          \and
          K. J. Luken\inst{3, 5}
          \and
          P. S\'{a}nchez-S\'{a}ez\inst{6}
          \and
          P. A. C. Cunha\inst{7, 8}
          \and
          A. Humphrey\inst{7, 9}
          \and
          H. Messias\inst{10, 11}
          \and
          S. Amarantidis\inst{12, 1}
          \and
          D. Barbosa\inst{1, 2}
          \and
          H. A. Cruz\inst{13}
          \and
          H. Miranda\inst{1, 2}
          \and
          A. Paulino-Afonso\inst{7}
          \and
          C. Pappalardo\inst{1, 2}
          }

   \institute{Instituto de Astrof\'{i}sica e Ci\^{e}ncias do Espa\c{c}o, Universidade de Lisboa, OAL, Tapada da Ajuda, 1349-018 Lisbon, Portugal\\ \email{racarvajal@ciencias.ulisboa.pt}
         \and
             Departamento de F\'{i}sica, Faculdade de Ci\^{e}ncias, Universidade de Lisboa, Edif\'{i}cio C8, Campo Grande, 1749-016 Lisbon, Portugal
        \and
            School of Science, Western Sydney University, Locked Bag 1797, Penrith, NSW 2751, Australia
        \and
            CSIRO Space \& Astronomy, Australia Telescope National Facility, P.O. Box 76, Epping, NSW 1710, Australia
        \and
            CSIRO Data61, P.O. Box 76, Epping, NSW 1710, Australia
        \and
            European Southern Observatory, Karl-Schwarzschild-Stra{\ss}e 2, 85748 Garching bei M\"{u}nchen, Germany.
        \and
            Instituto de Astrof\'{i}sica e Ci\^{e}ncias do Espa\c{c}o, Universidade do Porto, CAUP, Rua das Estrelas, 4150-762 Porto, Portugal
        \and
            Departamento de F\'{i}sica e Astronomia, Faculdade de Ci\^{e}ncias, Universidade do Porto, Rua do Campo Alegre 687, 4169-007 Porto, Portugal
        \and
            DTx - Digital Transformation CoLAB, Building 1, Azur\'{e}m Campus, University of Minho, 4800-058 Guimar\~{a}es, Portugal
        \and
            Joint ALMA Observatory, Alonso de C\'{o}rdova 3107, Vitacura 763-0355, Santiago, Chile
        \and
            European Southern Observatory, Alonso de C\'{o}rdova 3107, Vitacura, Casilla 19001, Santiago de Chile, Chile
        \and
            Institut de Radioastronomie Millim\'{e}trique (IRAM), Avenida Divina Pastora 7, Local 20, 18012 Granada, Spain
        \and
            Closer Consultoria Lda, Av. Eng. Duarte Pacheco, Torre 1, 15\textsuperscript{o}, 1070-101 Lisboa, Portugal
             }

   \date{Received ; accepted }

% \abstract{}{}{}{}{} 
% 5 {} token are mandatory
 
  \abstract
   {The study of active galactic nuclei (AGNs) is fundamental to discern the formation and growth of supermassive black holes (SMBHs) and their connection with star formation and galaxy evolution. Due to the significant kinetic and radiative energy emitted by powerful AGNs, they are prime candidates to observe the interplay between SMBH and stellar growth in galaxies.
   }
   {We aim to develop a method to predict the AGN nature of a source, its radio detectability, and redshift purely based on photometry. The use of such a method will increase the number of radio AGNs, allowing us to improve our knowledge of accretion power into an SMBH, the origin and triggers of radio emission, and its impact on galaxy evolution.   
   }
   {We developed and trained a pipeline of three machine learning (ML) models than can predict which sources are more likely to be an AGN and to be detected in specific radio surveys. Also, it can estimate redshift values for predicted radio-detectable AGNs. These models, which combine predictions from tree-based and gradient-boosting algorithms, have been trained with multi-wavelength data from near-infrared-selected sources in the Hobby-Eberly Telescope Dark Energy Experiment (HETDEX) Spring field. Training, testing, calibration, and validation were carried out in the HETDEX field. Further validation was performed on near-infrared-selected sources in the Stripe 82 field.
   }
   {In the HETDEX validation subset, our pipeline recovers $96\%$ of the initially labelled AGNs and, from AGNs candidates, we recover $50\%$ of previously detected radio sources. For Stripe 82, these numbers are $94\%$ and $55\%$. Compared to random selection, these rates are two and four times better for HETDEX, and $1.2$ and $12$ times better for Stripe 82. The pipeline can also recover the redshift distribution of these sources with $\sigma_{\mathrm{NMAD}} {=} 0.07$ for HETDEX ($\sigma_{\mathrm{NMAD}} {=} 0.09$ for Stripe 82) and an outlier fraction of $19\%$ ($25\%$ for Stripe 82), compatible with previous results based on broad-band photometry. Feature importance analysis stresses the relevance of near- and mid-infrared colours to select AGNs and identify their radio and redshift nature.
   }
   {Combining different algorithms in ML models shows an improvement in the prediction power of our pipeline over a random selection of sources. Tree-based ML models (in contrast to deep learning techniques) facilitate the analysis of the impact that features have on the predictions. This prediction can give insight into the potential physical interplay between the properties of radio AGNs (e.g. mass of black hole and accretion rate).
   }
   
   \keywords{Galaxies: active -- Radio continuum: galaxies -- Galaxies: high-redshift -- Catalogs -- Methods: statistical.
               }

   \maketitle
%
%-------------------------------------------------------------------

\section{Introduction}\label{sec:introduction}

Active galactic nuclei (AGNs) are instrumental in determining the nature, growth, and evolution of supermassive black holes (SMBHs). Their strong emission allows us to study the close environment within the hosting galaxies and, at a larger scale, the intergalactic medium \citep[e.g.][]{2017A&ARv..25....2P, 2022MNRAS.516.5775B}. Feedback due to AGN energetics, which most prominently manifest in the form of jetted radio emission, might play a fundamental role in regulating stellar growth and the overall evolution of hosts and their environments \citep{2015ApJ...798...31A, 2017MNRAS.472.4659V, 2020NewAR..8801539H}.

Although radio emission can trace high star formation (SF) in galaxies, above certain luminosities \citep[e.g. $\mathrm{log}L_{1.4\mathrm{GHz}} {>} 25\,\mathrm{W\, Hz^{-1}}$,][]{2021MNRAS.503.1780J}, it is a prime tracer of the powerful jet emission triggered by the SMBH in AGNs \citep[radio galaxies,][]{2014ARA&A..52..589H}. Traditionally, these powerful radio galaxies (RGs) were used to pinpoint AGN activity, but they have been superseded in the last decades by optical, near-infrared (NIR), and X-ray surveys. In fact, RGs in the high redshift Universe ($z > 2$) have been identified and studied mostly through the follow-up of AGNs selected at shorter wavelengths \citep[optical, NIR, millimetre, and X-rays, e.g.][]{2006ApJ...652..157M, 2020A&A...637A..84P, 2021MNRAS.501.3833D}.
The landscape is quickly changing and the advent of new radio instruments and surveys has allowed the detection of larger numbers of RGs \citep[e.g.][]{2018MNRAS.475.3429W, 2020A&A...642A.107C}. Some of these surveys are: the National Radio Astronomy Observatory (NRAO) Very Large Array (VLA) Sky Survey \citep[NVSS;][]{1998AJ....115.1693C}, the Faint Images of the Radio Sky at Twenty-Centimetres \citep[FIRST;][]{2015ApJ...801...26H}, the Evolutionary Map of the Universe \citep[EMU;][]{2011PASA...28..215N}, the Very Large Array Sky Survey \citep[VLASS;][]{2020RNAAS...4..175G}, and the Low Frequency Array (LOFAR) Two-metre Sky Survey \citep[LoTSS;][]{2019A&A...622A...1S}. 

One of the ultimate goals is to detect powerful RGs in the Epoch of Reionisation (EoR), which could be used to trace the neutral gas distribution during this critical phase of the Universe \citep[e.g.][]{2004NewAR..48.1029C, 2013MNRAS.435..460J}. Simulations have shown that as much as a few hundreds of RGs per deg$^2$ could be present in the EoR \citep[][]{2019MNRAS.485.2694A, 2019MNRAS.482....2B, 2021MNRAS.503.3492T} and detectable with present and future deep observations, for example the Square Kilometre Array (SKA), which is projected to have $\mu$Jy point-source sensitivity levels \citep[SKA1-Mid is expected to reach close to $2 \, \mu$Jy in $1$-hour continuum observations at $\nu\,{\gtrsim} 1$ GHz;][]{2015aska.confE..67P, 2019arXiv191212699B}. Most recent observational compilations \citep[e.g.][]{2020ARA&A..58...27I, 2020MNRAS.494..789R, sarah_e_i_bosman_2022_6039724, 2023ARA&A..61..373F} show that around $300$ AGNs have been confirmed to exist at redshifts higher than $z{\sim}6$ over thousands of square degrees. This disagreement highlights the uncertainties present in simulations, mainly due to our lack of knowledge of the triggering mechanisms and duty cycle for jetted emission in AGNs \citep{2015aska.confE..71A, 2022MNRAS.510.1163P}.

The selection of AGN candidates has had success in the X-rays and radio wavebands as they dominate the emission above certain luminosities. Unfortunately, deep X-ray surveys are limited in area and only of the order of $10\%$ of AGNs have strong radio emissions linked to jets (i.e. radio-loud sources) at any given time with variations, going from $\sim 6\%$ up to $\sim 30\%$, correlated to optical and X-ray luminosities, as well as with redshift \citep[e.g.][]{1993MNRAS.263..461P, 1994ApJ...430..533D, 2007ApJ...656..680J, 2019NatAs...3...48S, 2019A&A...622A..11G, 2021MNRAS.506.5888M, 2021A&A...656A.137G, 2022A&A...668A..27G, 2023MNRAS.tmp.1261B}.\footnote{Depending on the dataset, a random selection of AGNs would lead to a rate of radio-detectable AGNs in the range $6 - 30 \%$. We call this random choice a `no-skill' selection.}

The largest number of AGN candidates has been selected through the compilation of multi-wavelength spectral energy distributions (SED) for millions of sources \citep{2018ARA&A..56..625H, 2020PhDT.........3P}. Of particular relevance for AGNs are the mid-infrared (mid-IR) colours where \textit{Spitzer} \citep{2004ApJS..154....1W} and especially  the Wide-field Infrared Survey Explorer \citep[WISE;][]{2010AJ....140.1868W} have opened a window for the detection of AGNs over the whole sky, including the elusive fraction of heavily obscured ones \citep[e.g.][]{2012ApJ...753...30S, 2012MNRAS.426.3271M, 2017ApJ...836..182J, 2018ApJS..234...23A, 2021ApJ...922..179B}.

Currently, extensive spectroscopic follow-up measurements have allowed the confirmation of the estimated redshifts for more than $800\,000$ AGNs over large areas of the sky \citep{2021arXiv210512985F}. Spectroscopic surveys have also contributed to the detection of AGN activity through the analysis of the line ratio as is the case of the Baldwin-Phillips-Terlevich (BPT) diagram \citep*{1981PASP...93....5B}. However, their determination can take long integration times and require high-quality observations, rendering them ill-suited for most sources in large-sky catalogues. Photometric classification and redshifts (photo-$z$) are a viable option to understand the source nature and distribution across cosmic time \citep{1957AJ.....62....6B, 2019NatAs...3..212S}. Photometric redshift estimations have been obtained for galaxies \citep[e.g.][]{2021A&A...654A.101H} and AGNs \citep[e.g.][]{2017ApJ...850...66A}. Template-fitting photo-$z$ estimations are computationally expensive and require high-performance computing facilities for large catalogues \citep[${\gtrsim}10^{7}$ elements, ][]{2021ApJ...916...43G}. At the expense of redshift precision, the use of drop-out techniques offer a more computationally efficient solution to generate and study high-redshift sources or candidates that, otherwise, would not have enough information to produce a precise redshift value \citep[e.g.][]{2020ApJ...902..112B, 2020A&A...633A.160C, 2023MNRAS.519.4902S}.

Alternative statistical and computational methods can analyse a large number of elements and find relevant trends among their properties. One branch of these techniques is machine learning \citep[ML;][]{5392560}, which can, using previously modelled data, predict the behaviour that new data will have, that is, the values of their properties.
In astronomy, ML has been used with much success in a wide range of subjects, such as redshift determination, morphological classification, emission prediction, anomaly detection, observations planning, and more \citep[e.g.][]{2010IJMPD..19.1049B, 2019arXiv190407248B}. Traditional ML models are, in general, only fed with measurements and not with physical assumptions \citep{Desai2021}, and they do not need to check the consistency of the predictions or the results they provide. As a consequence, prediction times of traditional ML methods are typically less than those from physically based methods.

Despite the large number of applications it might have, one important criticism that ML has received is related to the lack of interpretability --or `explainability', as it is called in ML jargon-- of the derived models, trends, and correlations. Most ML models, after taking a series of measurements and properties as input, deliver a prediction of a different property. But they cannot provide coefficients or an analytical expression that might allow one to find an equation for future predictions \citep{goebel2018explainable}. An important counterexample of this fact is the use of symbolic regression \citep[e.g.][]{2020arXiv200611287C, 2021ApJ...915...71V, 2023arXiv230501582C}. This implies that, for most ML models, it is not a simple task to understand which properties, and to what extent they help predict and interpret another attribute. This fact hinders our capability of understanding the results in physical terms.

Recent work has been done to overcome the lack of explainability in ML models. The most widely used assessment is done with feature importance \citep{10.1007/978-3-030-10925-7_40, 9007737}, both global and local \citep{Saarela2021}. Game-theory-based analyses, such as the Shapley analysis \citep{Shapley_article}, have also been used to understand the importance of features in astrophysics \citep[e.g.][]{2021MNRAS.507.1468M, 2021Galax...9...86C, 2022MNRAS.515.5285D, 2022MNRAS.509.3441A, 2022MNRAS.516.4716A}. 

A further complication is that astronomical data can be very heterogeneous. Surveys and instruments gather data from many different areas in the sky with very different sensitivities and observational properties. This heterogeneity severely complicates most astronomical analyses, but in particular ML methods, as they are completely driven by data most of the time. This issue can be alleviated using observations in large, and homogeneous, surveys. Currently, among others, VLA, LOFAR, and the Giant Metrewave Radio Telescope (GMRT) allow such measurements to be obtained. Next-generation observatories and surveys --such as SKA and the Vera C. Rubin Observatory-- will also help in this regard, where observations will be carried out homogeneously over very large areas.

From a pure ML-based standpoint, several techniques used to lessen the effect of data heterogeneity have been developed (i.e. data cleansing and homogenisation). Some of them include discarding sources that add noise to the overall data distribution \citep{10.1145/3506712}. This can be extended to vetoing sources from specific areas in the sky (due to, for example, bad data reduction). Opposite to that, and when possible, previously mentioned techniques can be combined by increasing the survey area as a way to reduce possible biases.
After selecting the data sample to be used for modelling, it is also possible to homogenise the measured ranges of observed properties. This procedure implies, for instance, that normalising or standardising measured values can help ML models extract trends and connections among features more easily \citep{SINGH2020105524}.

Future observatories and surveys will deliver immense datasets. One option to analyse such observations and confirm their radio-AGN nature is through visual inspection \citep[e.g.][]{2015MNRAS.453.2326B}. The use of such a technique over large areas can have a very high cost. An alternative is using already-available multi-wavelength data and template-fitting tools to determine the likelihood of an AGN of being detected in radio wavelengths \citep[see, for instance,][]{2023ApJ...944..141P}. With the use of existing data, ML can help to speed this process up via the training of models that can detect counterparts in large radio surveys \citep[see, for  example, the efforts made to achieve this goal,][]{2015PASA...32...37H, 2021MNRAS.500.3821B}.

Building upon the work presented by \citet{2021Galax...9...86C}, we aim to identify candidates of high-redshift radio-detectable AGNs that can be extracted from heterogeneous large-area surveys. We developed a series of ML models to predict, separately, the detection of AGNs, the detection of the radio signal from AGNs, and the redshift values of radio-detectable AGNs using non-radio photometric data. In this way, it might be possible to avoid the direct analysis of large numbers of radio detections. Furthermore, we tested the performance of these models without applying a large number of previous cleaning steps, which might reduce the size of the training sets considerably. The compiled catalogue of candidates can help to use data from future large-sky surveys more efficiently, as observational and analytical efforts can be focussed on the areas in which AGNs have been predicted to exist. We seek, therefore, to test the generalisation power of such models by applying them in a different area from the training field with data that are not necessarily of the same quality.

The structure of this article is as follows. In Sect.~\ref{sec:data}, we present the data and its preparation for ML training. The selection of models and the metrics used to assess their results are shown in Sect.~\ref{sec:ML_training}. In Sect.~\ref{sec:results}, the results of model training and validation are provided as well as the predictions using the ML pipeline for radio AGN detections. We present the discussion of our results in Sect.~\ref{sec:discussion}. Finally, in Sect.~\ref{sec:summary_conclusions}, we summarise our work.

%--------------------------------------------------------------------
\section{Data}\label{sec:data}

A large area with deep and homogeneous quality radio observations is needed to train and validate our models and predictions for RGs with already existent observations. As training field we selected the area of the Hobby-Eberly Telescope Dark Energy Experiment Spring field \citep[HETDEX;][]{2008ASPC..399..115H} covered by the first data release of the LOFAR Two-metre Sky Survey \citep[LoTSS-DR1;][]{2019A&A...622A...1S}. The LoTSS-DR1 survey covers $424\, \mathrm{deg}^{2}$ in the HETDEX Spring field (hereafter, HETDEX field) with LOFAR \citep{2013A&A...556A...2V} $150\, \mathrm{MHz}$ observations that have a median sensitivity of $71\, \mu\mathrm{Jy}/\mathrm{beam}$ and an angular resolution of $6\arcsec$. HETDEX provides as well multi-wavelength homogeneous coverage as described below.

In order to test the performance of the models when applied to different areas of the sky, and with different coverages from radio surveys, we selected the Sloan Digital Sky Survey \citep[SDSS,][]{2000AJ....120.1579Y} Stripe 82 field \citep[S82,][]{2014ApJ...794..120A, 2014ApJS..213...12J}. For S82, we collected data from the same surveys as with the HETDEX (see the following section) field but with one important caveat: no LoTSS-DR1 data is available in the field and, thus, we gathered the radio information from the VLA SDSS Stripe 82 Survey \citep[VLAS82;][]{2011AJ....142....3H}. VLAS82 covers an area of $92\, \mathrm{deg}^{2}$ with a median rms noise of $52\,\mu\mathrm{Jy}/\mathrm{beam}$ at 1.4$\,$GHz with an angular resolution of $1\farcs 8$. We selected the S82 field (and, in particular, the area covered by VLAS82) given that it presents deep radio observations but taken with a different instrument than LOFAR. This difference allows us to test the suitability of our models and procedures in conditions that are different from the training circumstances.

\subsection{Data collection}\label{sec:data_collection}

The base survey from which all the studied sources have been drawn is the CatWISE2020 catalogue \citep[CW;][]{2021ApJS..253....8M}. It lists NIR-detected elements selected from WISE \citep{2010AJ....140.1868W} and the Near-Earth Object Wide-field Infrared Survey Explorer Reactivation Mission \citep[NEOWISE;][]{2011ApJ...731...53M, 2014ApJ...792...30M} over the entire sky at $3.4$ and $4.6$ $\mu$m (W1 and W2 bands, respectively). This catalogue includes sources detected at 5$\sigma$ in either of the used bands (i.e. W1${\sim} 17.43$ and W2${\sim} 16.47$ $\mathrm{mag}_{\mathrm{Vega}}$ respectively). The HETDEX field contains $15\,136\,878$ sources listed in CW. Conversely, in the S82 field, there are $3\,590\,306$ of them.

Multi-wavelength counterparts for CW sources were found on other catalogues applying a 5\arcsec search criteria. These catalogues include the Panoramic Survey Telescope and Rapid Response System \citep[Pan-STARRS DR1;][hereafter, PS1]{2016arXiv161205560C, 2020ApJS..251....7F}, the Two Micron All-Sky Survey \citep[2MASS All-Sky;][hereafter, 2M]{2006AJ....131.1163S, 2003tmc..book.....C, 2003yCat.2246....0C}, and AllWISE \citep[AW;][]{2013wise.rept....1C}\footnote{For the purposes of the analyses, and except when clearly stated otherwise, photometric measurements are converted to AB magnitudes.}. The adopted search radius corresponds to the distance that has been used by \citet{2010AJ....140.1868W} to match radio sources to Pan-STARRS and WISE observations. Nevertheless, the source density of the radio (LOFAR and VLA) and 2MASS catalogues imply a low statistical ($<1\%$) spurious counterpart association, this is not the case for PS1, where the source density is higher. For this reason, and to maintain a statistically low spurious association between CW and PS1, we limited our search radius to $1\farcs 1$. This distance corresponds to the smallest point-spread function (PSF) size of the bands included in PS1 \citep{2016arXiv161205560C}.

For the purposes of this work, observations in LoTSS and VLAS82 are only used to determine whether a source is radio detected, or not. In particular, no check has been performed on whether a selected source is extended or not in any of the radio surveys. A single Boolean feature is created from the radio measurements (see Sect.~\ref{sec:feature_creation}) and no further analyses were performed regarding the detection levels that might be found.

Additionally, we discarded the measurement errors of all bands. Traditionally, ML algorithms cannot incorporate uncertainties in a straightforward way and, thus, we opted to avoid attempting to use them for training. One significant counterexample corresponds to Gaussian processes \citep[GPs;][]{rasmussen2006gaussian}, where measurement uncertainties are needed by the algorithm to generate predictions. Additionally, the astronomical community has attempted to modify existing techniques to include uncertainties in their ML studies. Some examples include the works by \citet{2008ApJ...683...12B, 2019AJ....157...16R, 2022AJ....164....6S}. Furthermore, \citet{2023A&A...671A..99E} have shown that, in specific cases, the inclusion of measurement errors does not add new information to the training of the models and can be even detrimental to the prediction metrics. The degradation of the model by including uncertainties can likely be related to the fact that, by virtue of the large number of sources included in the training stages, the uncertainties are already encoded in the dataset in the form of scatter.

Following the same argument of measurement errors, upper limit values have been removed and a missing value is assumed instead. In general, ML methods (and their underlying statistical methods) cannot work with catalogues that have empty entries \citep{allison2001missing}. For that reason, we used single imputation \citep[a review on the use of this method, which is part of data cleansing, in astronomy can be seen in][]{ChattopadhyayData} to replace these missing values, and those fainter than $5{-}\sigma$ limits, with meaningful quantities that represent the lack of a measurement. 
We opted for the inclusion of the same $5{-}\sigma$ limiting magnitudes as the value to impute with. 
This method of imputation with some variations, has been successfully applied and tested, recently, by \citet{2020MNRAS.498.1750A, 2021Galax...9...86C, 2022MNRAS.512.2099C}, and \citet{2022MNRAS.514....1C}. In particular, \citet{2022MNRAS.512.2099C} tested several data imputation methods. Among those which replaced all missing values in a wavelength band with a single, constant value, using the $5{-}\sigma$ limiting magnitudes showed the best performance.

\begin{figure}
  \centering
  \includegraphics[width=0.90\columnwidth]{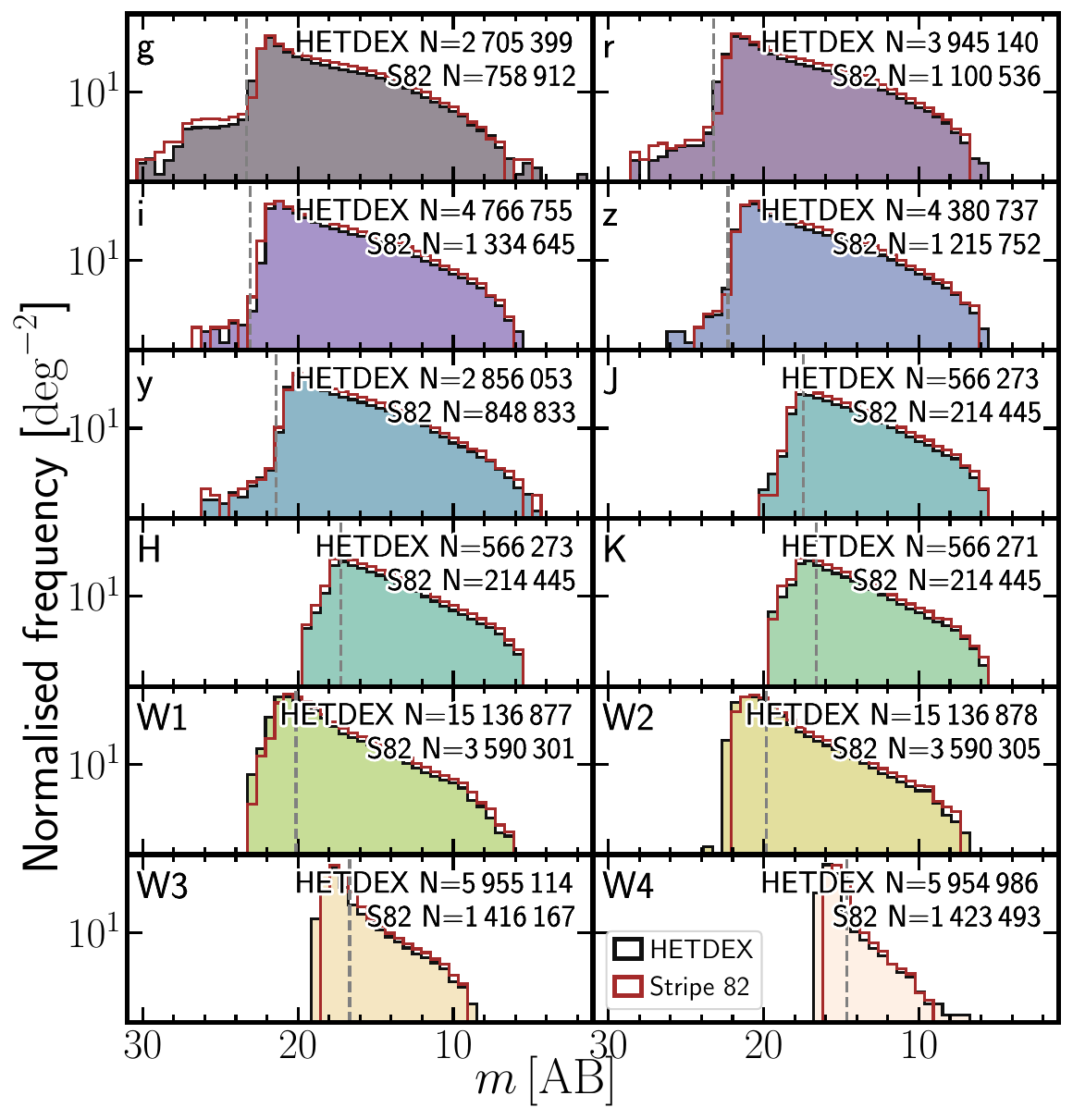}
  \caption{Histograms of base collected, non-imputed, non-radio bands for HETDEX (clean, background histograms) and S82 (empty, brown histograms). Each panel shows the distribution of measured magnitudes of detected sources divided by the total area of the field. Dashed, vertical lines represent the $5{-}\sigma$ magnitude limit for each band. The number in the upper right corner of each panel shows the number of measured magnitudes included in their corresponding histogram.}
  \label{fig:hists_bands_nonimp_HETDEX_S82}
\end{figure}

\begin{figure}
   \centering
   \includegraphics[width=0.90\columnwidth]{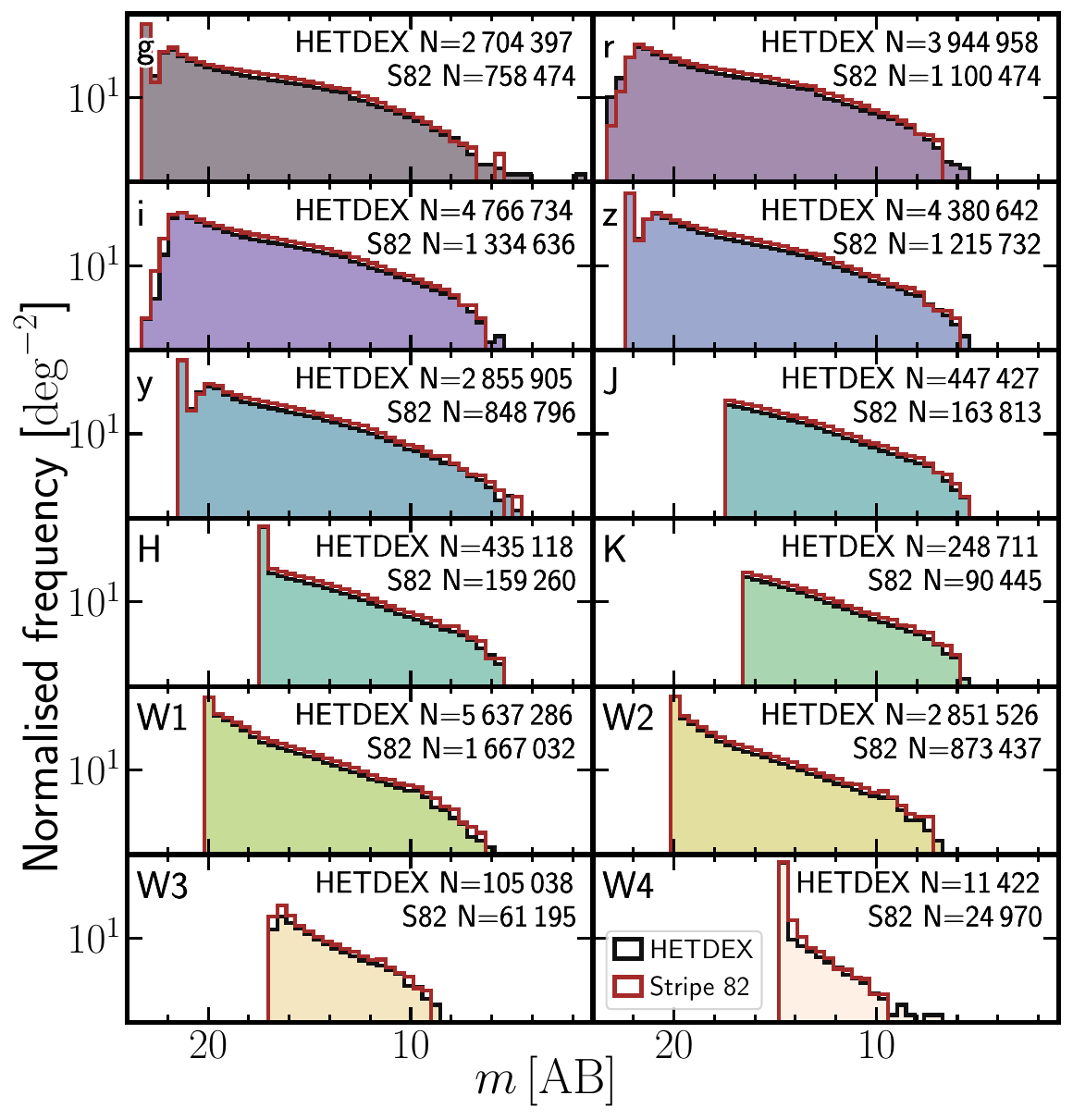}
   \caption{Histograms of base collected non-radio bands for HETDEX (clean, background histograms) and S82 (empty, brown histograms) fields. Description as in Fig.~\ref{fig:hists_bands_nonimp_HETDEX_S82}. The number in the upper right corner of each panel shows number of sources with magnitudes originally measured above the $5{-}\sigma$ limit included in their corresponding histogram for each field (i.e. sources that have not been imputed or replaced).}
   \label{fig:hists_bands_HETDEX_S82}
\end{figure}

\begin{figure}
   \centering
   \includegraphics[width=0.95\columnwidth]{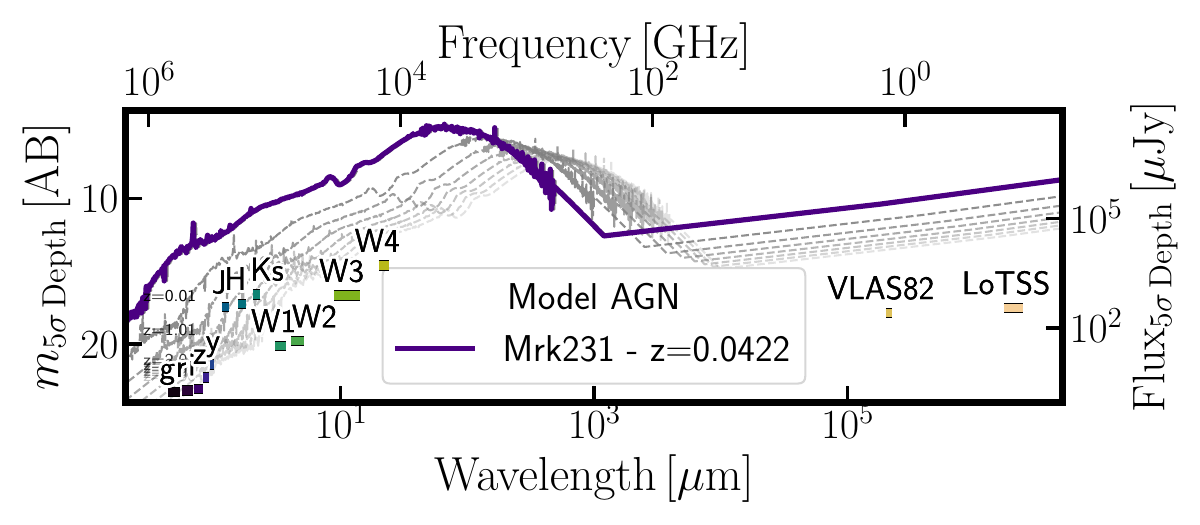}
   \caption{Flux and magnitude depths ($5{-}\sigma$) from the surveys and bands used in this work. Limiting magnitudes and fluxes were obtained from the description of the surveys, as referenced in Sect.~\ref{sec:data_collection}. In purple, rest-frame SED from Mrk231 \citep[$z = 0.0422$,][]{2019MNRAS.489.3351B} is displayed as an example AGN. Redshifted (from $z {=} 0.001$ to $z {=} 7$) versions of this SED are shown in dashed grey lines.}
   \label{fig:surveys_depth_HETDEX}
\end{figure}

In this way, observations from $12$ non-radio bands were gathered (as listed in Table~\ref{table:used_bands}). 
The magnitude density distribution for the sample from the HETDEX and S82 fields, without any imputation, is shown in Fig.~\ref{fig:hists_bands_nonimp_HETDEX_S82}. After imputation, the distribution of magnitudes changes, as shown in Fig.~\ref{fig:hists_bands_HETDEX_S82}. Each panel of the figure shows the number of sources which have a measurement above its $5{-}\sigma$ limit in such band. 
Additionally, a representation of the observational $5{-}\sigma$ limits of the bands and surveys used in this work is presented in Fig.~\ref{fig:surveys_depth_HETDEX}. 
It is worth noting the depth difference between VLAS82 and LoTSS-DR1 is ${\sim}1.5$\,mag for a typical synchrotron emitting source ($F_\nu \propto \nu^{\alpha}$ with $\alpha {=}-0.8)$, allowing the latter survey to reach fainter sources.

\begin{table}
\setlength{\tabcolsep}{2pt}
\caption{Bands available for model training in our dataset}             % title of Table
\label{table:used_bands}      % is used to refer this table in the text
\centering                          % used for centering table
\resizebox{0.75\columnwidth}{!}{
\begin{tabular}{c c}        % centered columns (2 columns)
\hline\hline                 % inserts double horizontal lines   
Survey             & Band (column name)\tablefootmark{a} \\
\hline
\multirow{2}{*}{Pan-STARRS (PS1)}       & g (\texttt{gmag}), r (\texttt{rmag}), i (\texttt{imag}),  \\
                 & z (\texttt{zmag}), y (\texttt{ymag})  \\[2pt]
2MASS (2M)            & J (\texttt{Jmag}), H (\texttt{Hmag}), Ks (\texttt{Kmag})  \\[2pt]
CatWISE2020 (CW) & W1 (\texttt{W1mproPM}), W2 (\texttt{W2mproPM})  \\[2pt]
AllWISE (AW)     & W3 (\texttt{W3mag}), W4 (\texttt{W4mag}) \\
\hline                                   %inserts single line
\end{tabular}
}
\tablefoot{
\tablefoottext{a}{In parentheses are shown the names of the columns or features in our dataset that represent each band.}
}
\end{table}

AGN labels and redshift information were obtained by cross-matching (with a $1\farcs 1$ search radius) the catalogue with the Million Quasar Catalog\footnote{\url{http://quasars.org/milliquas.htm}} \citep[MQC, v7.4d;][]{2021arXiv210512985F}, which lists information from more than $1\,500\,000$ objects that have been classified as optical quasi-stellar objects (QSOs), AGNs, or Blazars. Sources listed in the MQC may have additional counterpart information, including radio or X-ray associations. For the purposes of this work, only sources with secure spectroscopic redshifts were used. The matching yielded $50\,538$ spectroscopically confirmed AGNs in HETDEX and $17\,743$ confirmed AGNs in S82.

Similarly, the sources in our parent catalogue were cross-matched with the Sloan Digital Sky Survey Data Release 16 \citep[SDSS-DR16;][]{2020ApJS..249....3A}. This cross-match was done solely to determine which sources have been spectroscopically classified as galaxies (\verb|spClass == GALAXY|). 
For most of these galaxies, SDSS-DR16 lists a spectroscopic redshift value, which will be used in some stages of this work. In the HETDEX field, SDSS-DR16 provides $68\,196$ spectroscopically confirmed galaxies. In the S82 field, SDSS-DR16 identifies $4\,085$ galaxies spectroscopically. Given that MQC has access to more AGN detection methods than SDSS, when sources were identified as both galaxies (in SDSS-DR16) and AGNs (in the MQC), a final label of AGN was given. 
A description of the number of elements in each field and the multi-wavelength counterparts found for them is presented in Table~\ref{table:composition_catalogue}. From Table~\ref{table:composition_catalogue}, it is possible to see that the numbers and ratios of AGNs and galaxies in both fields are dissimilar. S82 has been subject to a larger number of observations, which have allowed the detection of a larger fraction of AGNs than in the HETDEX field \citep[see, for instance,][]{2020ApJS..250....8L}, which does not have such number of dedicated studies.

\begin{table}
\setlength{\tabcolsep}{3pt}
\caption{Composition of initial catalogue and number of cross matches with additional surveys and catalogues.}             % title of Table
\label{table:composition_catalogue}      % is used to refer this table in the text
\centering                          % used for centering table
\resizebox{0.50\columnwidth}{!}{
\begin{tabular}{c c c}        % centered columns (3 columns)
\hline\hline                 % inserts double horizontal lines  
                &  HETDEX        &  S82     \\
Survey          &                &               \\
\hline
CatWISE2020     & $15\,136\,878$ & $3\,590\,306$ \\
AllWISE         & $5\,955\,123$  & $1\,424\,576$ \\
Pan-STARRS      & $4\,837\,580$  & $1\,346\,915$ \\
2MASS           & $566\,273$     & $214\,445$    \\
LoTSS           & $187\,573$     & \ldots        \\
VLAS82          & \ldots         & $8\,747$      \\
MQC (AGNs)       & $50\,538$      & $17\,743$     \\
SDSS (galaxies)   & $68\,196$      & $4\,085$      \\
\hline
\hline                                   %inserts single line
\end{tabular}
}
\end{table}

Attending to the intrinsic differences between ML algorithms, not all of them have the same performance when being trained with features spanning a wide range of values (i.e. several orders of magnitude). For this reason, it is customary to re-scale the available values to either be contained within the range $[0, 1]$ or to have similar distributions. We applied a version of the latter transformation to our features (not the targets) as to have a mean value of $\mu = 0$ and a standard deviation of $\sigma = 1$ for each feature. Additionally, these new values were power-transformed to resemble a Gaussian distribution. This transformation helps the models avoid using the distribution of values as additional information for the training. For this work, a Yeo-Johnson transformation \citep{10.1093/biomet/87.4.954} was applied.

\subsection{Feature pool}\label{sec:feature_creation}

The initial pool of features that have been selected or engineered to use in our analysis is briefly described here. A full list of the features created for this work and their representation in the code and in some of the figures is presented in Table~\ref{table:feature_names_in_work}.

Most of the features used in this work come from photometry, both measured and imputed, in the form of AB magnitudes for a total of $12$ bands. Also, all available colours from measured and imputed magnitudes were considered. In total, there are $66$ colours, resulting from all available combinations of two magnitudes between the $12$ selected bands. These colours are labelled in the form \texttt{X\_Y} where \texttt{X} and \texttt{Y} are the respective magnitudes.

Additionally, the number of non-radio bands in which a source has valid measurements (\texttt{band\_num}) has been used. This feature could be, very loosely, attributed to the total flux a source can display. A higher \texttt{band\_num} will imply that such source can be detected in more bands, hinting a higher flux (regardless of redshift). The use of features with counting or aggregation of elements in the studied dataset is well established in ML \citep[see, for example,][]{zheng2018feature, duboue2020art, 2021AJ....161..141S, 2023A&A...671A..99E}.

Finally, as categorical features, we included an AGN-galaxy classification Boolean flag named \texttt{class} and a radio Boolean flag \texttt{LOFAR\_detect}. This feature flags whether sources have counterparts in the radio catalogues (LoTSS or VLAS82).

%--------------------------------------------------------------------
\section{Machine learning training}\label{sec:ML_training}

\begin{figure}
   \centering
   \includegraphics[width=0.48\columnwidth]{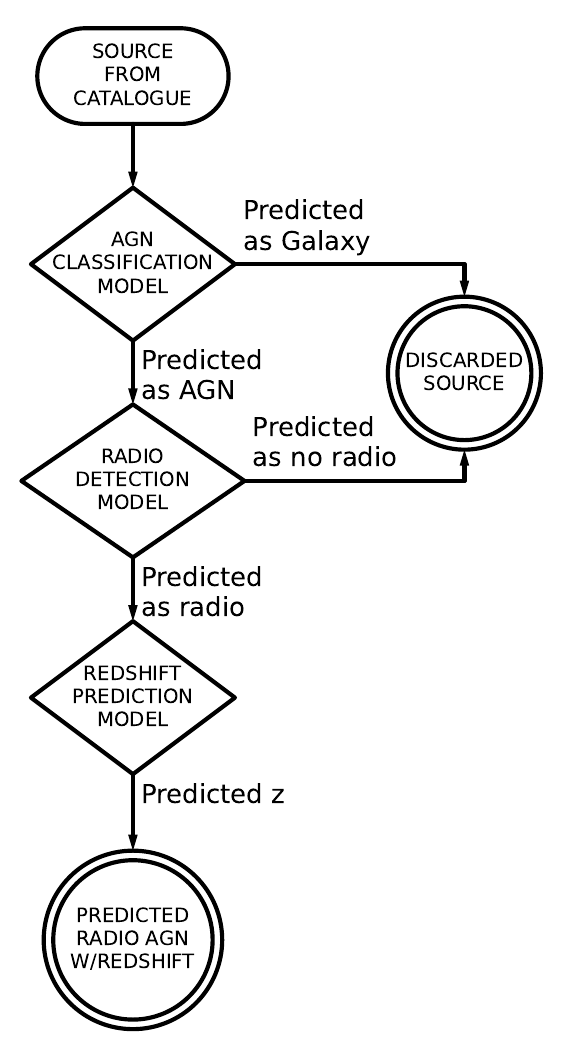}
   \caption{Flowchart representing the prediction pipeline used to predict the presence of radio-detectable AGNs and their redshift values. At the beginning of each model step, the most relevant features are selected as described in Sect.~\ref{sec:feat_selection}.}
   \label{fig:pipeline_flowchart}
\end{figure}

In an attempt to extract the largest available amount of information from the data, and let ML algorithms improve their predictions, we decided to perform our training and predictions through a series of sequential steps, which we refer to as `models' henceforth. We started with the training and prediction of the class of sources (AGNs or galaxies). The next model predicts whether an AGN could be detected in radio at the depth used during training (LoTSS). A final model will predict the redshift values of radio-predicted AGNs. A visual representation of this process can be seen in Fig.~\ref{fig:pipeline_flowchart}. Creating separate models gives us the opportunity to select the best subset of features for training as well as the best combination of ML algorithms for training in each step.

In broad terms, our goal with the classification models is to recover the largest number of elements from the positive classes (i.e. \texttt{class = 1} and \texttt{LOFAR$\_$detect = 1}). For the regression model, we aim to retrieve predictions as close as the originally fed redshift values.

In general, classification models provide a final score in the range [$0$, $1$], which can only be associated with a true probability after a careful calibration  \citep{10.1214/17-EJS1338SI, pmlr-v54-kull17a}. Calibration of these scores can be done by applying a transformation to their values. For our work, we decided to apply a Beta transformation\footnote{Beta transformation functions have the general form $\mu_{beta}(S;a,b,c) = 1/\left(1 + 1 / \left(e^{c} \frac{S^{a}}{(1 - S)^{b}}\right)\right)$, with $S$ being the score from the classifier and $a,b,c$, free parameters to be optimised.}. This type of transformation allows us to re-distribute the scores of an uncalibrated classifier allowing them to get closer to the definition of probability. Further details of the calibration process are given in the Appendix~\ref{app:calibration_models}.

Given that we need to be able to compare the results from the training and application of the ML models with values obtained independently (i.e. ground truth), we divided our dataset into labelled and unlabelled sources. Labelled sources are all elements of our catalogue that have been classified as either AGNs or galaxies. Unlabelled sources are those which lack such classification and that will only be subject to the prediction of our models, not taking part in any training step.

\begin{figure}
  \centering
  \begin{minipage}{0.65\columnwidth}
    \centering
    \includegraphics[width=0.99\textwidth]{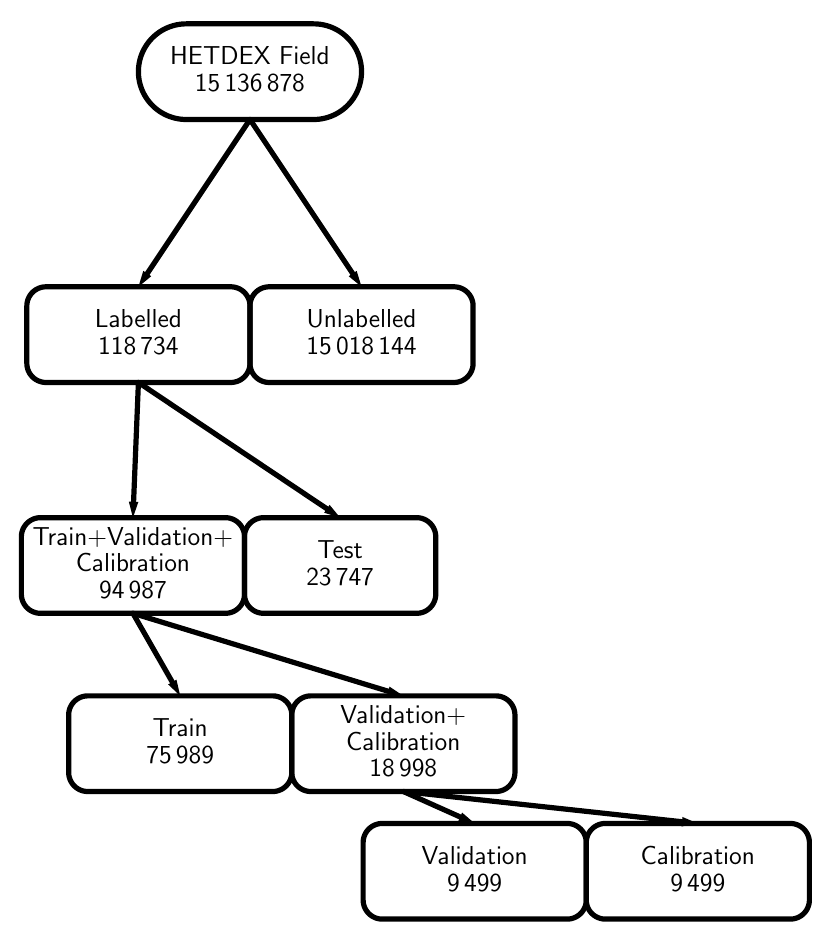}\hfill\break
    {(a) HETDEX field}
  \end{minipage}
  \hfill
  \begin{minipage}{0.34\columnwidth}
    \centering
    \includegraphics[width=0.99\textwidth]{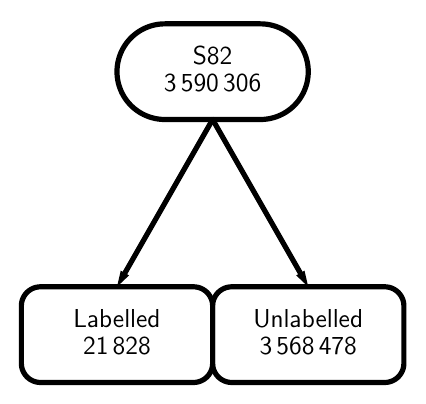}
    \vspace{3.6cm}\hfill\break
    {(b) S82 field}
  \end{minipage}
  \caption{Composition of datasets used for the different steps of this work. (a) HETDEX field. (b) S82.}
  \label{fig:dataset_sizes}
\end{figure}

Before any calculation or transformation is applied to the data from the HETDEX field, we split the labelled dataset into training, validation, calibration, and testing subsets. The early creation of these subsets helps avoid information leakage from the test subset into the models. Initially, a $20 \%$ of the dataset has been reserved as testing data. Of the remaining elements, an $80 \%$ of them have been used for training, and the rest of the data has been divided equally between calibration and validation subsets (i.e. a $10 \%$ each). The splitting process and the number of elements for each subset are shown in Fig.~\ref{fig:dataset_sizes}. Depending on the model, the needed sources are selected from each of the subsets that have been already created. The training set will be used to select algorithms for each step and to optimise their hyperparameters. The inclusion of the validation subset helps in the parameter optimisation of the models. The probability calibration of the trained model is performed over the calibration subset and, finally, the completed models are tested on the test subset. The use of these subsets will be expanded in Sects.~\ref{sec:model_selection} and \ref{sec:models_training}.

All the following transformations (feature selection, standardisation, and power transform of features) were applied to the training and validation subsets before the training of the algorithms and models. 
The calibration and testing subsets were subject to the same transformations after the modelling stage.

\subsection{Feature selection}\label{sec:feat_selection}

Machine learning algorithms, as with most data analysis tools, require execution times which increase at least linearly with the size of the datasets. In order to reduce training times without losing relevant information for the model, the most important features were selected at each step through a process called feature selection. 

To avoid redundancy, the process starts discarding features that have a high correlation with another property of the dataset. For discarding features, we calculated Pearson's correlation matrix for the full train+validation dataset only and selected the pairs of features that showed a correlation factor higher than $\rho = 0.75$, in absolute values\footnote{A value of $\rho = 0.75$ is a compromise between stringent thresholds (e.g. $\rho = 0.5$) and more relaxed values (e.g. $\rho \approx 0.9$). For an explanation on the selection of correlation values, see, for instance \citet{Ratner2009}.}. From each pair, we discarded the feature with the lowest relative standard deviation \citep[RSD;][]{johnson1964statistics}. The RSD is defined as the ratio between the standard deviation of a set and its mean value. A feature which covers a small portion of its probable values (i.e. low coverage of parameter space, and lower RSD) will give less information to a model than one with largely spread values.

For each model, the process of feature selection begins with $79$ base features and three targets (\verb|class|, \verb|LOFAR_detect|, and \verb|Z|). Feature selection is run, independently, for each trained model (i.e. AGN-galaxy classification, radio detection, and redshift predictions), delivering three different sets of features.

\subsection{Metrics}\label{sec:metrics}

A set of metrics will be used to understand the reliability of the results and put them in context with results in the literature. 
Since our work includes the use of classification and regression models, we briefly discuss the appropriate metrics in the following sections.

\subsubsection{Classification metrics}\label{sec:metrics_classfication}

The main tool to assess the performance of classification methods is the confusion (or error) matrix. It is a two-dimension (predicted vs. true) matrix where the true and predicted class(es) are compared and results stored in cells with the rate of true positives (TP), true negative (TN), false positives (FP), and false gegatives (FN). As mentioned earlier in Sect.~\ref{sec:ML_training}, we seek to maximise the number of positive-class sources that are recovered as such. Using the elements of the confusion matrix, this aim can be translated into the maximisation of TP and, consequently, the minimisation of FN.

From the elements of the confusion matrix, we can obtain additional metrics, such as the F1 and $\mathrm{F}_{\beta}$ scores \citep{10.2307/1932409, sorenson1948method, van1979information}, and the Matthews correlation coefficient \citep[MCC;][]{10.2307/2340126, nla.cat-vn81100, MATTHEWS1975442} which are better suited for unbalanced data as they take into account the behaviour and correlations among all elements of the confusion matrix.
As such, the F1 coefficient is defined as the following: 

\begin{equation}\label{eq:f1}
\mathrm{F1} = \frac{2 \mathrm{TP}}{2 \mathrm{TP} + \mathrm{FN} + \mathrm{FP}}\,.
\end{equation}

\noindent 
The values of F1 can go from $0$ (no prediction of positive instances) to $1$ (perfect prediction of elements with positive labels). This definition assigns equal weight (importance) to both the number of FN and FP. An extension to the F1 score, which adds a non-negative parameter, $\beta$, to increase, the importance given to each one of them is the F score ($\mathrm{F}_{\beta}$), defined as follows:

\begin{equation}\label{eq:f_beta}
\mathrm{F}_{\beta} = \frac{(1 + \beta^{2}) \times \mathrm{TP}}{(1 + \beta^{2}) \times \mathrm{TP} + \beta^{2} \times \mathrm{FN} + \mathrm{FP}}\,.
\end{equation}

Using ${\beta > 1}$, more relevance is given to the optimisation of FN. When ${0 \leq \beta < 1}$, the optimisation of FP is more relevant. If $\beta = 1$, the initial definition of F1 is recovered. As with F1, $\mathrm{F}_{\beta}$ values can be in the range ${[0, 1]}$. As we seek to minimise the number of FN detection, we adopt a conservative value of ${\beta = 1.1}$, giving more significance to their reduction without removing the aim for FP. Also, this value is close enough to $\beta = 1$, which will allow us to compare our scores to those produced in previous works.

The MCC metric is defined as

\begin{equation}\label{eq:mcc}
\mathrm{MCC} = \frac{\mathrm{TP} \times \mathrm{TN} - \mathrm{FP} \times \mathrm{FN}}{\sqrt{(\mathrm{TP} + \mathrm{FP}) (\mathrm{TP} + \mathrm{FN}) (\mathrm{TN} + \mathrm{FP}) (\mathrm{TN} + \mathrm{FN})}}\,,
\end{equation}

\noindent which includes also the information about the TN elements. The values of MCC can range from $-1$ (total disagreement between true and predicted values) to $+1$ (perfect prediction) with $0$ representing a prediction analogous to a random guess.

The recall, or true positive rate \citep[TPR, also called completeness, or sensitivity;][]{10.2307/4586294} corresponds to the rate of relevant, or correct, elements that have been recovered by a process. Using the elements from the confusion matrix, it can be defined as the following:

\begin{equation}\label{eq:recall}
\mathrm{recall} = \mathrm{TPR} = \frac{\mathrm{TP}}{\mathrm{TP} + \mathrm{FN}}\,,
\end{equation}

\noindent and it can go from $0$ to $1$, with a value of $1$ meaning that the model can recover all the true instances.

The last metric used is precision (also known as purity), which can be defined as the ratio between the number of correctly classified elements and the number of sources in the positive class (AGN or radio detectable): 

\begin{equation}\label{eq:precision}
\mathrm{precision} = \frac{\mathrm{TP}}{\mathrm{TP} + \mathrm{FP}}\,,
\end{equation}

\noindent and their values can range from $0$ to $1$ where higher values show that more real positive instances of the studied set were retrieved as such by the model.

In order to establish a baseline from which the aforementioned metrics can be assessed, it is possible to obtain them in the case of a random, or no-skill prediction. Following, for instance, the derivations and notation from \citet{https://doi.org/10.1111/2041-210X.14071}, no-skill versions of classification metrics (Eqs.~\ref{eq:f_beta}--\ref{eq:precision}) are the following:

\begin{align}
\mathrm{F}_{\beta}^{\mathrm{no{-}skill}} &= p\,,\label{eq:no_skill_Fb}\\
\mathrm{MCC}^{\mathrm{no{-}skill}} &= 0\,,\label{eq:no_skill_MCC}\\
\mathrm{recall}^{\mathrm{no{-}skill}} &= p\,,\label{eq:no_skill_recall}\\
\mathrm{precision}^{\mathrm{no{-}skill}} &= p\,,\label{eq:no_skill_precision}
\end{align}

\noindent where $p$ corresponds to the ratio between the elements of the positive class and the total number of elements involved in the prediction.

\subsubsection{Regression metrics}\label{sec:metrics_regression}

For the case of individual redshift value determination, two commonly used metrics are the difference between predicted and true redshift,

\begin{equation}
\Delta z = z_{\mathrm{True}} - z_{\mathrm{Predicted}}\,,
\end{equation}

\noindent and its normalised difference,

\begin{equation}\label{eq:delta_z_N}
\Delta z^{\mathrm{N}} = \frac{z_{\mathrm{True}} - z_{\mathrm{Predicted}}}{1 + z_{\mathrm{True}}}\,.
\end{equation}

\noindent If the comparison is made over a larger sample of elements, the bias of the redshift is used \citep{2013ApJ...775...93D}, with the median of the quantities instead of its mean to avoid the strong influence of extreme values. This bias can be written as

\begin{align}
\Delta z_{\mathrm{Total}} &= \mathrm{median}\left(z_{\mathrm{True}} - z_{\mathrm{Predicted}}\right) = \mathrm{median}(\Delta z)\,,\\
\Delta z_{\mathrm{Total}}^{\mathrm{N}} &= \mathrm{median}\left(\frac{z_{\mathrm{True}} - z_{\mathrm{Predicted}}}{1 + z_{\mathrm{True}}}\right) = \mathrm{median}(\Delta z^{\mathrm{N}})\,.
\end{align}

Using the previous definitions, four additional metrics can be calculated. These are the median absolute deviation (MAD, $\sigma_{\mathrm{MAD}}$) and normalised median absolute deviation \citep[NMAD, $\sigma_{\mathrm{NMAD}}$;][]{hoaglin1983understanding, 2009ApJ...690.1236I}, which are less sensitive to outliers. Also, the standard deviation of the predictions, $\sigma_{z}$, and its normalised version, $\sigma_{z}^{\mathrm{N}}$ are typically used. They are defined as

\begin{align}
\sigma_{\mathrm{MAD}} &= 1.48 \times \mathrm{median}\left(|\Delta z|\right)\,,\\
\sigma_{\mathrm{NMAD}} &= 1.48 \times \mathrm{median}\left(\left|\Delta z^{\mathrm{N}}\right|\right)\,,\\
\sigma_{z} &= \sqrt{\frac{1}{\mathrm{d}} \sum_{i}^{\mathrm{d}} \left(\Delta z\right)^{2}}\,,\\
\sigma_{z}^{\mathrm{N}} &= \sqrt{\frac{1}{\mathrm{d}} \sum_{i}^{\mathrm{d}} \left(\Delta z^{\mathrm{N}}\right)^{2}}\,,
\end{align}

\noindent with d being the number of elements in the studied sample (i.e. its size).

Also, the outlier fraction \citep[$\eta$, as used in][]{2013ApJ...775...93D, 2022A&C....3800510L} is considered, which is defined as the fraction sources with a predicted redshift difference ($\left|\Delta z^{\mathrm{N}}\right|$, Eq.~\ref{eq:delta_z_N}) larger than a previously set value. Taking the results from \citet{2009ApJ...690.1236I} and \citet{2010A&A...523A..31H}, we selected this threshold to be $0.15$, leaving the definition of the outlier fraction as follows:

\begin{equation}\label{eq:outlier_fraction}
\eta = \frac{\# \left( \left|\Delta z^{\mathrm{N}}\right| > 0.15 \right)}{d}\,,
\end{equation}

\noindent where $\#$ symbolises the number of sources fulfilling the described relation, and $d$ corresponds to the size of the selected sample.

\subsubsection{Calibration metrics}\label{sec:metrics_calibration}

One of the most used analytical metrics to assess calibration of a model is the Brier score \citep[BS;][]{Brier_1950}. It measures the mean square difference between the predicted probability of an element and its true class. If the total number of elements in the studied sample is $d$, the BS can be written (for binary classification problems, as the ones studied in this work) as

\begin{equation}\label{eq:brier_score}
\mathrm{BS} = \frac{1}{d} \sum_{i}^{d}(\mathbb{C} - \mathtt{class})^{2}\,,
\end{equation}

\noindent where $\mathbb{C}$ is the predicted class and \texttt{class} corresponds the true class of each of the elements in the sample ($0$ or $1$).
The BS can range between $0$ and $1$ with $0$ representing a model that is completely reliable in its predictions. Additionally, the BS can be used to compare the reliability (or calibration) between a model and a reference using the Brier skill score \citep[BSS; e.g.][]{Glahn_1970}, which can be defined as the following:

\begin{equation}\label{eq:brier_skill_score}
\mathrm{BSS} = 1 - \frac{\mathrm{BS}}{\mathrm{BS}_{\mathrm{ref}}}\,.
\end{equation}

In our case, $\mathrm{BS}_{\mathrm{ref}}$ corresponds to the value calculated from the uncalibrated model. The BSS can take values between $-1$ and $+1$. The closer the BSS gets to $1$, the more reliable the analysed model is. These values include the case where ${\mathrm{BSS} {\approx} 0}$, in which both models perform similarly in terms of calibration.

For our pipeline, after a model has been fully trained, a calibrated version of their scores will be obtained. With both of them, the BSS will be calculated and, if it is not much lower than $0$, that calibrated transformation will be used as the final scores from the prediction.

\subsection{Model selection}\label{sec:model_selection}

By design, each ML algorithm has been developed and tuned to work better with certain data conditions. For instance, balance of target categories and ranges of base features. 
The predicting power of different algorithms can be combined with the use of meta learners \citep{Vanschoren2019}. Meta learners use the properties or predictions from other algorithms (base learners) as additional information during their training stages. A simple implementation of this procedure is called generalised stacking \citep{WOLPERT1992241} which can be interpreted
as the addition of priors to the model training stage. Generalised stacking has been applied in several astrophysical problems. That is the case of \citet{2016MNRAS.460.3152Z}, \citet{2022A&A...666A..87C}, and \citet{2023MNRAS.520.3529E}, \citet{2023A&A...671A..99E}.

Base and meta learners have been selected based upon the metrics described in Sect.~\ref{sec:metrics}. We trained five algorithms with the training subset and calculated the metrics for all of them using a $10$-fold cross-validation approach \citep[e.g.][]{https://doi.org/10.1111/j.2517-6161.1974.tb00994.x, doi:10.1080/00401706.1974.10489157} over the same training subset. For each metric, the learners have been given a rank (from $1$ to $5$) and a mean value has been obtained from them. Out of the analysed algorithms, the one with the best overall performance (i.e. best mean rank) is selected to be the meta learner while the remaining four are used as base learners.

For the AGN-galaxy classification and radio detection problems, we tested five classification algorithms: Random Forest \citep[\texttt{RF};][]{Breiman2001}, Gradient Boosting Classifier \citep[\texttt{GBC};][]{10.1214/aos/1013203451}, Extra Trees \citep[\texttt{ET};][]{Geurts2006}, Extreme Gradient Boosting \citep[\texttt{XGBoost}, \texttt{v1.5.1};][]{Chen:2016:XST:2939672.2939785}, and Category Boosting \citep[\texttt{CatBoost}, \texttt{v1.0.5};][]{NEURIPS2018_14491b75, DBLP:journals/corr/abs-1810-11363}.
For the redshift prediction problem, we tested five regressors as well: \texttt{RF}, \texttt{ET}, \texttt{XGBoost}, \texttt{CatBoost}, and Gradient Boosting Regressor \citep[\texttt{GBR};][]{10.1214/aos/1013203451}.
We used the Python implementations of these algorithms and, in particular for \texttt{RF}, \texttt{ET}, \texttt{GBC}, and \texttt{GBR}, the versions offered by the package \texttt{scikit-learn}\footnote{\url{https://scikit-learn.org}} \citep[\texttt{v0.23.2};][]{scikit-learn}. These algorithms were selected given that they offer tools to interpret the global and local influence of the input features in the training and predictions (cf. Sects.~\ref{sec:introduction} and \ref{sec:model_explain}).

All the algorithms selected for this work fall into the broad family of tree-based models. Forest models (\texttt{RF} and \texttt{ET}) rely on a collection of decision trees to, after applying a majority vote, predict either a class or a continuum value. Each of these decision trees uses a different, randomly selected subset of features to make a decision on the training set \citep{Breiman2001}. Opposite to forests, gradient boosting models (\texttt{GBC}, \texttt{GBR}, \texttt{XGBoost} and \texttt{CatBoost}) apply decision trees sequentially to improve the quality of the previous predictions \citep{10.1214/aos/1013203451, FRIEDMAN2002367}.

\subsection{Training of models}\label{sec:models_training}

The procedure described in Sect.~\ref{sec:model_selection} includes an initial fit of the selected algorithms to the training data (including the selected features) to optimise their parameters. The stacking step includes a new optimisation of the parameters of the meta learner using $10$-fold cross-validation on the training data with the addition of the output from the base learners, which are treated as regular features. Then, the hyperparameters of the stacked models are optimised over the training subset (a brief description of this step is presented in Appendix~\ref{sec:app_hyperpars}).

The final step involves a last parameter fitting instance but using, this time, the combined train+validation subset, which includes the output of the base algorithms, to ensure wider coverage of the parameter space and better-performing models. Consequently, only the testing set is available for assessing the quality of the predictions made by the models.

\subsection{Probability calibration}\label{sec:prob_calibration}

The calibration procedure was performed in the calibration subset. In this way, we avoid influencing the process with information from the training and validation steps. A broader description of the calibration process and the results obtained for our models are presented in Appendix~\ref{app:calibration_models}. Thus, from this point onwards, and with the sole exception of some of the outcomes shown in Sect.~\ref{sec:model_explain}, all results from classifications will be based on the calibrated probabilities.

\subsection{Optimisation of classification thresholds}\label{sec:threshold_opt}

As mentioned in the first paragraphs of Sect.~\ref{sec:ML_training}, classification models deliver a range of probabilities for which a threshold is needed to separate their predictions between negative and positive classes. By default, these models set a threshold at $0.5$ in score\footnote{Throughout this work, we call this a naive threshold.} but, in principle, and given the characteristics of the problem, a different optimal threshold might be needed.

In our case, we want to optimise (increase) the number of recovered elements in each model (i.e. AGNs or radio-detectable sources). This maximisation corresponds to obtaining thresholds that optimise the recall given a specific precision limit. We did that with the use of the statistical tool called precision-recall (PR) curve. A deeper description of this method and the results obtained from our work are presented in Appendix~\ref{sec:app_pr_curve}\footnote{Thresholds derived from the PR curves are labelled as PR.}.

%----------------------------------------------------------------

\section{Results}\label{sec:results}
In the present section, we report the results from the training of the different models in the HETDEX field. All metrics are evaluated using the testing subset. The metrics are also computed on labelled AGNs in the S82 field. As no training is done on S82 data, it offers a way to test the validity of the pipeline on data that, despite having similar optical-to-NIR photometric properties, presents distinct radio information and location in the sky.

The three models are chained afterwards in sequential mode to create a pipeline, and related metrics, for the prediction of radio-AGN activity. Novel predictions were obtained from the application of such pipeline to unlabelled sources from both the HETDEX and S82 fields. 

\subsection{AGN-galaxy classification}\label{sec:results_agn}
Feature selection was applied to the train+validation subset with $85\,488$ confirmed elements (galaxies from SDSS DR16 and AGNs from MQC, i.e. \texttt{class == 0} or \texttt{class == 1}). After the selection procedure described in Sect.~\ref{sec:feat_selection}, $18$ features were selected for training: \verb|band_num|, \verb|W4mag|, \verb|g_r|, \verb|r_i|, \verb|r_J|, \verb|i_z|, \verb|i_y|, \verb|z_y|, \verb|z_W2|, \verb|y_J|, \verb|y_W1|, \verb|y_W2|, \verb|J_H|, \verb|H_K|, \verb|H_W3|, \verb|W1_W2|, \verb|W1_W3|, and \verb|W3_W4|. The target feature is \verb|class|.

The results of model testing for the AGN-galaxy classification are reported in Table~\ref{table:fit_AGN_models}. The \verb|CatBoost| algorithm provides the best metric values (highest mean rank) and is therefore selected as the  meta model. \verb|XGBoost|, \verb|RF|, \verb|ET|, and \verb|GBC| were used as base learners.

\begin{table}
  \setlength{\tabcolsep}{2pt}
  \caption{Best performing models for the AGN-galaxy classification}             % title of Table
  \label{table:fit_AGN_models}      % is used to refer this table in the text
  \centering                          % used for centering table
  \resizebox{0.90\columnwidth}{!}{
  \begin{tabular}{c c c c c c}        % centered columns (6 columns)
  \hline\hline                 % inserts double horizontal lines   
  Model             & F$_{\beta}$         & MCC                 & Precision           & Recall              & Rank  \\
                    & $(\times 100)$      & $(\times 100)$      & $(\times 100)$      & $(\times 100)$      &       \\
  \hline
  \texttt{CatBoost} & $95.70 {\pm} 0.28$  & $92.46 {\pm} 0.48$  & $95.45 {\pm} 0.32$  & $95.91 {\pm} 0.37$  & 1.00  \\
  \texttt{XGBoost}  & $95.67 {\pm} 0.27$  & $92.40 {\pm} 0.48$  & $95.41 {\pm} 0.39$  & $95.88 {\pm} 0.34$  & 2.00  \\
  \texttt{RF}       & $95.52 {\pm} 0.36$  & $92.14 {\pm} 0.63$  & $95.28 {\pm} 0.46$  & $95.71 {\pm} 0.40$  & 3.00  \\
  \texttt{ET}       & $95.40 {\pm} 0.40$  & $91.94 {\pm} 0.69$  & $95.13 {\pm} 0.43$  & $95.63 {\pm} 0.47$  & 4.00  \\
  \texttt{GBC}      & $95.26 {\pm} 0.31$  & $91.66 {\pm} 0.54$  & $94.82 {\pm} 0.41$  & $95.63 {\pm} 0.35$  & 5.00  \\
  \hline                                   %inserts single line
  \end{tabular}
  }
  \tablefoot{Metrics obtained using the default probability threshold of $0.5$.\\
  Algorithms are sorted by decreasing recall values.\\
   For display purposes, all metrics have been multiplied by $100$.\\ 
   Uncertainties show standard deviation of metrics obtained across all $10$ training folds (cf. Sect.~\ref{sec:model_selection}).
  }
  \end{table}

The optimisation of the PR curve for the calibrated predictor provides an optimal threshold for this algorithm of $0.34895$. This value was used for the AGN-galaxy model throughout this work.

\begin{figure}
  \centering
  \begin{minipage}{0.49\columnwidth}
    \centering
    \includegraphics[width=0.9\textwidth]{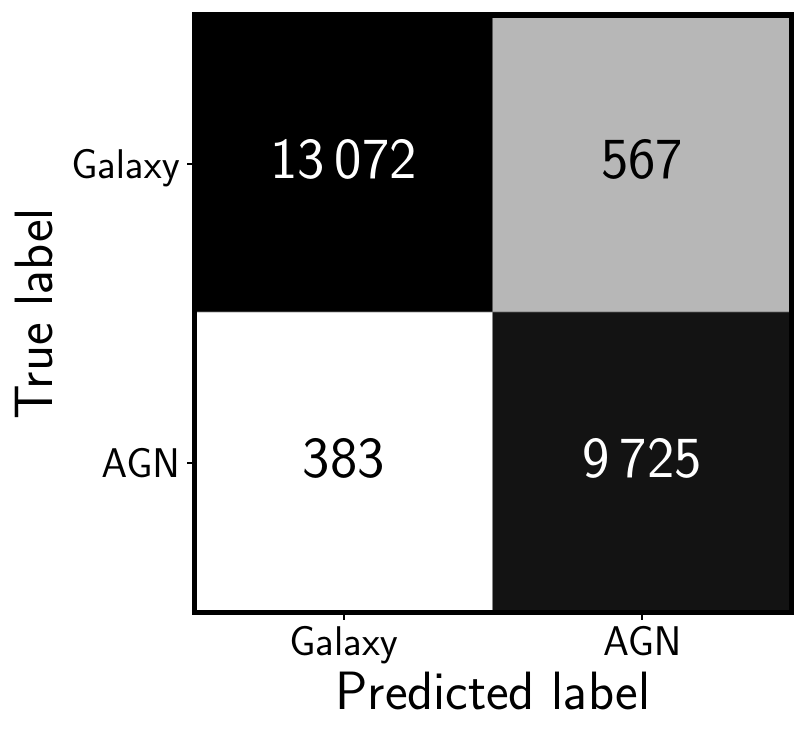}\hfill\break
    {(a) AGN-galaxy classification}
  \end{minipage}%\\%
  \begin{minipage}{0.49\columnwidth}
    \centering
    \includegraphics[width=0.9\textwidth]{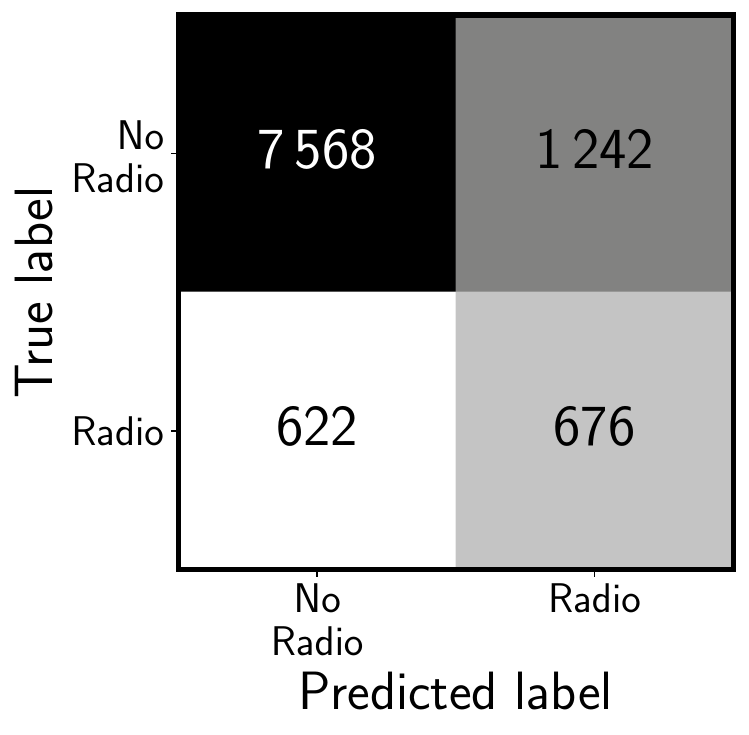}\hfill\break
    {(b) Radio detection on AGNs}
  \end{minipage}\hfill\break%\\%
  \begin{minipage}{0.70\columnwidth}
    \centering
    \includegraphics[width=0.9\textwidth]{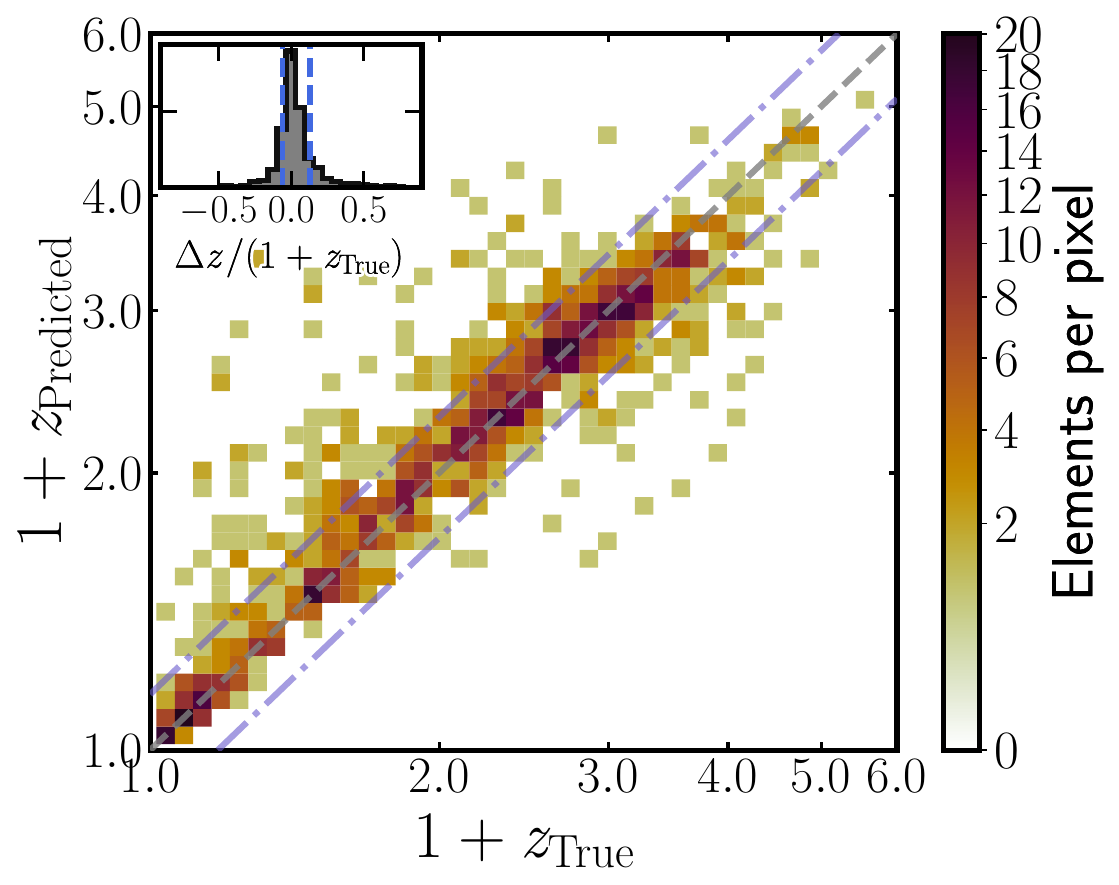}\hfill\break
    {(c) Redshift on radio AGNs}
  \end{minipage}%
  \caption{Performance of individual models (AGN-galaxy classification, radio-detectability classification and redshift regression) when applied to the HETDEX test subset. (a): confusion matrix for AGN-galaxy classification. 
  (b): Same as (a), but for radio detection. (c): Density plot comparison between original and the predicted redshifts. 
  Grey, dashed line shows the 1:1 relation while dot-dashed lines show the limits for outliers (cf. Eq.~\ref{eq:outlier_fraction}). Inset displays the distribution of $\Delta z^{\mathrm{N}}$ with a ${{<}\Delta z^{\mathrm{N}}{>} = 0.0442}$.}
  \label{fig:results_models_test}
\end{figure}

The results of the application of the stacked and calibrated model for the testing subset and the labelled sources in S82 are presented in Table~\ref{table:fit_AGN_results}. The metrics are shown for the use of two different thresholds, the naive value of $0.5$ and the  PR-derived value of $0.34895$. The confusion matrix (calculated on the testing dataset) is shown in the upper left panel of Fig.~\ref{fig:results_models_test}. 

Overall, the model is able to separate AGNs from galaxies with a very high (recall ${\gtrsim} 94\%$) success rate. A comparison with traditional colour-colour criteria for AGNs selection is presented in Sect.~\ref{sec:previous_AGN_detection}. In particular, Table~\ref{table:previous_AGN_methods} displays metrics for such criteria. Our classification model can recover, in the HETDEX field, $15\%$ and $59\%$ more AGNs than said formulae. In the S82 field, these differences range between $17\%$ and $61\%$. Such differences highlight the fact that most of the information that separates AGNs from galaxies is traced by the selected features (mostly colours). Also, the increase in the recovery rate underlines the importance of using photometric information from several bands for such task, as opposed to traditional colour-colour criteria.

\begin{table}
\setlength{\tabcolsep}{2pt}
\caption{Resulting metrics of AGN-galaxy classification model for the test subset and the labelled sources in S82 using two different threshold values, as described in Sect.~\ref{sec:results_agn}. HETDEX and S82 pipeline results are described in Sect.~\ref{sec:results_prediction_pipeline}.}             % title of Table
\label{table:fit_AGN_results}      % is used to refer this table in the text
\centering                          % used for centering table
\resizebox{0.99\columnwidth}{!}{
\begin{tabular}{c c c c c c}        % centered columns (6 columns)
\hline\hline                 % inserts double horizontal lines   
Subset                        & Threshold & F$_{\beta}$           & MCC                   & Precision             & Recall                \\
                              &           & $(\times 100)$        & $(\times 100)$        & $(\times 100)$        & $(\times 100)$        \\
\hline
\multirow{2}{*}{HETDEX-test}  & Naive     & $95.37 {\pm} 0.36$ & $91.81 {\pm} 0.67$ & $97.47 {\pm} 0.69$ & $95.89 {\pm} 2.27$ \\
                              & PR        & $95.42 {\pm} 0.38$ & $91.85 {\pm} 0.70$ & $94.49 {\pm} 0.65$ & $96.21 {\pm} 0.43$ \\[0.1em]
\multirow{2}{*}{S82-label}    & Naive     & $94.15 {\pm} 0.44$ & $70.54 {\pm} 2.02$ & $95.16 {\pm} 0.41$ & $93.33 {\pm} 0.66$ \\
                              & PR        & $94.37 {\pm} 0.36$ & $70.67 {\pm} 1.72$ & $94.81 {\pm} 0.40$ & $94.01 {\pm} 0.59$ \\[0.1em]
\multirow{2}{*}{HETDEX-pipe}  & Naive     & $95.37 {\pm} 0.36$ & $91.81 {\pm} 0.67$ & $97.47 {\pm} 0.69$ & $95.89 {\pm} 2.27$ \\
                              & PR        & $95.42 {\pm} 0.38$ & $91.85 {\pm} 0.70$ & $94.49 {\pm} 0.65$ & $96.21 {\pm} 0.43$ \\[0.1em]
\multirow{2}{*}{S82-pipe}     & Naive     & $94.15 {\pm} 0.44$ & $70.54 {\pm} 2.02$ & $95.16 {\pm} 0.41$ & $93.33 {\pm} 0.66$ \\
                              & PR        & $94.37 {\pm} 0.36$ & $70.67 {\pm} 1.72$ & $94.81 {\pm} 0.40$ & $94.01 {\pm} 0.59$ \\
\hline                                   %inserts single line
\end{tabular}
}
\tablefoot{All metrics have been multiplied by $100$.\\ 
Uncertainties show standard deviation of metrics obtained across all $10$ training folds (cf. Sect.~\ref{sec:model_selection}).
}
\end{table}

\subsection{Radio detection}\label{sec:results_radio}

Training of the radio detection model was applied only to sources confirmed to be AGN (\texttt{class == 1}).
Feature selection was applied to the train+validation subset, with $36\,387$ confirmed AGNs. 
The target feature is \verb|LOFAR_detect| and the base of selected features are:  \verb|band_num|, \verb|W4mag|, \verb|g_r|, \verb|g_i|, \verb|r_i|, \verb|r_z|, \verb|i_z|, \verb|z_y|, \verb|z_W1|, \verb|y_J|, \verb|y_W1|, \verb|J_H|, \verb|H_K|, \verb|K_W3|, \verb|K_W4|, \verb|W1_W2|, and \verb|W2_W3|.

The performance of the tested algorithms is shown in Table~\ref{table:fit_radio_models}. 
In this case, \verb|GBC| shows the highest mean rank. For this reason, we used it as the meta learner and \verb|XGBoost|, \verb|CatBoost|, \verb|RF|, and \verb|ET| were selected as base learners. The optimal threshold for this model is found to be ${\sim}0.20460$. Finally, the stacked model metrics and confusion matrix are shown in Table~\ref{table:fit_radio_results}, for PR-optimised and naive thresholds, and in Fig.~\ref{fig:results_models_test} respectively.

\begin{table}
\setlength{\tabcolsep}{2pt} 
\caption{Best performing models the radio detection classification.}             % title of Table
\label{table:fit_radio_models}      % is used to refer this table in the text
\centering                          % used for centering table
\resizebox{0.90\columnwidth}{!}{
\begin{tabular}{c c c c c c}        % centered columns (6 columns)
\hline\hline                 % inserts double horizontal lines   
Model             & F$_{\beta}$         & MCC                 & Precision           & Recall              & Rank  \\
                  & $(\times 100)$      & $(\times 100)$      & $(\times 100)$      & $(\times 100)$      &       \\
\hline
\texttt{XGBoost}  & $29.98 {\pm} 2.29$  & $29.81 {\pm} 2.17$  & $56.74 {\pm} 2.93$  & $21.61 {\pm} 2.00$  & 2.75      \\
\texttt{CatBoost} & $29.57 {\pm} 1.62$  & $30.56 {\pm} 1.71$  & $60.10 {\pm} 2.85$  & $20.85 {\pm} 1.36$  & 2.25      \\
\texttt{GBC}      & $29.60 {\pm} 1.66$  & $31.31 {\pm} 1.93$  & $62.55 {\pm} 3.95$  & $20.66 {\pm} 1.40$  & 1.75      \\
\texttt{RF}       & $29.16 {\pm} 2.47$  & $30.26 {\pm} 2.65$  & $60.03 {\pm} 3.73$  & $20.48 {\pm} 1.96$  & 3.75      \\
\texttt{ET}       & $28.40 {\pm} 1.27$  & $29.73 {\pm} 1.47$  & $60.06 {\pm} 2.85$  & $19.80 {\pm} 1.05$  & 4.50      \\
\hline                                   %inserts single line
\end{tabular}
}
\tablefoot{Values and uncertainties as in Table~\ref{table:fit_AGN_models}.}
\end{table}

\begin{table}
\setlength{\tabcolsep}{2pt}
\caption{Resulting metrics of the radio detection model on the test subset and the labelled sources in S82 using two different threshold values, as explained in Sect.~\ref{sec:results_radio}. HETDEX and S82 pipeline results shown as part of the discussion in Sect.~\ref{sec:results_prediction_pipeline}.}             % title of Table
\label{table:fit_radio_results}      % is used to refer this table in the text
\centering                          % used for centering table
\resizebox{0.99\columnwidth}{!}{
\begin{tabular}{c c c c c c}        % centered columns (6 columns)
\hline\hline                 % inserts double horizontal lines   
Subset                        & Threshold & F$_{\beta}$           & MCC                   & Precision             & Recall                \\
                              &           & $(\times 100)$        & $(\times 100)$        & $(\times 100)$        & $(\times 100)$        \\
\hline
\multirow{2}{*}{HETDEX-test}  & Naive     & $24.87 {\pm} 2.94$ & $27.36 {\pm} 3.46$ & $60.61 {\pm} 8.18$ & $16.72 {\pm} 2.31$ \\
                              & PR        & $42.88 {\pm} 2.93$ & $32.47 {\pm} 3.49$ & $35.28 {\pm} 2.74$ & $52.16 {\pm} 3.59$ \\[0.1em]
\multirow{2}{*}{S82-label}    & Naive     & $27.15 {\pm} 2.28$ & $23.36 {\pm} 2.27$ & $25.72 {\pm} 1.91$ & $28.47 {\pm} 3.24$ \\
                              & PR        & $21.62 {\pm} 1.20$ & $19.37 {\pm} 1.64$ & $12.29 {\pm} 0.73$ & $58.16 {\pm} 3.06$ \\[0.1em]
\multirow{2}{*}{HETDEX-pipe}  & Naive     & $24.37 {\pm} 3.53$ & $26.93 {\pm} 4.18$ & $59.36 {\pm} 7.17$ & $16.38 {\pm} 2.63$ \\
                              & PR        & $41.57 {\pm} 4.16$ & $31.67 {\pm} 4.81$ & $34.65 {\pm} 3.24$ & $49.80 {\pm} 5.85$ \\[0.1em]
\multirow{2}{*}{S82-pipe}     & Naive     & $26.52 {\pm} 5.44$ & $23.29 {\pm} 5.73$ & $25.71 {\pm} 5.89$ & $27.72 {\pm} 5.21$ \\
                              & PR        & $20.19 {\pm} 2.84$ & $18.40 {\pm} 4.07$ & $11.45 {\pm} 1.58$ & $54.78 {\pm} 8.44$ \\
\hline                                   %inserts single line
\end{tabular}
}
\tablefoot{Values and uncertainties as in Table~\ref{table:fit_AGN_results}.}
\end{table}

\subsection{Redshift predictions}\label{sec:results_redshift}

The redshift value prediction model was applied to sources confirmed to be radio-detected AGN (i.e. \texttt{class~==~1} and \texttt{radio\_detect~==~1}).
Feature selection (cf. Sect.~\ref{sec:feat_selection}) was applied to the train+validation subset, with $4\,612$ sources, leading to the selection of $17$ features. The target feature is \verb|Z| and the selected base features are \verb|band_num|, \verb|W4mag|, \verb|g_r|, \verb|g_W3|, \verb|r_i|, \verb|r_z|, \verb|i_z|, \verb|i_y|, \verb|z_y|, \verb|y_J|, \verb|y_W1|, \verb|J_H|, \verb|H_K|, \verb|K_W3|, \verb|K_W4|, \verb|W1_W2|, and \verb|W2_W3|.
\\
\begin{table}
\setlength{\tabcolsep}{2pt}
\caption{Results of initial fit for redshift value prediction.}             % title of Table
\label{table:fit_redshift_models}      % is used to refer this table in the text
\centering                          % used for centering table
\resizebox{0.99\columnwidth}{!}{
\begin{tabular}{c c c c c c c}        % centered columns (7 columns)
\hline\hline                 % inserts double horizontal lines   
Model               & $\sigma_{\mathrm{MAD}}$ & $\sigma_{\mathrm{NMAD}}$  & $\sigma_{z}$        & $\sigma_{z}^{\mathrm{N}}$ & $\eta$              & Rank  \\
                    & $(\times 100)$          & $(\times 100)$            & $(\times 100)$      & $(\times 100)$            & $(\times 100)$      &       \\
\hline
\texttt{RF}         & $17.88 {\pm} 1.41$      & $07.95 {\pm} 0.50$        & $42.02 {\pm} 5.28$  & $19.38 {\pm} 2.44$        & $19.51 {\pm} 1.98$  & 2.0   \\[0.1em]
\texttt{ET}         & $18.53 {\pm} 1.03$      & $08.42 {\pm} 0.43$        & $41.12 {\pm} 4.16$  & $18.65 {\pm} 2.26$        & $19.24 {\pm} 1.16$  & 1.8   \\[0.1em]
\texttt{CatBoost}   & $21.71 {\pm} 1.38$      & $10.08 {\pm} 0.47$        & $40.35 {\pm} 3.03$  & $18.52 {\pm} 1.39$        & $21.93 {\pm} 1.55$  & 2.2   \\[0.1em]
\texttt{XGBoost}    & $22.89 {\pm} 1.05$      & $10.84 {\pm} 0.78$        & $43.14 {\pm} 3.99$  & $19.62 {\pm} 1.78$        & $24.15 {\pm} 1.84$  & 4.0   \\[0.1em]
\texttt{GBR}        & $27.73 {\pm} 1.57$      & $12.72 {\pm} 0.74$        & $44.82 {\pm} 3.80$  & $20.41 {\pm} 1.67$        & $28.67 {\pm} 2.25$  & 5.0   \\[0.1em]
\hline                                   %inserts single line
\end{tabular}
}
\tablefoot{Algorithms sorted by increasing $\sigma_{\mathrm{MAD}}$ values.\\
Uncertainties as in Table~\ref{table:fit_AGN_models}.}
\end{table}

For the redshift prediction, the tested algorithms performed as shown in Table~\ref{table:fit_redshift_models}. 
Based on their mean rank values, \verb|RF|, \verb|CatBoost|, \verb|XGBoost|, and \verb|GBR| were selected as base learners and \verb|ET| (which shows the best $\sigma_{\mathrm{MAD}}$ value of the two models with the best rank) was used as meta learner.
The redshift regression metrics of the stacked model are presented in Table~\ref{table:fit_redshift_results}. 
Likewise, the comparison between the original and predicted redshifts is shown in the lower panel of Fig.~\ref{fig:results_models_test}.

\begin{table}
\setlength{\tabcolsep}{2pt}
\caption{Redshift prediction metrics for the test subset from HETDEX and S82 labelled sources as discussed in Sect.~\ref{sec:results_prediction_pipeline}.}             % title of Table
\label{table:fit_redshift_results}      % is used to refer this table in the text
\centering                          % used for centering table
\resizebox{0.99\columnwidth}{!}{
\begin{tabular}{c c c c c c}        % centered columns (6 columns)
\hline\hline                 % inserts double horizontal lines   
Subset            & $\sigma_{\mathrm{MAD}}$ & $\sigma_{\mathrm{NMAD}}$  & $\sigma_{z}$        & $\sigma_{z}^{\mathrm{N}}$ & $\eta$              \\
                  & $(\times 100)$          & $(\times 100)$            & $(\times 100)$      & $(\times 100)$            & $(\times 100)$      \\
\hline
HETDEX-test       & $16.54 {\pm} 2.55$      & $7.27 {\pm} 0.99$         & $41.14 {\pm} 09.97$ & $20.56 {\pm} 5.98$        & $19.03 {\pm} 3.35$  \\[0.1em]
S82-label         & $18.66 {\pm} 2.26$      & $9.28 {\pm} 1.37$         & $51.08 {\pm} 11.62$ & $24.69 {\pm} 4.36$        & $24.29 {\pm} 4.68$  \\[0.1em]
HETDEX-pipe-Naive & $08.11 {\pm} 3.95$      & $5.42 {\pm} 2.19$         & $32.00 {\pm} 12.27$ & $20.97 {\pm} 9.69$        & $19.01 {\pm} 8.22$  \\[0.1em]
HETDEX-pipe-PR    & $15.86 {\pm} 1.77$      & $7.17 {\pm} 0.81$         & $37.80 {\pm} 03.06$ & $22.93 {\pm} 2.73$        & $18.91 {\pm} 1.59$  \\[0.1em]
S82-pipe-Naive    & $15.17 {\pm} 2.70$      & $9.14 {\pm} 1.23$         & $43.05 {\pm} 07.20$ & $24.32 {\pm} 5.00$        & $24.09 {\pm} 4.52$  \\[0.1em]
S82-pipe-PR       & $20.71 {\pm} 1.23$      & $9.84 {\pm} 0.56$         & $45.14 {\pm} 04.42$ & $26.14 {\pm} 3.77$        & $25.18 {\pm} 2.26$  \\[0.1em]
\hline                                   %inserts single line
\end{tabular}
}
\tablefoot{Values and uncertainties as in Table~\ref{table:fit_AGN_results}.}
\end{table}

\subsection{Prediction pipeline}\label{sec:results_prediction_pipeline}

The sequential combination of the models described in Sect.~\ref{sec:ML_training} defines the pipeline for the prediction of radio-detectable AGNs and their redshift. As separate tasks, the pipeline was applied to the labelled sources in the HETDEX testing subset, to the labelled sources in S82, and to all unlabelled sources across both fields. S82 provides an independent test of the pipeline as no data in this field was used for training the different models. A full candidate catalogue is extracted from this exercise and based on the unlabelled datasets.

\begin{figure*}
  \centering
  \begin{minipage}{0.20\textwidth}
    \centering
    \includegraphics[width=0.99\textwidth]{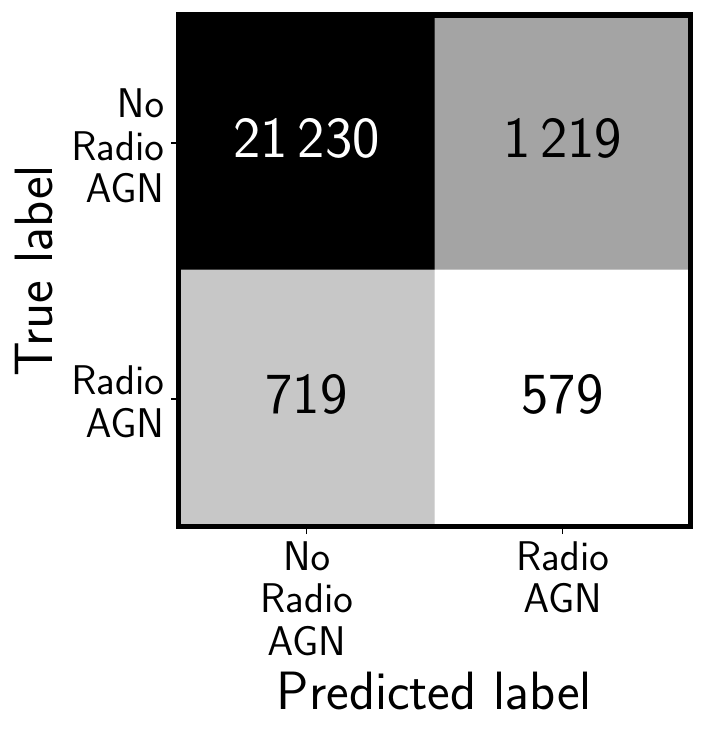}\hfill\break
    {(a) Radio-AGN confusion matrix}
  \end{minipage}%\\%
      \centering
  \begin{minipage}{0.30\textwidth}
    \centering
    \includegraphics[width=0.99\textwidth]{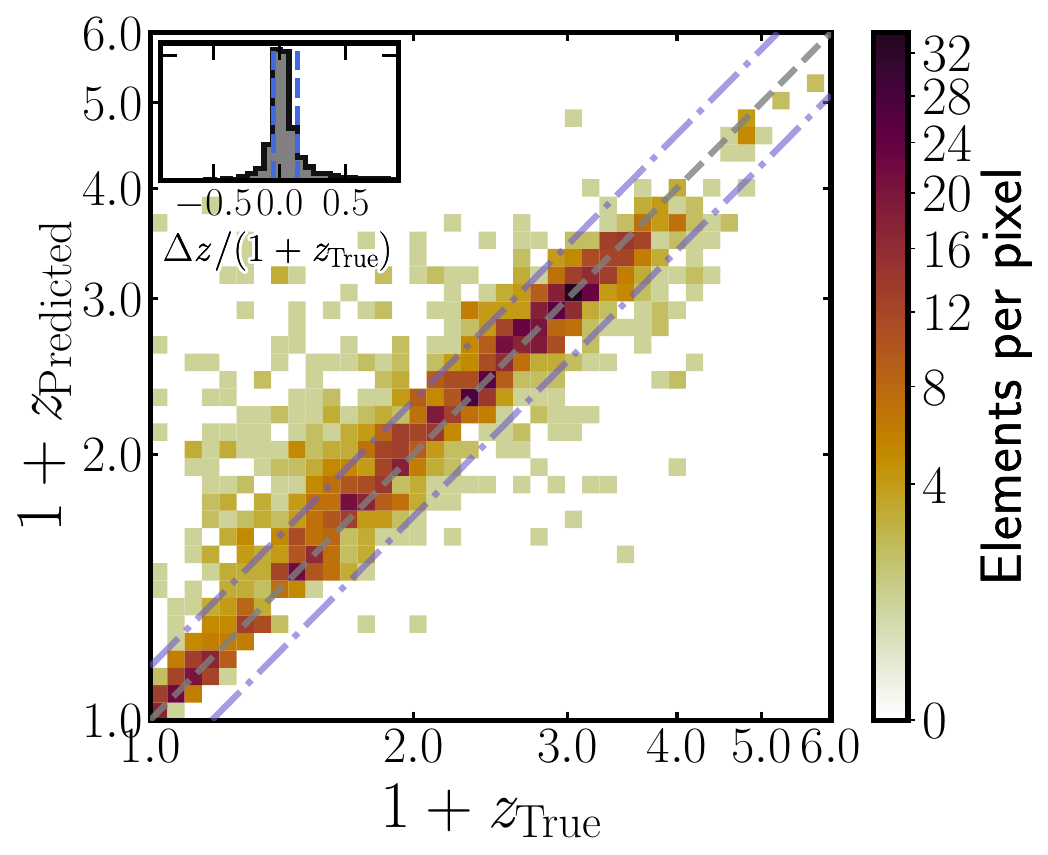}\hfill\break
    {(b) Predicted-True $z$ comparison}
  \end{minipage}%\\%
  \begin{minipage}{0.20\textwidth}
    \centering
    \includegraphics[width=0.99\textwidth]{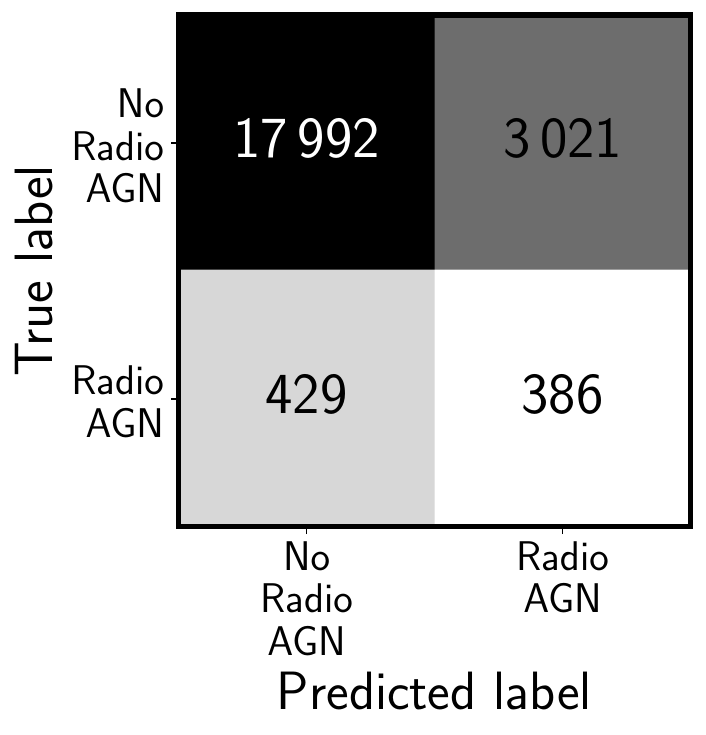}\hfill\break
    {(c) Radio-AGN confusion matrix}
  \end{minipage}%\\%
  \begin{minipage}{0.30\textwidth}
    \centering
    \includegraphics[width=0.99\textwidth]{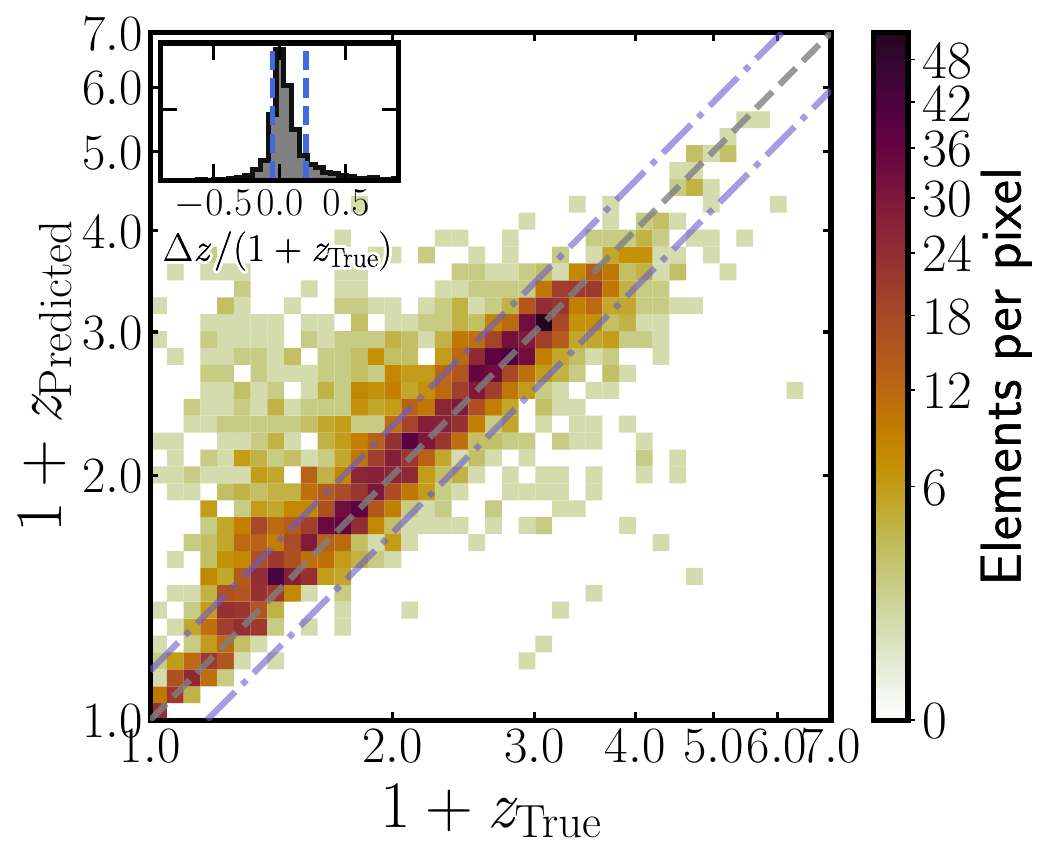}\hfill\break
    {(d) Predicted-True $z$ comparison}
  \end{minipage}%\\%
  \caption{Combined confusion matrices and True/predicted redshift density plot for the full radio AGN detection prediction computed using the testing subset from HETDEX (panels (a) and (b)) and the known labelled sources from S82 (panels (c) and (d)).
  }
  \label{fig:conf_matx_results_radio_AGN}
\end{figure*}

As the metrics discussed in the previous sections correspond to each individual model, new --combined-- metrics, based on the knowledge for labelled sources, are calculated for HETDEX and S82 and presented in Fig.~\ref{fig:conf_matx_results_radio_AGN} and Tables~\ref{table:fit_redshift_results} and \ref{table:fit_radio_AGN_results}. Overall, we observe worse combined metrics  with respect to the ones calculated for individual models (e.g. recall of $45\%$ for HETDEX and $47\%$ for S82). This degradation might be understood by the fact that the pipeline is composed of three sequential models. Each additional step is fed with sources classified by the previous algorithm. And some of these sources might not be similar, in terms of features, to those used for training, thus adding noise to the output of such model. A small sample of the output of the pipeline for five high-$z$ labelled radio AGN sources in HETDEX and S82 are shown in Tables~\ref{table:pred_radio_AGN_known_HETDEX} and \ref{table:pred_radio_AGN_known_S82} respectively.

\begin{figure}
  \centering
    \includegraphics[width=0.90\columnwidth]{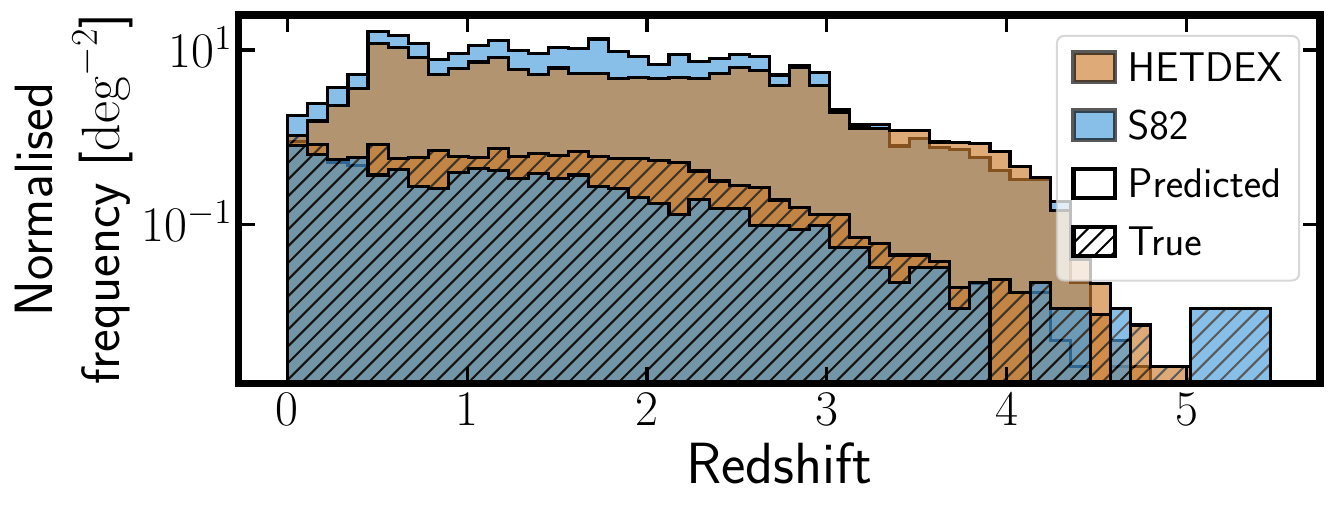}
  \caption{Redshift density distribution of the predicted radio AGNs within the unlabelled sources (clean histograms) in HETDEX (ochre histograms) and S82 (blue histograms) and true redshifts from labelled radio AGNs (dashed histograms).}
  \label{fig:hist_pred_z_unlabel_true_z_label}
\end{figure}

The application of the prediction pipeline to the unlabelled sources from the HETDEX field led to $9\,974\,990$ predicted AGNs, from which $68\,252$ were predicted to be radio detectable. The pipeline predicts, as well, $2\,073\,997$ AGNs in the unlabelled data from S82, being $22\,445$ of them candidates to be detected in the radio (to the detection level of LoTSS). 
The distribution of the predicted redshifts for radio AGNs in HETDEX and S82 is presented in  Fig.~\ref{fig:hist_pred_z_unlabel_true_z_label}. The pipeline outputs for a small sample of the predicted radio AGNs are presented in Tables~\ref{table:pred_radio_AGN_unknown_HETDEX} and \ref{table:pred_radio_AGN_unknown_S82} for HETDEX and S82 respectively. Section~\ref{sec:discussion} explores the comparison of these results with previous works in the literature and discusses the main drivers (i.e. features) for the detection of these radio AGNs.

\begin{table}
\setlength{\tabcolsep}{2pt}
\caption{Results of application of radio AGN prediction pipeline to the labelled sources in the HETDEX and S82 fields.} % title of Table
\label{table:fit_radio_AGN_results}      % reference of table in the text
\centering                          % used for centering table
\resizebox{0.99\columnwidth}{!}{
\begin{tabular}{c c c c c c}        % centered columns (6 columns)
\hline\hline                 % inserts double horizontal lines   
Subset                        & Threshold & F$_{\beta}$           & MCC                   & Precision             & Recall                \\
                              &           & $(\times 100)$        & $(\times 100)$        & $(\times 100)$        & $(\times 100)$        \\
\hline
\multirow{2}{*}{HETDEX-test}  & Naive     & $20.68 {\pm} 3.17$ & $24.93 {\pm} 3.72$ & $52.34 {\pm} 6.56$ & $13.79 {\pm} 2.27$ \\
                              & PR        & $37.99 {\pm} 2.59$ & $33.66 {\pm} 2.79$ & $32.20 {\pm} 2.72$ & $44.61 {\pm} 2.46$ \\[0.1em]
\multirow{2}{*}{S82-label}    & Naive     & $24.08 {\pm} 3.44$ & $21.43 {\pm} 3.53$ & $25.44 {\pm} 3.64$ & $23.07 {\pm} 3.72$ \\
                              & PR        & $19.42 {\pm} 2.31$ & $17.23 {\pm} 3.08$ & $11.33 {\pm} 1.32$ & $47.36 {\pm} 6.22$ \\
\hline                                   %inserts single line
\end{tabular}
}
\tablefoot{Values and uncertainties as in Table~\ref{table:fit_AGN_models}.}
\end{table}

\begin{table*}
\setlength{\tabcolsep}{3pt}
\caption{Predicted and original properties for the $5$ sources in testing subset with the highest redshift predicted radio AGNs. Sources are sorted by decreasing predicted redshift. A description of the columns is presented in Appendix~\ref{sec:app_prediction_results}.}\label{table:pred_radio_AGN_known_HETDEX}
\centering
\resizebox{0.90\textwidth}{!}{
\begin{tabular}{lrrrrrrrrrrrcr}
\hline
\hline
ID &     RA\_ICRS &    DE\_ICRS &  band\_num &  class &  Score\_AGN &  Prob\_AGN &  LOFAR\_detect &  Score\_radio &  Prob\_radio &  Score\_rAGN &  Prob\_rAGN &      $z$ &  pred\_$z$ \\
   &     (deg) &    (deg) &    &    &    &    &    &    &    &    &    &        &    \\
\hline
9898717  &  203.016113 &  55.518097 &         9 &    1.0 &   0.500082 &  0.954114 &             0 &     0.390861 &    0.375122 &    0.195462 &   0.357909 &  4.738 &  4.3679 \\
168686   &  164.769135 &  45.806320 &         8 &    1.0 &   0.500048 &  0.858157 &             0 &     0.450279 &    0.418719 &    0.225161 &   0.359326 &  4.893 &  4.1733 \\
14437074 &  213.226517 &  54.236343 &         9 &    1.0 &   0.500090 &  0.965187 &             0 &     0.251632 &    0.263746 &    0.125839 &   0.254564 &  4.326 &  4.0475 \\
10408176 &  188.163651 &  52.880898 &         9 &    1.0 &   0.500012 &  0.622448 &             0 &     0.604838 &    0.526003 &    0.302426 &   0.327410 &  4.340 &  3.9553 \\
12612753 &  227.216370 &  51.941029 &         9 &    1.0 &   0.500055 &  0.887909 &             0 &     0.364423 &    0.355080 &    0.182231 &   0.315278 &  3.795 &  3.8797 \\
\hline
\end{tabular}
}
\end{table*}

\begin{table*}
\setlength{\tabcolsep}{3pt}
\caption{Predicted and original properties for the $5$ sources in S82 with the highest predicted redshift on the labelled sources predicted to be radio AGNs. Sources are sorted by decreasing predicted redshift. A description of the columns is presented in Appendix~\ref{sec:app_prediction_results}.}\label{table:pred_radio_AGN_known_S82}
\centering
\resizebox{0.90\textwidth}{!}{
\begin{tabular}{lrrrrrrrrrrrrr}
\hline
\hline
ID &     RA\_ICRS &   DE\_ICRS &  band\_num &  class &  Score\_AGN &  Prob\_AGN &  radio\_detect &  Score\_radio &  Prob\_radio &  Score\_rAGN &  Prob\_rAGN &      $z$ &  pred\_$z$ \\
  &     (deg) &   (deg) &    &    &    &   &    &    &    &    &    &       &    \\
\hline
1406323 &   32.679794 & -0.305035 &         6 &    1.0 &   0.500050 &  0.866373 &             1 &     0.185842 &    0.204867 &    0.092930 &   0.177491 &  4.650 &  4.4986 \\
326139  &   33.580879 & -1.121398 &         8 &    1.0 &   0.500040 &  0.822622 &             0 &     0.208769 &    0.225946 &    0.104393 &   0.185868 &  4.600 &  4.3785 \\
633752  &   12.526446 & -0.888660 &         9 &    1.0 &   0.500035 &  0.793882 &             0 &     0.206182 &    0.223600 &    0.103098 &   0.177512 &  4.310 &  4.2946 \\
2834844 &  344.101440 &  0.789000 &         7 &    1.0 &   0.500062 &  0.909395 &             0 &     0.375735 &    0.363709 &    0.187891 &   0.330756 &  4.099 &  4.0635 \\
3191865 &   31.881712 &  1.063655 &         9 &    1.0 &   0.500087 &  0.962260 &             0 &     0.264210 &    0.274477 &    0.132128 &   0.264118 &  3.841 &  4.0509 \\
\hline
\end{tabular}
}
\end{table*}

\begin{table*}
\setlength{\tabcolsep}{3pt}
\caption{Predicted and original properties for the $5$ sources in the HETDEX field with the highest predicted redshift on the unlabelled sources predicted to be radio AGNs. A description of the columns is presented in Appendix~\ref{sec:app_prediction_results}.}
\label{table:pred_radio_AGN_unknown_HETDEX}
\centering
\resizebox{0.89\textwidth}{!}{
\begin{tabular}{lrrrrrrrrrrr}
\hline
\hline
ID &   RA\_ICRS &  DE\_ICRS &  band\_num &  Score\_AGN &  Prob\_AGN &  radio\_detect &  Score\_radio &  Prob\_radio &  Score\_rAGN &  Prob\_rAGN &  pred\_$z$ \\
  &     (deg) &   (deg) &    &    &    &    &    &   &   &    &   \\
\hline
9544254  &  201.309235 &  53.746429 &         6 &   0.500007 &  0.578804 &             0 &     0.351672 &    0.345250 &    0.175838 &   0.199832 &  4.7114 \\
12355845 &  220.838120 &  50.319016 &         5 &   0.500007 &  0.578804 &             0 &     0.937123 &    0.794128 &    0.468568 &   0.459644 &  4.6064 \\
13814216 &  219.839142 &  52.660328 &         7 &   0.500015 &  0.650248 &             0 &     0.213846 &    0.230529 &    0.106926 &   0.149901 &  4.5622 \\
6698239  &  184.694901 &  49.063766 &         5 &   0.499995 &  0.467527 &             0 &     0.799085 &    0.662753 &    0.399538 &   0.309855 &  4.5483 \\
2951011  &  175.882446 &  55.497799 &         5 &   0.500008 &  0.589419 &             0 &     0.823295 &    0.681768 &    0.411654 &   0.401847 &  4.5320 \\
\hline
\end{tabular}
}
\end{table*}

\begin{table*}
\setlength{\tabcolsep}{3pt}
\caption{Predicted and original properties for the $5$ sources in S82 with the highest predicted redshift on the unlabelled sources predicted to be radio AGNs. A description of the columns is presented in Appendix~\ref{sec:app_prediction_results}.}\label{table:pred_radio_AGN_unknown_S82}
\centering
\resizebox{0.89\textwidth}{!}{
\begin{tabular}{lrrrrrrrrrrr}
\hline
\hline
ID &     RA\_ICRS &   DE\_ICRS &  band\_num &  Score\_AGN &  Prob\_AGN &  radio\_detect &  Score\_radio &  Prob\_radio &  Score\_rAGN &  Prob\_rAGN &  pred\_$z$ \\
  &   (deg) &  (deg) &    &   &    &   &    &    &    &   &    \\
\hline
3244450 &  26.276423 &  1.104065 &         7 &   0.500002 &  0.531172 &             0 &     0.542061 &    0.483128 &    0.271031 &   0.256624 &  4.3938 \\
1062270 &  11.744675 & -0.562642 &         7 &   0.499982 &  0.356043 &             0 &     0.196326 &    0.214586 &    0.098159 &   0.076402 &  4.3563 \\
3261269 &  28.882526 &  1.117103 &         7 &   0.500011 &  0.608660 &             0 &     0.354936 &    0.347777 &    0.177472 &   0.211678 &  4.3153 \\
1466227 &  18.157259 & -0.258997 &         5 &   0.500013 &  0.630968 &             0 &     0.456207 &    0.422973 &    0.228110 &   0.266882 &  4.3146 \\
1134866 &  11.304936 & -0.507943 &         7 &   0.500011 &  0.616439 &             0 &     0.226178 &    0.241539 &    0.113091 &   0.148894 &  4.3140 \\
\hline
\end{tabular}
}
\end{table*}

\subsection{No-skill classification}\label{sec:random_classification}

As presented in Sect.~\ref{sec:metrics_classfication}, Eqs.~\ref{eq:no_skill_Fb}--\ref{eq:no_skill_precision} show the base results for a classification with no skill. Table~\ref{table:random_classification} presents the scores generated by using this technique. These values are the base from which any improvement can be assessed.

\begin{table}
  \setlength{\tabcolsep}{2pt}
  \caption{Results of no-skill selection of sources in different stages of pipeline to the labelled sources in the HETDEX test subset and S82 fields.} % title of Table
  \label{table:random_classification}      % reference of table in the text
  \centering                          % used for centering table
  \resizebox{0.90\columnwidth}{!}{
  \begin{tabular}{c c c c c c}        % centered columns (6 columns)
  \hline\hline                 % inserts double horizontal lines   
  Subset                  & Prediction              & F$_{\beta}$     & MCC             & Precision       & Recall          \\
                          &                         & $(\times 100)$  & $(\times 100)$  & $(\times 100)$  & $(\times 100)$  \\
  \hline
  \multirow{4}{*}{HETDEX} & AGN-galaxy              & $42.57$        & $0.00$        & $42.57$        & $42.57$        \\
  %                        & Radio-detection (pipe)  & $11.82$        & $0.00$        & $11.82$        & $11.82$        \\
                          & Radio-detection (label) & $12.84$        & $0.00$        & $12.84$        & $12.84$        \\
                          & Radio AGN               & $05.47$        & $0.00$        & $05.47$        & $05.47$        \\[0.1em]
  \multirow{4}{*}{S82}    & AGN-galaxy              & $81.29$        & $0.00$        & $81.29$        & $81.29$        \\
  %                        & Radio-detection (pipe)  & $03.94$        & $0.00$        & $03.94$        & $03.94$        \\
                          & Radio-detection (label) & $04.59$        & $0.00$        & $04.59$        & $04.59$        \\
                          & Radio AGN               & $03.73$        & $0.00$        & $03.73$        & $03.73$        \\
  \hline                                   %inserts single line
  \end{tabular}
  }
  \tablefoot{All metrics have been multiplied by $100$.}
\end{table}

Subsets and prediction modes displayed in Table~\ref{table:random_classification} coincide with those exhibited in Tables~\ref{table:fit_AGN_results}, \ref{table:fit_radio_results}, and \ref{table:fit_radio_AGN_results}. For instance, in the test HETDEX sub-sample, ${\sim} 43 \%$ of sources are labelled as AGNs. From all AGNs, ${\sim} 13 \%$ of them have radio detections. This can be summarised stating that ${\sim} 6 \%$ of all sources in the test sub-sample are radio-detected AGNs.

%--------------------------------------------------------------------
\section{Discussion}\label{sec:discussion}

\subsection{Comparison with previous prediction or detection works}\label{sec:compare_previous_works}

In this subsection, we provide a few examples of related published works as well as plausible explanations for observed discrepancies when these are present. This comparison attempts to be representative of the literature on the subject but does not intends to be complete in any way.

\subsubsection{AGN detection prediction}\label{sec:previous_AGN_detection}

In order to understand the significance of our results and ways for future improvement, we separate the comparison with previous works in two parts. First, we present previously published results from traditional methodologies. In second place, we offer a comparison with ML methods.

Traditional AGN selection methods are based on the comparison of the measured SED photometry to a template library \citep{2011Ap&SS.331....1W}. A recent example of its application is presented by  \citet{2022MNRAS.509.4940T} where best fit classifications were calculated for more than $700\,000$ galaxies in the D10 field of the Deep Extragalactic VIsible Legacy Survey \citep[DEVILS;][]{2018MNRAS.480..768D} and the Galaxy and Mass Assembly survey \citep[GAMA;][]{2011MNRAS.413..971D, 2015MNRAS.452.2087L}.
The $91 \%$ recovery rate of AGNs, selected through various means (X-ray measurements, narrow and broad emission lines, and mid-IR colours), is very much in line with our findings in S82, where our rate (recall) reaches $89 \%$.

Traditional methods also encompass the colour-based selection of AGNs. While less precise, they provide access to a much larger base of candidates with a very low computational cost. We implemented some of the most common colour criteria on the data from S82.
Of particular interest is the predicting power of the mid-IR colour selection due to its potential to detect hidden or heavily obscured AGN activity.

\begin{figure}
   \centering
   \includegraphics[width=0.99\columnwidth]{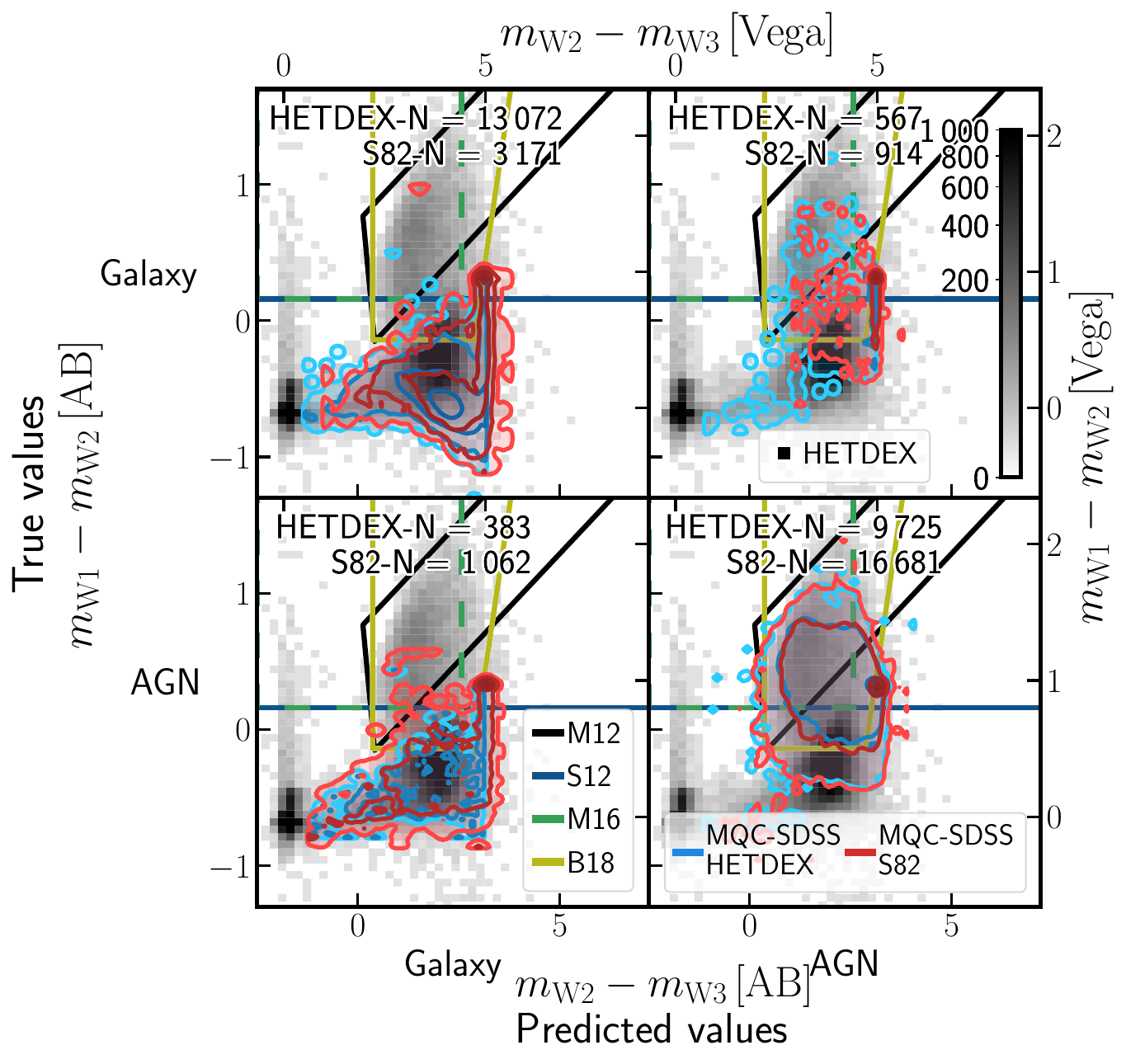}
   \caption{W1~-~W2, W2~-~W3 colour-colour diagrams for sources in the testing subset, from HETDEX, and labelled sources from S82 given their position in the AGN-galaxy confusion matrix (see, for HETDEX, rightmost panel of Fig.~\ref{fig:conf_matx_results_radio_AGN}). In the background, density plot of all CW-detected sources in the full HETDEX field sample is displayed. Colour of each square represents the number of sources in that position of parameter space, with darker squares having more sources (as defined in the colourbar of the upper-right panel). Contours represent distribution of sources for each of the aforementioned subsets at $1$, $2$, $3$, and $4\,\sigma$ levels (shades of blue, for testing set and shades of red for labelled S82 sources). Coloured, solid lines display limits from the criteria for the detection of AGNs described in Sect.~\ref{sec:previous_AGN_detection}.}
   \label{fig:W1_W2_W2_W3_AGN_pred_HETDEX_S82}
\end{figure}

Based on WISE \citep{2010AJ....140.1868W} data, \citet[][S12]{2012ApJ...753...30S} proposed a threshold at W1~-~W2 $\geq 0.8$ to separate AGNs from non-AGNs using data from AGNs in the Cosmic Evolution Survey (COSMOS) field \citep{2007ApJS..172....1S}.
A more stringent criterion was developed by \citet[][M12]{2012MNRAS.426.3271M}, the AGN wedge, which can be defined by the sources located inside the region defined by the relations W1~-~W2 $< 0.315 \times($W2~-~W3$)+ 0.791$, W1~-~W2 $> 0.315 \times($W2~-~W3$)- 0.222$, and W1~-~W2 $> -3.172 \times($ W2~-~W3 $)+ 7.624$. In order to define this wedge, they used data from X-ray selected AGNs over an area of $44.43\, \mathrm{deg}^{2}$ in the northern sky.
\citet[][M16]{2016MNRAS.462.2631M} cross-correlated data from WISE observations with X-ray and radio surveys creating a sample of star-forming galaxies and AGNs in the northern sky. They developed individual relations to separate classes of galaxies and AGNs in the W1~-~W2, W2~-~W3 space and, for AGNs the criterion, the relation is  W1~-~W2 $\geq 0.5$ and W2~-~W3 $< 4.4$.
More recently, \citet[][B18]{2018MNRAS.478.3056B} analysed the quality of mid-IR colour selection methods for the identification of obscured AGNs involved in mergers. Using hydrodynamic simulations for the evolution of AGNs in galaxy mergers, they developed a selection criterion from WISE colours which is shown to be able to separate, with high reliability, starburst galaxies from AGNs. The expressions have the form W1~-~W2 $> 0.5$, W2~-~W3 $> 2.2$, and W1~-~W2 $> 2 \times($ W2~-~W3$) -8.9$. The results from the application of these criteria to our samples in the testing subset and in the labelled sources of S82 field are summarised in Table~\ref{table:previous_AGN_methods} and a graphical representation of the boundaries they create in their respective parameter spaces is presented in Fig.~\ref{fig:W1_W2_W2_W3_AGN_pred_HETDEX_S82}. 

Table~\ref{table:previous_AGN_methods} shows that previous colour-colour criteria have been designed and calibrated to have very high precision values. Most of the sources deemed to be AGN by them are, indeed, of such class. Despite being tuned to maximise their recall (and $\mathrm{F}_{\beta}$ to a lesser extent), our classifier, and the criterion derived from it, still show precision values compatible with those of such criteria. This result underlines the power of ML methods. They can be on a par with traditional colour-colour criteria and excel in additional metrics.

Figure~\ref{fig:W1_W2_W2_W3_AGN_pred_HETDEX_S82} is constructed as a confusion matrix, plotting in each quadrant the whole WISE population in the background and in colour contours the corresponding fraction of the testing set (TP, TN, FP, and FN, see Fig.~\ref{fig:results_models_test}a and Sect.~\ref{sec:metrics_classfication}). As expected, our pipeline is able to separate with high confidence sources which are closer to the AGN or the galaxy loci (TP and TN) while sources in the FN and FP quadrant show a different situation. Active galactic nuclei predicted to be galaxies (FN, $1.6\%$ of sources for HETDEX, and $4.9\%$ for S82) are located in the galaxy region of the colour-colour diagram. On the opposite corner of the plot, galaxies predicted to be AGNs (FP, $2.4\%$ of sources for HETDEX, and $4.2\%$ for S82) cover the areas of AGNs and galaxies uniformly. False negative sources might be sources that are identified as AGNs by means not included in our feature set (e.g. X-ray, radio emission). Sources in the FP quadrant, alternatively, might be galaxies with extreme properties, similar to AGNs.

\begin{table}
\setlength{\tabcolsep}{3pt}
\caption{Results of application of several AGN detection criteria to our testing subset and the labelled sources from the S82 field.}             % title of Table
\label{table:previous_AGN_methods}      % is used to refer this table in the text
\centering                          % used for centering table
\resizebox{0.62\columnwidth}{!}{
\begin{tabular}{c c c c c}        % centered columns (5 columns)
\hline\hline  
\multicolumn{5}{c}{HETDEX test set} \\
Method\tablefootmark{a} & F$_{\beta}$       & MCC              & Precision         & Recall         \\
                        & $(\times 100)$    & $(\times 100)$   & $(\times 100)$    & $(\times 100)$ \\
\hline
S12                     & 86.10             & 78.78            & 93.98             & 80.51 \\
M12                     & 51.80             & 49.71            & 98.87             & 37.18 \\
M16                     & 67.21             & 61.30            & 97.48             & 53.48 \\
B18                     & 82.14             & 75.76            & 97.54             & 72.66 \\
This work               & 92.71             & 87.64            & 94.00             & 91.67 \\
\hline\\[0.5em]
\hline\hline
\multicolumn{5}{c}{S82 (labelled)} \\
Method      & F$_{\beta}$       & MCC               & Precision         & Recall \\
            & $(\times 100)$    & $(\times 100)$    & $(\times 100)$    & $(\times 100)$ \\
\hline
S12         & 83.59             & 45.47             & 93.93             & 76.62 \\
M12         & 46.80             & 28.22             & 99.59             & 32.54 \\
M16         & 64.69             & 37.76             & 98.80             & 50.32 \\
B18         & 79.71             & 51.07             & 98.72             & 68.77 \\
This work   & 90.63             & 58.53             & 94.15             & 87.91 \\
\hline
\end{tabular}
}
\tablefoot{
\tablefootmark{a}{Naming codes for the used methods are described in the main text (cf. Sect.~\ref{sec:previous_AGN_detection}). Last row of each sub-table corresponds to the criterion derived in this work (as described in Sect~\ref{sec:feat_importances}). All metrics have been multiplied by $100$.}
}
\end{table}

For the case of ML-based models for AGN-galaxy classification, several analyses have been published in recent years. An example of their application is provided in \citet{2020A&A...639A..84C} where a random forest model for the classification of stars, galaxies and AGNs using photometric data was trained from more than $3\,000\,000$ sources in the SDSS \citep[DR15;][]{2019ApJS..240...23A} and WISE with associated spectroscopic observations. Close to $400\,000$ sources have a quasar spectroscopic label and from the application of their model to a validation subset, they obtain a recall of $0.929$ and F1 score of $0.943$ for the quasar classification. These scores are of the same order as the ones obtained when applying our AGN-galaxy model to the testing set (see Table~\ref{table:fit_AGN_results}). Thus, and despite using an order of magnitude fewer sources for the full training and validation process, our model can achieve equivalently good scores. 

Expanding on \citet{2020A&A...639A..84C}, \citet{2022A&A...666A..87C} built a ML pipeline, \texttt{SHEEP}, for the classification of sources into stars, galaxies and QSOs. In contrast to \citet{2020A&A...639A..84C} or the pipeline described here, the first step in their analysis is the redshift prediction, which is used as part of the training features by the subsequent classifiers. They extracted WISE and SDSS \citep[DR15;][]{2019ApJS..240...23A} photometric data for almost $3\,500\,000$ sources classified as stars, galaxies or QSOs. The application of their pipeline to sources predicted to be QSOs leads to a recall of $0.960$ and an F1 score of $0.967$. The improved scores in their pipeline might be a consequence not only of the slightly larger pool of sources, but also the inclusion of the coordinates of the sources (right ascension, declination) and the predicted redshift values as features in the training.

A test with a larger number of ML methods was performed by \citet{2021A&A...651A.108P}. For training, they used optical and infrared data from close to $1\,500$ sources (galaxies and AGNs) located at the AKARI North Ecliptic Pole (NEP) Wide-field \citep{2009PASJ...61..375L, 2012A&A...548A..29K} covering a $5.4\, \mathrm{deg}^{2}$ area. They tested  \verb|LR|, \verb|SVM|, \verb|RF|, \verb|ET|, and \verb|XGBoost| including the possibility of generalised stacking. In general, they obtain results with F1 scores between $0.60$~--~$0.70$ and recall values in the range of $50\%$~--~$80\%$. 
These values, lower than the works described here, can be fully understood given the small size of their training sample. A larger photometric sample covers a wider range of the parameter space which significantly helps the metrics of any given model.

\subsubsection{Radio detection prediction}\label{sec:previous_radio_detection}

We have not found in the literature any work attempting the prediction of AGN radio detection at any level and therefore this is the first attempt at doing so. In the literature we do find several correlations between the AGN radio emission (flux) and that at other wavelengths \citep[e.g. with infrared emission,][]{1985ApJ...298L...7H, 1992ARA&A..30..575C} and substantial effort has been done towards classifying RGs based upon their morphology \citep[e.g.][]{2017ApJS..230...20A, 2019MNRAS.482.1211W} and its connection to environment \citep{2008A&ARv..15...67M, 2022A&ARv..30....6M}. None of these extensive works has directly focussed on the a priori presence or absence of radio emission above a certain threshold. Therefore, the results presented here are the first attempt at such an effort.

The ${\sim} 2$x success rate of the pipeline to identify radio emission in AGNs (${\sim} 44.61\%$ recall and ${\sim} 32.20\%$ precision; see Table~\ref{table:fit_radio_AGN_results}) with the respect to a no-skill selection (${\lessapprox}30 \%$), provides the opportunity to understand what the model has learned from the data and, therefore, gain some insight into the nature or triggering mechanisms of the radio emission. We, therefore, reserve the discussion of the most important features, and the linked physical processes, driving the pipeline improved predictions to Sect.~\ref{sec:feat_importances}.

\subsubsection{Redshift value prediction}\label{sec:previous_z_values}

We compare our results to that of \citet[][Stripe 82X]{2017ApJ...850...66A} where the authors analysed multi-wavelength data from more than $6\,100$ X-ray detected AGNs from the $31.3\, \mathrm{deg}^{2}$ of the Stripe 82X survey. They obtained photometric redshifts for almost $6\,000$ of these sources using the template-based fitting code \verb|LePhare| \citep{1999MNRAS.310..540A, 2006A&A...457..841I}. Their results present a normalised median absolute deviation of $\sigma_{\mathrm{NMAD}} {=} 0.062$ and an outlier fraction of $\eta {=} 13.69 \%$, values which are similar to our results in HETDEX and S82 except for a better outlier fraction (as shown in Table~\ref{table:fit_redshift_results}, we obtain $\eta_{S82}=25.18\%$, $\sigma_{\mathrm{NMAD}}^{\mathrm{HETDEX}} {=} 0.071$, and $\eta^{\mathrm{HETDEX}} {=} 18.9$\%).

On the ML side, we compare our results to those produced by \citet{2021Galax...9...86C} in S82, with $\sigma_{\mathrm{NMAD}} = 0.1197$ and $\eta = 29.72 \%$, and find that our redshift prediction model improves by at least $25 \%$ for any given metric.
The source of improvement is probably many-fold. First, it might be related to the different sets of features used (colours vs ratios) and, second, the more specific population of radio AGN used to train our models. \citet{2021Galax...9...86C} used a limited set of colours to train their model, while we allowed the use of all available combinations of magnitudes (Sect.~\ref{sec:feature_creation}). Additionally, their redshift model was trained on all available AGNs in HETDEX, while we trained (and tested) it only with radio-detected AGNs. Using a more constrained sample reduces the likelihood of handling sources that are too different in the parameter space.

Another example of the use of ML for AGN redshift prediction has been presented by \citet{2019PASP..131j8003L}. They studied the use of the k-nearest neighbours algorithm \verb|KNN| \citep{1053964}, a non-parametric supervised learning approach, to derive redshift values for radio-detectable sources. They combined $1.4$ GHz radio measurements, infrared, and optical photometry in the European Large Area Infrared Space Observatory (ISO) Survey-South 1 \citep[ELAIS-S1;][]{2000MNRAS.316..749O} and extended Chandra Deep Field South \citep[eCDFS;][]{2005ApJS..161...21L} fields, matching their sensitivities and depths to the expected values in the Evolutionary Map of the Universe \citep[EMU;][]{2011PASA...28..215N}. From the different experiments they run, their resulting NMAD values are in the range ${\sigma_{\mathrm{NMAD}} = 0.05 - 0.06}$, and their outlier fraction can be found between ${\eta = 7.35 \%}$ and ${\eta = 13.88 \%}$. 
As an extension to the previous results, \citet{LUKEN2022100557} analysed multi-wavelength data from radio-detected sources the eCDFS and the ELAIS-S1 fields. Using \texttt{KNN} and \texttt{RF} methods to predict the redshifts of more than $1\,300$ RGs, they developed regression methods that show NMAD values between ${\sigma_{\mathrm{NMAD}} = 0.03}$ and ${\sigma_{\mathrm{NMAD}} = 0.06}$, ${\sigma_{z} = 0.10 - 0.19}$, and outlier fractions of ${\eta = 6.36 \%}$ and ${\eta = 12.75 \%}$.

In addition to the previous work, \citet{2019PASP..131j8004N} compared a number of methodologies, mostly related with ML but also \texttt{LePhare}, for predicting redshift values for radio sources. They used more than $45$ photometric measurements (including $1.4$ GHz fluxes) from different surveys in the COSMOS field. From several settings of features, sensitivities, and parameters, they retrieve redshift predictions with NMAD values between ${\sigma_{\mathrm{NMAD}} = 0.054}$ and ${\sigma_{\mathrm{NMAD}} = 0.48}$ and outlier fractions that range between ${\eta = 7 \%}$ and ${\eta = 80 \%}$. The broad span of obtained values might be due to the combinations of properties for each individual training set (including the use of radio or X-ray measurements, the selection depth, and others) and to the size of these sets, which was small for ML purposes (less than $400$ sources). The slightly better results can be understood given the heavily populated photometric data available in COSMOS.

Specifically related to HETDEX, it is possible to compare our results to those from \citet{2019A&A...622A...3D}. They use a hybrid photometric redshift approach combining  traditional template fitting redshift determination and ML-based methods. In particular, they implemented a GP algorithm, which is able to model both the intrinsic noise and the uncertainties of the training features. Their redshift prediction analysis of 
AGN sources with a spectroscopic redshift detected in the LoTSS DR1 ($6,811$ sources) recovers a NMAD value of ${\sigma_{\mathrm{NMAD}} = 0.102}$ and an outlier fraction of ${\eta = 26.6 \%}$.
The differences between these results and those obtained from the application of our models (individually and as part of the prediction pipeline) might be due to the differences in the creation of the training sets. \citet{2019A&A...622A...3D} used information from all available sources in the HETDEX field for training the redshift GP whilst our redshift model has been only trained on radio-detected AGNs, giving it the opportunity to focus its parameter exploration only on these sources.

Finally, \citet{2022A&A...666A..87C} also produced photometric redshift predictions for almost $3\,500\,000$ sources (stars, galaxies, and QSOs) as part of their pipeline (see Sect.~\ref{sec:previous_AGN_detection}). They combined three algorithms for their predictions: \texttt{XGBoost}, \texttt{CatBoost}, and \texttt{LightGBM} \citep{NIPS2017_6449f44a}. This procedure leads to ${\sigma_{\mathrm{NMAD}} = 0.018}$ and ${\eta = 2 \%}$. As with previous examples, the differences with our results can be a consequence of the number of training samples. Also, in the case of \citet{2022A&A...666A..87C}, they applied an additional post-processing step to the redshift predictions attempting to predict and understand the appearance of catastrophic outliers.

\subsection{Influence of data imputation}\label{sec:band_num_trends}

\begin{figure}
  \centering
    \includegraphics[width=0.90\columnwidth]{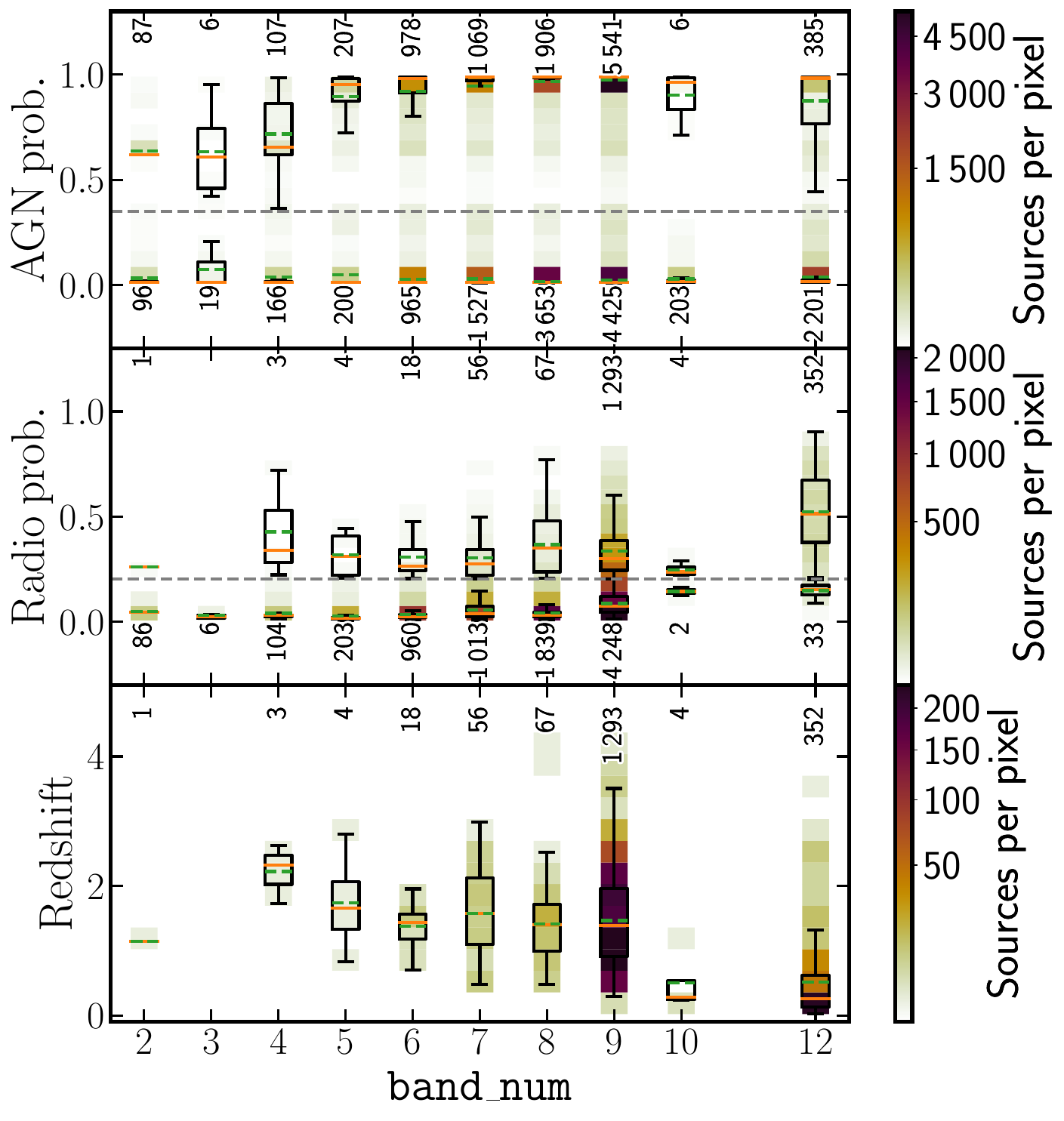}
  \caption{Evolution of predicted probabilities (top: probability to be AGN, middle: probability of AGNs to be detected in radio) and redshift values for radio-detectable AGNs (bottom panel) as a function of the number of observed bands for sources in test set. In top panel, sources have been divided between those predicted to be AGN and galaxy. In middle panel, sources are divided between predicted AGN that are predicted to be detected in radio and those predicted to not have radio detection. Background density plots (following colour coding in colourbars) show location of predicted values. Overlaid boxplots display main statistics for each number of measured bands. Black rectangles encompass sources in second and third quartiles. Vertical lines show the place of sources from first and fourth quartiles. Orange lines represent median value of sample and dashed, green lines indicate their mean values. Dashed, grey lines show PR thresholds for AGN-galaxy and radio detection classifications. Close to each boxplot, written values correspond to the number of sources considered to create each set of statistics.}
  \label{fig:probs_band_num_test}
\end{figure}

One effect which might influence the training of the models and, consequently, the prediction for new sources is related to the imputation of missing values (cf. Sect.~\ref{sec:data_collection}). In Fig.~\ref{fig:probs_band_num_test}, we plotted the distributions of predicted scores (for classification models) and predicted redshift values as a function of the number of measured bands (\texttt{band\_num}) for each step of the pipeline as applied to sources predicted to be of each class in the test subset.

The top panel of Fig.~\ref{fig:probs_band_num_test} shows the influence of the degree of imputation in the classification between AGNs and galaxies. For most of the bins, probabilities for predicted galaxies are distributed close to $0.0$, without any noticeable trend. In the case of predicted AGNs, the combination of low number of sources and high degree of imputation ($\mathtt{band\_num} < 5$) lead to low mean probabilities.

The case of radio detection classification is somewhat different. Given the number and distribution of sources per bin, it is not possible to extract any strong trend for the probabilities of radio-predicted sources. The absence of evolution with the number of observed bands is stronger for sources predicted to be devoid of radio detection.

Finally, a stronger effect can be seen with the evolution of predicted redshift values for radio-detectable AGNs. Despite the lower number of available sources, it is possible to recognise that sources with higher number of available measurements are predicted to have lower redshift values. Sources that are closer to us have higher probabilities to be detected in a large number of bands. Thus, it is expected that our model predicts lower redshift values for the most measured sources in the field.

In consequence, Fig.~\ref{fig:probs_band_num_test} allows us to understand the influence of imputation over the predictions. The most highly affected quantity is the redshift, where large fractions of measured magnitudes are needed to obtain scores that are in line with previous results (cf. Sect.~\ref{sec:previous_z_values}). The AGN-galaxy and radio detection classifications show a mild influence of imputation in their results.

\subsection{Model explanations}\label{sec:model_explain}

Given the success of the models and pipeline in classifying AGNs, their radio detectability and redshift with the provided set of observables, knowing the relative weights that they have in the decision-making process is of utmost relevance. In this way, physical insight might be gained about the triggers of AGN and radio activity and its connection to their host.
Therefore, we estimated both local and global feature importances for the individual models and the combined pipeline. Global importances were retrieved using the so-called `decrease in impurity' approach \citep[see, for example,][]{Breiman2001}. Local importances have been determined via Shapley values. A more detailed description of what these importances are and how they are calculated is given in the following sections.

\subsubsection{Global feature importances}\label{sec:feat_importances}

Overall, mean or global feature importances can be retrieved from models that are based on decision trees \citep[e.g. random forests and boosting models,][]{Breiman2001, breiman2003manual}. All algorithms selected in this work (\verb|RF|, \verb|CatBoost|, \verb|XGBoost|, \verb|ET|, \verb|GBR|, and \verb|GBC|) belong to these two classes. For each feature, the decrease in impurity (a term frequently used in the literature related to machine learning) of the dataset is calculated for all the nodes of the tree in which that feature is used. Features with the highest impurity decrease will be more important for the model \citep{NIPS2013_e3796ae8}\footnote{For some models not based on decision trees, feature importances can be obtained from the coefficients delivered by the training process. These coefficients are related to the level to which each quantity is scaled to obtain a final prediction (as in the coefficients from a polynomial regression).}.

Insight into the decision-making of the pipeline can only rely on the specific weight of the original set of features (see Sect.~\ref{sec:feat_selection}). Table ~\ref{table:feat_importances} presents the ranked combined importances from the observables selected in each of the three sequential models that compose the pipeline. They have been combined using the importances from the meta learner (as shown in Table~\ref{table:base_feat_importances}) and that of base learners. The derived importances will be dependent on the dataset used, including any imputation for the missing data, and the details of the models (i.e. algorithms used and stacking procedure). 
We first notice in Table~\ref{table:feat_importances} that the order of the features is different for all three models. This difference reinforces the need, as stated in Sect.~\ref{sec:ML_training}, of developing separate models for each of the prediction stages of this work that would evaluate the best feature weights for the related classification or regression task. 

\begin{table}
\setlength{\tabcolsep}{2.9pt}
\caption{Relative importances (rescaled to add to $100$) for observed features from the three models combined between meta and base models.}             % title of Table
\label{table:feat_importances}      % is used to refer this table in the text
\centering                          % used for centering table
\resizebox{0.97\columnwidth}{!}{
\begin{tabular}{c c c c c c}        % centered columns (6 columns)
\hline\hline                 % inserts double horizontal lines   
\multicolumn{6}{c}{AGN-galaxy (meta model: \texttt{CatBoost})} \\
Feature             & Importance    & Feature               & Importance    & Feature         & Importance \\
\hline
\texttt{W1\_W2}     & 68.945        & \texttt{H\_K}         & 1.715         & \texttt{z\_W2}  & 1.026      \\
\texttt{W1\_W3}     &  4.753        & \texttt{y\_W1}        & 1.659         & \texttt{z\_y}   & 0.722      \\
\texttt{g\_r}       &  4.040        & \texttt{y\_W2}        & 1.513         & \texttt{W3\_W4} & 0.669      \\
\texttt{r\_J}       &  4.006        & \texttt{i\_y}         & 1.441         & \texttt{W4mag}  & 0.558      \\
\texttt{r\_i}       &  3.780        & \texttt{i\_z}         & 1.366         & \texttt{H\_W3}  & 0.408      \\
\texttt{band\_num}  &  1.842        & \texttt{y\_J}         & 1.187         & \texttt{J\_H}   & 0.371      \\[0.5em]
\hline\hline
\multicolumn{6}{c}{Radio detection (meta model: \texttt{GBC})} \\
Feature             & Importance    & Feature               & Importance    & Feature             & Importance \\
\hline
\texttt{W2\_W3}     &  9.609        & \texttt{y\_W1}        & 7.150         & \texttt{W4mag}      & 4.759      \\
\texttt{y\_J}       &  8.102        & \texttt{g\_r}         & 7.123         & \texttt{K\_W4}      & 2.280      \\
\texttt{W1\_W2}     &  8.010        & \texttt{z\_W1}        & 7.076         & \texttt{J\_H}       & 1.283      \\
\texttt{g\_i}       &  7.446        & \texttt{r\_z}         & 6.981         & \texttt{H\_K}       & 1.030      \\
\texttt{K\_W3}      &  7.357        & \texttt{i\_z}         & 6.867         & \texttt{band\_num}  & 1.018      \\
\texttt{z\_y}       &  7.321        & \texttt{r\_i}         & 6.588         &                     &            \\[0.5em]
\hline\hline
\multicolumn{6}{c}{Redshift prediction (meta model: \texttt{ET})} \\
Feature               & Importance    & Feature           & Importance    & Feature           & Importance \\
\hline
\texttt{y\_W1}        & 35.572        & \texttt{y\_J}     & 3.018         & \texttt{i\_z}     & 1.215    \\
\texttt{W1\_W2}       & 13.526        & \texttt{r\_z}     & 3.000         & \texttt{J\_H}     & 1.162    \\
\texttt{W2\_W3}       & 12.608        & \texttt{r\_i}     & 2.896         & \texttt{g\_W3}    & 1.000    \\
\texttt{band\_number} &  6.358        & \texttt{z\_y}     & 2.827         & \texttt{K\_W3}    & 0.925    \\
\texttt{H\_K}         &  4.984        & \texttt{W4mag}    & 2.784         & \texttt{K\_W4}    & 0.762    \\
\texttt{g\_r}         &  4.954        & \texttt{i\_y}     & 2.408         &                   &          \\
\hline
\end{tabular}
}
\end{table}

\begin{table}
  \setlength{\tabcolsep}{2.9pt}
  \caption{Relative feature importances (rescaled to add to $100$) for base algorithms in each prediction step.}             % title of Table
  \label{table:base_feat_importances}      % is used to refer this table in the text
  \centering                          % used for centering table
  \resizebox{0.65\columnwidth}{!}{
  \begin{tabular}{c c c c}        % centered columns (4 columns)
  \hline\hline                 % inserts double horizontal lines   
  \multicolumn{4}{c}{AGN-galaxy model (\texttt{CatBoost})} \\
  Feature             & Importance    & Feature           & Importance  \\
  \hline
  \texttt{gbc}        & 49.709        & \texttt{xgboost}  & 14.046      \\
  \texttt{et}         & 19.403        & \texttt{rf}       &  8.981      \\
  \multicolumn{3}{r}{Remaining feature importances:} & 7.861 \\[0.2em]
  \hline\hline
  \multicolumn{4}{c}{Radio detection model (\texttt{GBC})} \\
  
  Feature             & Importance    & Feature           & Importance \\
  \hline
  \texttt{rf}         & 12.024        & \texttt{catboost} & 7.137      \\
  \texttt{et}         &  7.154        & \texttt{xgboost}  & 6.604      \\
  %\hline
  \multicolumn{3}{r}{Remaining importances:} & 67.081 \\[0.2em]
  \hline\hline
  \multicolumn{4}{c}{Redshift prediction model (\texttt{ET})} \\
  
  Feature           & Importance    & Feature           & Importance \\
  \hline
  \texttt{xgboost}  & 25.138        & \texttt{catboost} & 21.072     \\
  \texttt{gbr}      & 21.864        & \texttt{rf}       & 13.709     \\
  %\hline
  \multicolumn{3}{r}{Remaining importances:} & 18.217 \\
  \hline
  \end{tabular}
  }
  \end{table}

\begin{figure}
    \centering
    \begin{minipage}{0.80\columnwidth}
    \includegraphics[width=\textwidth]{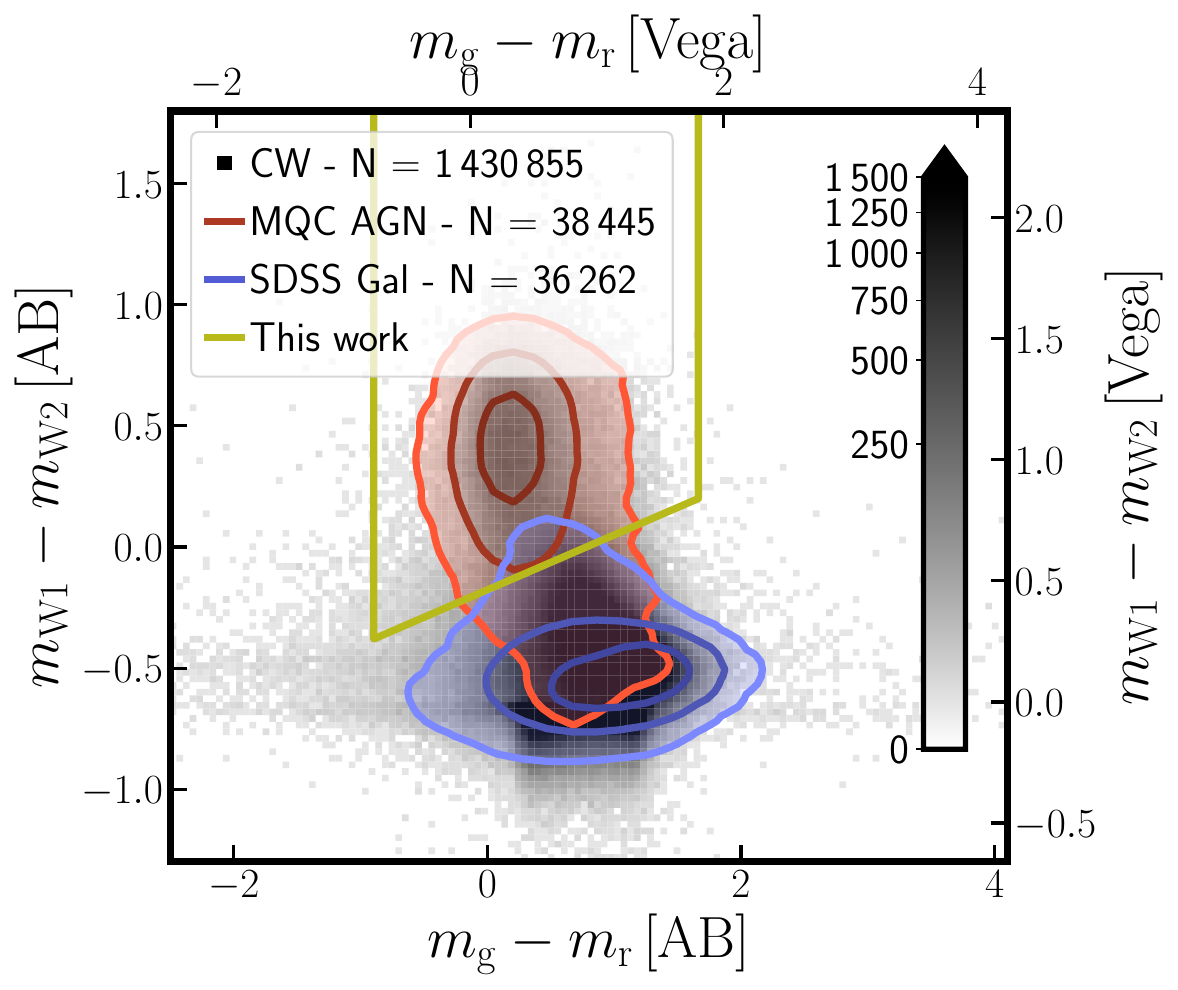}
    \end{minipage}%\\%
    \caption{AGN classification colour-colour plot in the HETDEX field using CW (W1, W2) and PS1 (g, r) passbands. Grey-scale density plot include all CW detected and non-imputed sources. Red contours highlight the density distribution of the AGNs in the Million QSO catalogue (MQC) and blue contours show the density distribution for the galaxies from SDSS DR16. Contours are located at 1, 2, and 3 $\sigma$ levels.}
   \label{fig:HETDEX_gr_W1W2_AGN_gal_class}
\end{figure}

For the AGN-galaxy classification model, it is very interesting to note that the most important feature for the predicted probability of a source to be an AGN is the WISE colour W1~-~W2 (as well as W1~-~W3). This colour is indeed one of the axes of the widely used WISE colour-colour selection, with the second axis being the W2~-~W3 colour (cf. Sect~\ref{sec:previous_AGN_detection}). The WISE W3 photometry is though significantly less sensitive than W1, W2 or PS1 (see Fig.~\ref{fig:surveys_depth_HETDEX}) and a significant number of sources will be represented as upper limits in such plot (see Table~\ref{table:composition_catalogue}). From the importances in Table~\ref{table:feat_importances} and the values presented in Fig.~\ref{fig:hists_bands_nonimp_HETDEX_S82} we infer that using optical colours could in principle create selection criteria with metrics equivalent to those shown in Table~\ref{table:previous_AGN_methods} but for a much larger number of sources ($100\,000$ sources for colour plots using W3 vs $4\,700\,000$ sources for colours based in r, i or z magnitudes). We tested this hypothesis and derived a selection criterion in the g~-~r vs W1~-~W2 colour-colour plot shown in Fig.~\ref{fig:HETDEX_gr_W1W2_AGN_gal_class} using the labelled sources in the test subset of the HETDEX field. The results of the application of this criterion to the testing data and to the labelled sources in S82 is presented in the last row of Table~\ref{table:previous_AGN_methods}. Their limits are defined by the following expressions:

\begin{eqnarray}
g - r &>& -0.76\,,\\
g - r &<& 1.8\,,\\
W1 - W2 &>& 0.227 \times (g - r) + 0.43\,,
\end{eqnarray}

\noindent where W1, W2, g, and r are Vega magnitudes. Our colour criteria provides better and more homogeneous scores across the different metrics with purity (precision) and completeness (recall) above $87\%$. Avoiding the use of the longer WISE wavelength (W3 and W4), the criteria can be applied to a much larger dataset.

\begin{figure}
  \centering
    \includegraphics[width=0.90\columnwidth]{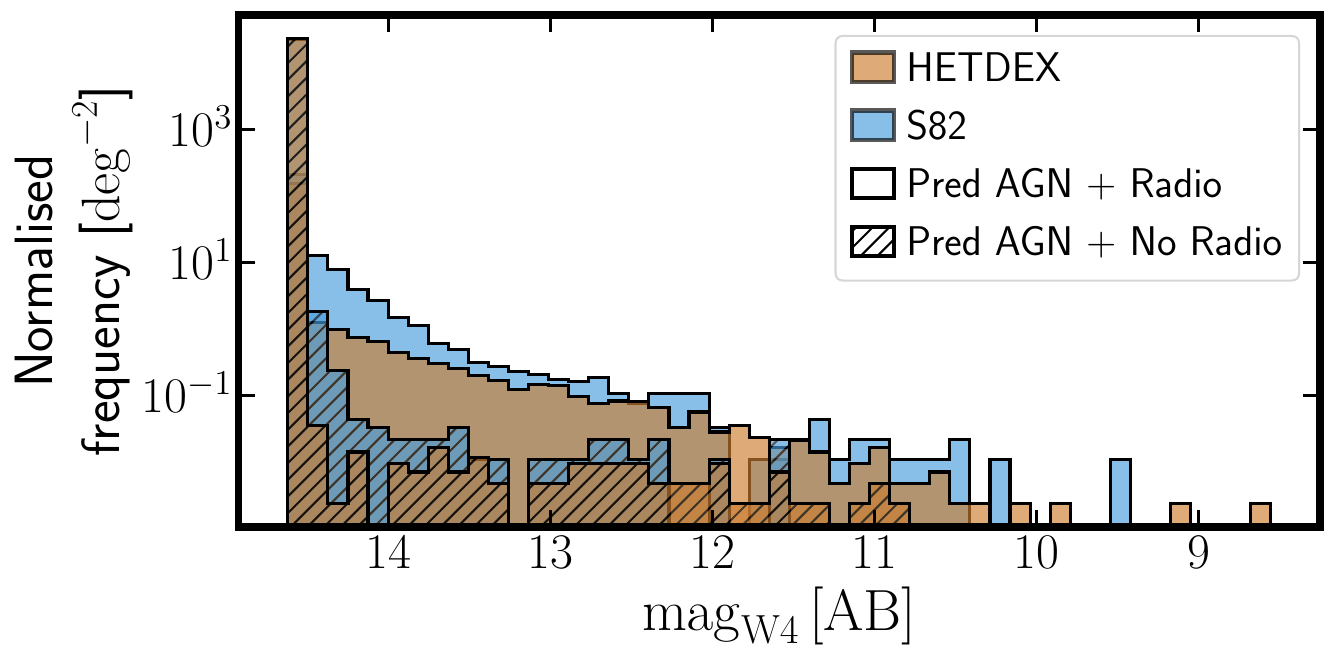}
  \caption{W4 magnitudes density distribution of the newly predicted radio AGNs (clean histograms) in HETDEX (ochre histograms) and S82 (blue histograms) and W4 magnitudes from predicted AGNs that are predicted to not have radio detection (dashed histograms).}
  \label{fig:hist_W4_nonimputed_pred_radio_non_radio_AGN}
\end{figure}

One of the main potential uses of the pipeline is its capability to pinpoint radio-detectable AGNs. The global features analysis for the radio detection model shows a high dependence on the near- and mid-IR magnitudes and colours, especially those coming from WISE. As a useful outcome similar to the AGN-galaxy classification, we can use the most relevant features to build useful plots for the pre-selection of these sources and get insight into the origin of the radio emission. This is the case for the W4 histogram, shown in Fig.~\ref{fig:hist_W4_nonimputed_pred_radio_non_radio_AGN}, where sources predicted to be radio-emitting AGNs extend to brighter measured W4 magnitudes. This added mid-IR flux might be simply due to an increased star formation rates (SFR) in these sources. In fact the $24\mu m$ flux is often used, together with that of H$\alpha$ as a proxy for SFR \citep{2009ApJ...703.1672K}. The radio detection for these sources might have a strong component linked to the ongoing SF, especially for the sources with real or predicted redshift below ${z {\sim} 1.5}$. A detailed exploration of the implications that these dependencies might have in our understanding of the triggering of radio emission on AGNs, whether related to SF or jets, is left for a future publication (Carvajal et al. in preparation).

Finally, the redshift prediction model shows again that the final estimate is mostly driven by the results of the base learners, accounting for ${\sim} 82\%$ of the predicting power. The overall combined importance of features shows also in this case a strong dependence on several near-IR colours of which {y~-~W1} and {W1~-~W2} are the most relevant ones. 
The model still relies, to a lesser extent, on a broad range of optical features needed to trace the broad range of redshift possibilities ($z \in [0,6]$).

\subsubsection{Local feature importances: Shapley values}\label{sec:shapley_values}

As opposed to the global (mean) assessment of feature importances derived from the decrease in impurity, local (i.e. source by source) information on the performance of such features can be obtained from Shapley values. This is a method from coalitional game theory that tells us how to fairly distribute the dividends (the prediction in our case) among the features \citep{Shapley_article}. The previous statement means that the relative influence of each property from the dataset can be derived for individual predictions in the decision made by the model \citep[which is not the same as obtaining causal correlations between features and the target;][]{2020arXiv200805052M}. 
The combination of Shapley values with several other model explanation methods  was used by \citet[][]{NIPS2017_7062} to create the SHapley Additive exPlanations (SHAP) values. In this work, SHAP values were calculated using the python package \verb|SHAP|\footnote{\url{https://github.com/slundberg/shap}} and, in particular, its module for tree-based predictors \citep{lundberg2020local2global}.
To speed calculations up, the package \verb|FastTreeSHAP|\footnote{\url{https://github.com/linkedin/fasttreeshap}} \citep[\texttt{v0.1.2};][]{2021arXiv210909847Y} was also used, which allows the user to run multi-thread computations. 

\begin{figure}[t]
    \centering
    \begin{minipage}{0.70\columnwidth}
    \includegraphics[width=\textwidth]{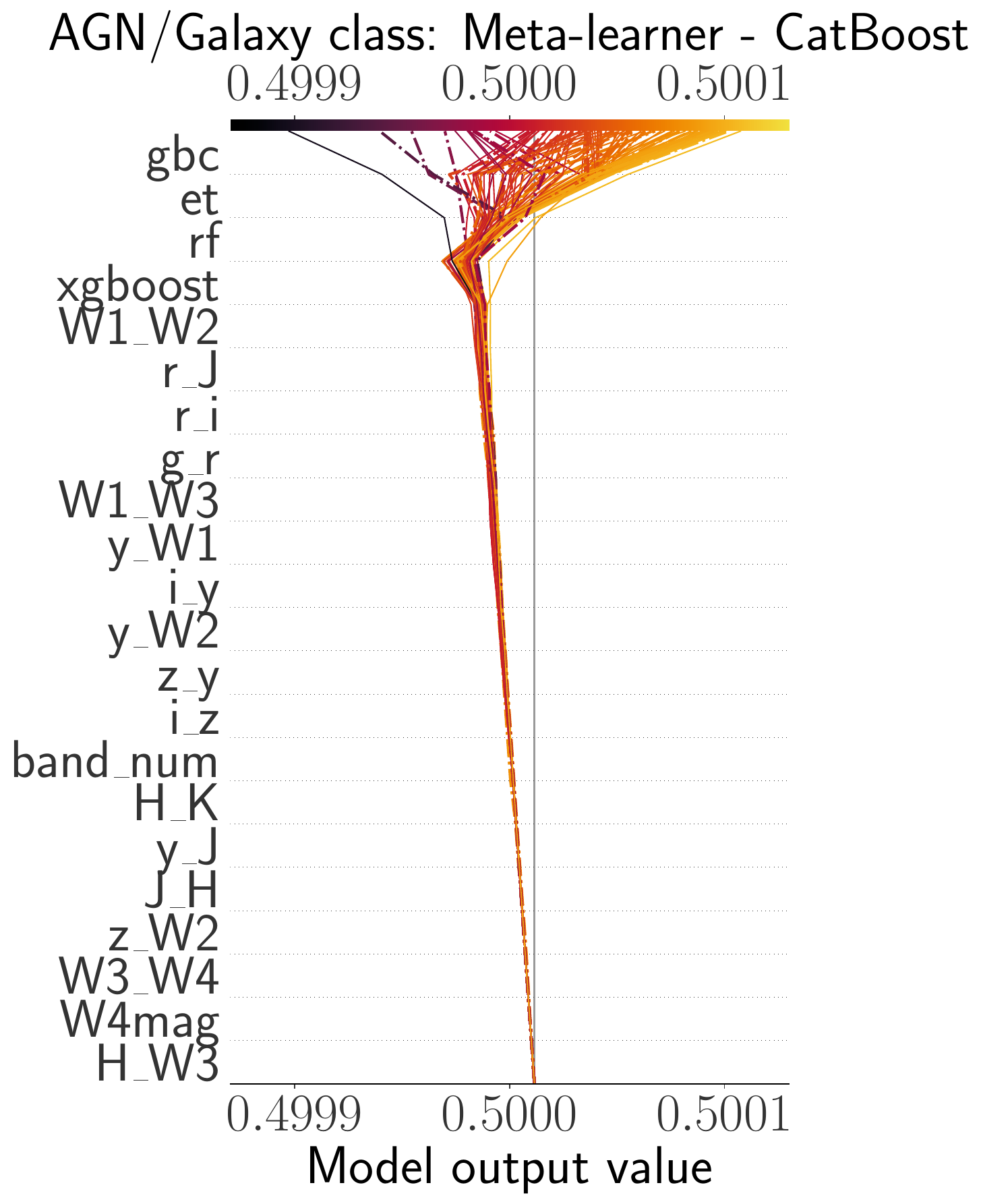}
    \end{minipage}%\\%
    \caption{Decision plot from SHAP values for AGN-galaxy classification from the $121$ high redshift ($z \geq 4$) spectroscopically confirmed AGNs in HETDEX. Horizontal axis represents the model's output with a starting value for each source centred on the selected naive threshold for classification. Vertical axis shows features used in the model sorted, from top to bottom, by decreasing mean absolute SHAP value. Each prediction is represented by a coloured line corresponding to its final predicted value as shown by the colourbar at the top. Moving from the bottom of the plot to the top, SHAP values for each feature are added to the previous value in order to highlight how each feature contributes to the overall prediction. Predictions for sources detected by LOFAR are highlighted with a dotted, dashed line.}
   \label{fig:SHAP_decision_AGN_meta_HETDEX_high_z}
\end{figure}

\begin{figure}
   \centering
   \begin{minipage}{0.70\columnwidth}
   \includegraphics[width=\textwidth]{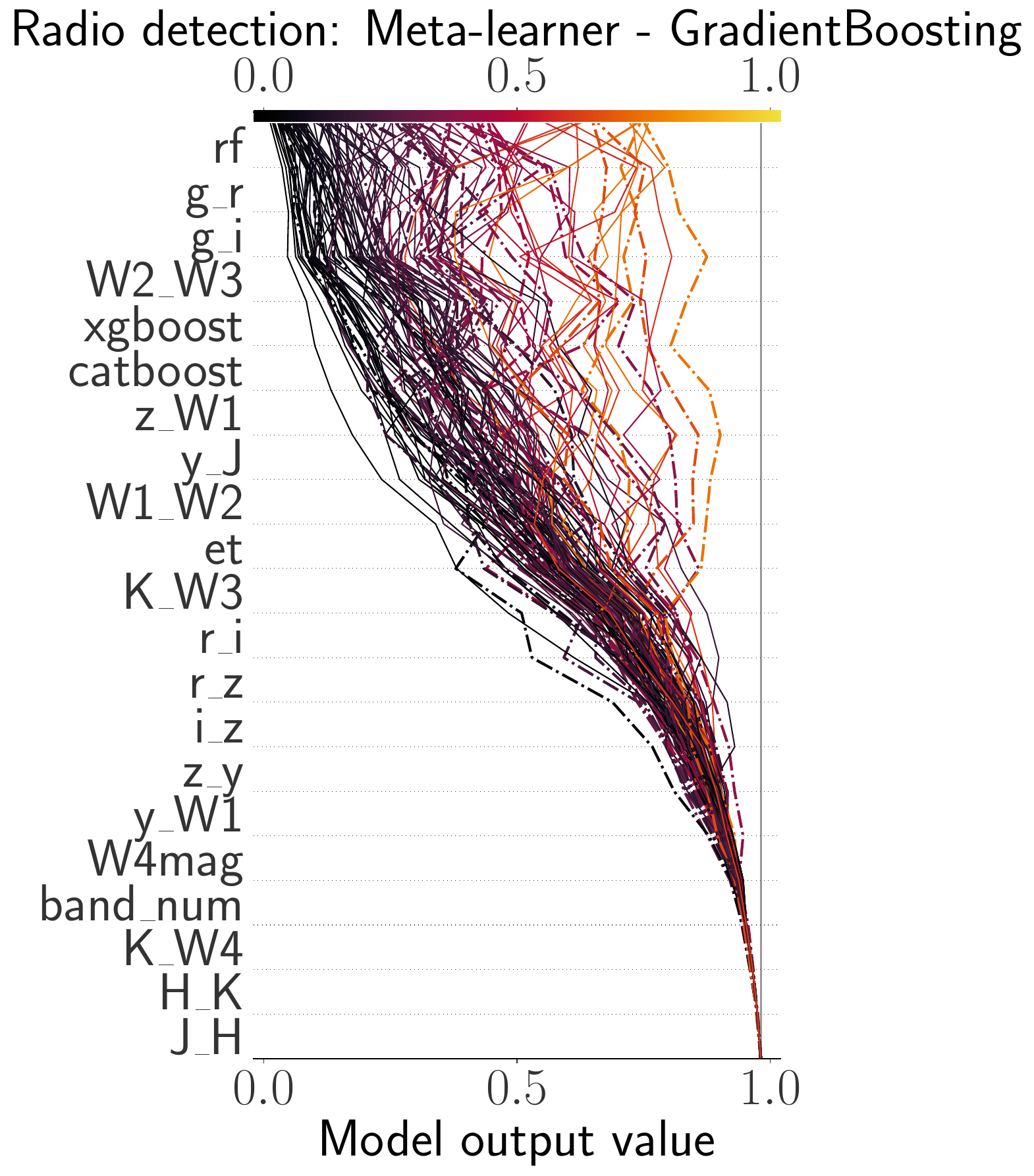}
   \end{minipage}
   \caption{Decision plot from the SHAP values for all features from the radio detection model in the $121$ high redshift ($z \geq 4$) spectroscopically confirmed AGNs from HETDEX. Description as in Fig.~\ref{fig:SHAP_decision_AGN_meta_HETDEX_high_z}.}
   \label{fig:SHAP_decision_radio_meta_HETDEX_high_z}
\end{figure}

\begin{figure}
   \centering
   \begin{minipage}{0.70\columnwidth}
   \includegraphics[width=\textwidth]{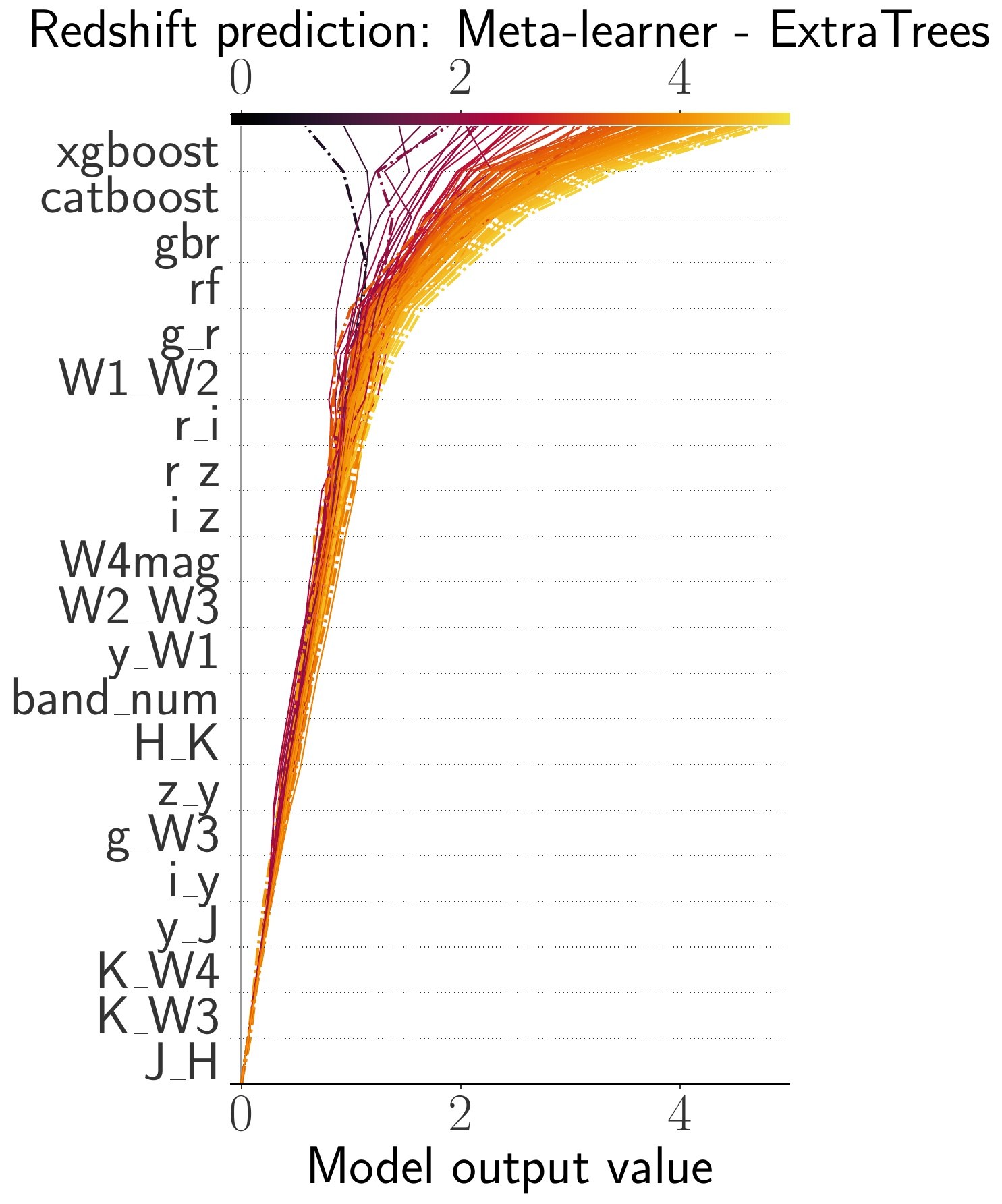}
   \end{minipage}
   \caption{Decision plot from the SHAP values for all features from the redshift prediction model in the $121$ high redshift ($z \geq 4$) spectroscopically confirmed AGNs from HETDEX. Description as in Fig.~\ref{fig:SHAP_decision_AGN_meta_HETDEX_high_z}.}
   \label{fig:SHAP_decision_z_meta_HETDEX_high_z}
\end{figure}

One way to display these SHAP values is through the so-called decision plots. They can show how individual predictions are driven by the inclusion of each feature. Besides determining the most relevant properties that help the model make a decision, it is possible to detect sources that follow different prediction paths which could be, eventually and upon further examination, labelled as outliers. An example of this decision plot, linked to the AGN-galaxy classification, is shown in Fig.~\ref{fig:SHAP_decision_AGN_meta_HETDEX_high_z} for a subsample of the high-redshift (${z \geq 4.0}$) spectroscopically classified AGNs in the HETDEX field ($121$ sources, regardless of them being part of any subset involved in the training or validation of the models). The different features used by the meta learner are stacked on the vertical axis with increasing weight and these final weight are summarised in Table~\ref{table:base_shap_values}. Similarly, SHAP decision plots for the radio detection and redshift prediction are presented in Figs.~\ref{fig:SHAP_decision_radio_meta_HETDEX_high_z} and \ref{fig:SHAP_decision_z_meta_HETDEX_high_z}, respectively.

\begin{table}[b]
\setlength{\tabcolsep}{2.9pt}
\caption{SHAP values (rescaled to add to $100$) for base algorithms in each prediction step for observed features using $121$ spectroscopically confirmed AGNs at high redshift values ($z > 4$).}             % title of Table
\label{table:base_shap_values}      % is used to refer this table in the text
\centering                          % used for centering table
\resizebox{0.65\columnwidth}{!}{
\begin{tabular}{c c c c}        % centered columns (4 columns)
\hline\hline                 % inserts double horizontal lines   
\multicolumn{4}{c}{AGN-galaxy model (\texttt{CatBoost})}             \\
Feature             & SHAP value    & Feature           & SHAP value \\
\hline
\texttt{gbc}        & 36.250        & \texttt{rf}       & 21.835     \\
\texttt{et}         & 30.034        & \texttt{xgboost}  &  7.198      \\
\multicolumn{3}{r}{Remaining SHAP values:} & 4.683 \\[0.2em]
\hline\hline
\multicolumn{4}{c}{Radio detection model (\texttt{GBC})}        \\

Feature             & SHAP value    & Feature           & SHAP value \\
\hline
\texttt{rf}         & 11.423        & \texttt{catboost} &  5.696     \\
\texttt{xgboost}    &  7.741        & \texttt{et}       &  5.115     \\
%\hline
\multicolumn{3}{r}{Remaining SHAP values:} & 70.025 \\[0.2em]
\hline\hline
\multicolumn{4}{c}{Redshift prediction model (\texttt{ET})}          \\

Feature             & SHAP value    & Feature           & SHAP value \\
\hline
\texttt{xgboost}    & 41.191        & \texttt{gbr}      & 13.106     \\
\texttt{catboost}   & 20.297        & \texttt{rf}       & 11.648     \\
%\hline
\multicolumn{3}{r}{Remaining SHAP values:} & 13.758                  \\
\hline
\end{tabular}
}
\end{table}

As it can be seen, for the three models, base learners are amongst the features with the highest influence. This result raises the question of what drives these individual base predictions. Appendix~\ref{sec:app_shap_base} includes SHAP decision plots for all base learners used in this work. Additionally, and to be able to compare these results with the features importances from Sect.~\ref{sec:feat_importances}, we constructed Table~\ref{table:shap_values_combined}, which displays the combined SHAP values of base and meta learners but, in this case, for the same $121$ high-redshift confirmed AGNs (with $29$ of them detected by LoTSS). Table~\ref{table:shap_values_combined} shows, as Table~\ref{table:feat_importances}, that the colour W1~-~W2 is the most important discriminator between AGNs and galaxies for this specific set of sources. The importance of the rest of the features is mixed: similar colours are located on the top spots (e.g. g~-~r, W1~-~W3 or r~-~i).

\begin{table}
\setlength{\tabcolsep}{2.9pt}
\caption{Combined and normalised (rescaled to add to $100$) mean absolute SHAP values for observed features from the three models using $121$ spectroscopically confirmed AGNs at high redshift values ($z \geq 4$).}             % title of Table
\label{table:shap_values_combined}      % is used to refer this table in the text
\centering                          % used for centering table
\resizebox{0.95\columnwidth}{!}{
\begin{tabular}{c c c c c c}        % centered columns (6 columns)
\hline\hline                 % inserts double horizontal lines   
\multicolumn{6}{c}{AGN-galaxy model} \\
Feature             & SHAP value    & Feature             & SHAP value    & Feature         & SHAP value \\
\hline
\texttt{W1\_W2}     & 32.458        & \texttt{i\_y}       & 5.086         & \texttt{z\_y}   & 1.591      \\
\texttt{g\_r}       & 11.583        & \texttt{y\_W1}      & 4.639         & \texttt{H\_W3}  & 1.048      \\
\texttt{W1\_W3}     &  8.816        & \texttt{band\_num}  & 4.050         & \texttt{W4mag}  & 0.514      \\
\texttt{r\_i}       &  7.457        & \texttt{y\_W2}      & 3.228         & \texttt{H\_K}   & 0.466      \\
\texttt{i\_z}       &  6.741        & \texttt{z\_W2}      & 2.348         & \texttt{W3\_W4} & 0.466      \\
\texttt{r\_J}       &  6.613        & \texttt{y\_J}       & 1.718         & \texttt{J\_H}   & 0.178      \\[0.5em]
\hline\hline
\multicolumn{6}{c}{Radio detection model} \\
Feature             & SHAP value    & Feature             & SHAP value    & Feature             & SHAP value \\
\hline
\texttt{g\_i}     & 14.120        & \texttt{z\_W1}        & 6.751         & \texttt{W4mag}      & 2.691      \\
\texttt{W2\_W3}   & 13.201        & \texttt{r\_i}         & 5.577         & \texttt{band\_num}  & 2.661      \\
\texttt{g\_r}     & 12.955        & \texttt{r\_z}         & 5.161         & \texttt{K\_W4}      & 0.939      \\
\texttt{y\_J}     &  8.224        & \texttt{i\_z}         & 4.512         & \texttt{H\_K}       & 0.719      \\
\texttt{K\_W3}    &  7.441        & \texttt{z\_y}         & 4.121         & \texttt{J\_H}       & 0.190      \\
\texttt{W1\_W2}   &  6.874        & \texttt{y\_W1}        & 3.864         &                     &            \\[0.5em]
\hline\hline
\multicolumn{6}{c}{Redshift prediction model} \\
Feature             & SHAP value    & Feature             & SHAP value    & Feature         & SHAP value \\
\hline
\texttt{g\_r}       & 32.594        & \texttt{z\_y}       & 3.557         & \texttt{W4mag}  & 1.639    \\
\texttt{y\_W1}      & 20.770        & \texttt{y\_J}       & 3.010         & \texttt{g\_W3}  & 1.479    \\
\texttt{W2\_W3}     & 12.462        & \texttt{band\_num}  & 2.595         & \texttt{K\_W3}  & 0.853    \\
\texttt{W1\_W2}     &  5.692        & \texttt{i\_y}       & 2.381         & \texttt{K\_W4}  & 0.451    \\
\texttt{r\_i}       &  4.381        & \texttt{H\_K}       & 2.230         & \texttt{J\_H}   & 0.146    \\
\texttt{r\_z}       &  3.755        & \texttt{i\_z}       & 2.005         &                 &          \\
\hline
\end{tabular}
}
\end{table}

For the radio classification step of the pipeline, we find that features linked to those $121$ high-$z$ AGNs perform at the same level as for the overall population.
The improved metrics with respect to those obtained from the no-skill selection do indicate that the model has learned some connections between the data and the radio emission. Feature importance has changed when compared to the overall population. If the radio emission observed from these sources were exclusively due to SF, this connection would imply SFR of several hundred $M_\sun$\,yr$^{-1}$. This explanation can not be completely ruled out from the model side but some contribution of radio emission from the AGN is expected. The detailed analysis of the exact contribution for the SF and AGN component will be left for a forthcoming publication (Carvajal et al. in preparation).
 
%--------------------------------------------------------------------
\section{Summary and conclusions}\label{sec:summary_conclusions}

With the ultimate intention of better understanding the triggering of radio emission in AGNs, in this paper, we have shown that it is possible to build a pipeline to detect AGNs, determine their detectability in radio, within a given flux limit, and predict their redshift value. Most importantly, we have described a series of methodologies to understand the driving properties of the different decisions, in particular for the radio detection which is, to our best knowledge, the first attempt at doing so.

We have trained the models using multi-wavelength photometry from almost $120\,000$ spectroscopically identified infrared-detected sources in the HETDEX field and created stacked models with them. These models were applied, sequentially, to $15\,018\,144$ infrared detections in the HETDEX Spring field, leading to the creation of $68\,252$ radio AGNs candidates with their corresponding predicted redshift values. Additionally, we applied the models to $3\,568\,478$ infrared detections in the S82 field, obtaining $22\,445$ new radio AGNs candidates with their predicted redshift values.

We then applied a number of analyses to the models to understand the influence of the observed properties over the predictions and their confidence levels. In particular, the use of SHAP values gives the opportunity to extract the influence that the feature set has for each individual prediction. From the application of the prediction pipeline on labelled and unlabelled sources and the analysis of the predictions and the models themselves, the following conclusions can be drawn.

\begin{itemize}
\item Generalised stacking is a useful procedure which collects results from individual ML algorithms into a single model that can outperform each of the individual models, while preventing the inclusion of biases from individual algorithms. Proper selection of models and input features, together with detailed probability and threshold calibration, maximises the metrics of the final model.
\item Classification between AGNs and galaxies derived from our model is in line with previous works. Our pipeline is able to retrieve a high fraction of previously classified AGNs from HETDEX (recall $= 0.9621$, precision $= 0.9449$) and from the S82 field (recall $= 0.9401$, precision $= 0.9481$).
\item Radio detection classification for predicted AGNs has proven to be highly demanding in terms of the data needed for creating the models. Thanks to the use of the techniques shown in this article (i.e. feature creation and selection, generalised stacking, probability calibration, and threshold optimisation), we were able to retrieve previously known radio-detectable AGNs in the HETDEX field (recall $= 0.5216$, precision $= 0.3528$) and in the S82 field (recall $= 0.5816$, precision $= 0.1229$). These rates improve significantly upon a purely random selection ($4$ times better for the HETDEX field and $13$ times better for S82), showing the power of ML methods for obtaining new RG candidates.
\item The prediction of redshift values for sources classified as radio-detectable AGNs can deliver results that are in line with works that use either traditional or ML methods. The good quality of these predictions is achieved despite the fact of them being produced after two previous ML steps (the two classifications of the pipeline), which might introduce large uncertainties to their values.
\item Our models (classification and regression) can be applied to areas of the sky that have different radio coverage from that used for training without a strong degradation of the prediction results. This feature can lead to the use of our pipeline over very distinct datasets (in radio and multi-wavelength coverage) expecting to recover the sources predicted to be radio-detectable AGNs with a high probability.
\item Machine-learning models cannot be only used for a direct prediction of a value (or a set of values). They can also be subject to analyses that allow additional results to be extracted. We took advantage of this fact by using global and local feature importances to derive novel colour-colour AGN selection methods.
\end{itemize}

The next generation of observatories is already producing source catalogues with an order of magnitude better sensitivity over large areas of the sky than was previously the case. Some examples of these catalogues and surveys include the Rapid ASKAP Continuum Survey \citep[RACS;][]{2020PASA...37...48M}, EMU \citep{2011PASA...28..215N}, and the MeerKAT International GHz Tiered Extragalactic Exploration \citep[MIGHTEE;][]{2016mks..confE...6J}. With the increased number of radio detections, the need to understand the fraction of those detections related to AGNs and to determine counterparts across wavelengths is more necessary than ever.

Although we developed the pipeline as a tool to better understand the aforementioned issues, we foresee additional possibilities in which the pipeline can be of great use. The first possibility involves the use of the pipeline to assist with the selection of radio-detectable AGNs within any set of observations. This application might turn out to be particularly valuable in recent surveys carried out with MeerKAT \citep{2016mks..confE...1J} or the future SKA where the population at the faintest sources will be dominated by star-forming galaxies. This change needs to use the corresponding data in the training set.

Future developments of the pipeline will concentrate on minimising the existent biases in the training sample as well as in increasing the coverage of the parameter space. We also plan to generalise the pipeline to make it useful for non-radio or galaxy-related research communities. These developments include, for instance, the capability to carry the full analysis for the galactic and stellar populations (i.e. models to determine if a galaxy can be detected in the radio and to predict redshift values for galaxies and non-radio AGNs).

In order to increase the parameter space of our training sets, we plan to include information from radio surveys with different properties in terms of covered area and multi-wavelength coverage. In particular, we aim to include far-IR, X-ray, and multi-survey radio measurements from larger areas. The inclusion of a larger, and possibly deeper, set of measurements makes part of our goal to improve detections, not only in radio, but in additional wavelengths.

\begin{acknowledgements}
We thank the anonymous referee for their valuable comments and constructive suggestions which have greatly improved the manuscript. The authors would also like to thank insightful comments from P. Papaderos and B. Arsioli.
This work was supported by Funda\c{c}\~{a}o para a Ci\^{e}ncia e a Tecnologia (FCT) through research grants PTDC/FIS-AST/29245/2017, EXPL/FIS-AST/1085/2021, UID/FIS/04434/2019, UIDB/04434/2020, and UIDP/04434/2020. RC acknowledges support from the Funda\c{c}\~{a}o para a Ci\^{e}ncia e a Tecnologia (FCT) through the Fellowship PD/BD/150455/2019 (PhD:SPACE Doctoral Network PD/00040/2012) and POCH/FSE (EC). IM acknowledges support from DL 57/2016 (P2461) from the `Departamento de F\'{i}sica, Faculdade de Ci\^{e}ncias da Universidade de Lisboa'. AH acknowledges support from contract DL 57/2016/CP1364/CT0002 and an FCT-CAPES funded Transnational Cooperation project ``Strategic Partnership in Astrophysics Portugal-Brazil''. PACC acknowledges financial support by the Funda\c{c}\~{a}o para a Ci\^{e}ncia e a Tecnologia (FCT) through the grant 2022.11477.BD. DB acknowledges support from the Funda\c{c}\~{a}o para a Ci\^{e}ncia e a Tecnologia (FCT) through the Fellowship UI/BD/152315/2021. HM acknowledges support from the Funda\c{c}\~{a}o para a Ci\^{e}ncia e a Tecnologia (FCT) through the PhD Fellowship 2022.12891.BD. APA acknowledges support from the Funda\c{c}\~{a}o para a Ci\^{e}ncia e a Tecnologia (FCT) through the work Contract No. 2020.03946.CEECIND. CP acknowledges support from DL 57/2016 (P2460) from the `Departamento de F\'{i}sica, Faculdade de Ci\^{e}ncias da Universidade de Lisboa'.
This publication makes use of data products from the Wide-field Infrared Survey Explorer, which is a joint project of the University of California, Los Angeles, and the Jet Propulsion Laboratory/California Institute of Technology, funded by the National Aeronautics and Space Administration.
LOFAR data products were provided by the LOFAR Surveys Key Science project (LSKSP\footnote{\url{https://lofar-surveys.org/}}) and were derived from observations with the International LOFAR Telescope (ILT). LOFAR \citep{2013A&A...556A...2V} is the Low Frequency Array designed and constructed by ASTRON. It has observing, data processing, and data storage facilities in several countries, which are owned by various parties (each with their own funding sources), and which are collectively operated by the ILT foundation under a joint scientific policy. The efforts of the LSKSP have benefited from funding from the European Research Council, NOVA, NWO, CNRS-INSU, the SURF Co-operative, the UK Science and Technology Funding Council and the J\"{u}lich Supercomputing Centre.
The Pan-STARRS1 Surveys (PS1) and the PS1 public science archive have been made possible through contributions by the Institute for Astronomy, the University of Hawaii, the Pan-STARRS Project Office, the Max-Planck Society and its participating institutes, the Max Planck Institute for Astronomy, Heidelberg and the Max Planck Institute for Extraterrestrial Physics, Garching, The Johns Hopkins University, Durham University, the University of Edinburgh, the Queen's University Belfast, the Harvard-Smithsonian Center for Astrophysics, the Las Cumbres Observatory Global Telescope Network Incorporated, the National Central University of Taiwan, the Space Telescope Science Institute, the National Aeronautics and Space Administration under Grant No. NNX08AR22G issued through the Planetary Science Division of the NASA Science Mission Directorate, the National Science Foundation Grant No. AST-1238877, the University of Maryland, E\"{o}tv\"{o}s Lor\'{a}nd University (ELTE), the Los Alamos National Laboratory, and the Gordon and Betty Moore Foundation.
This publication makes use of data products from the Two Micron All Sky Survey, which is a joint project of the University of Massachusetts and the Infrared Processing and Analysis Center/California Institute of Technology, funded by the National Aeronautics and Space Administration and the National Science Foundation.
This work made use of public data from the Sloan Digital Sky Survey, Data Release 16. Funding for the Sloan Digital Sky Survey IV has been provided by the Alfred P. Sloan Foundation, the U.S. Department of Energy Office of Science, and the Participating Institutions. 
SDSS-IV acknowledges support and resources from the Center for High Performance Computing  at the University of Utah. The SDSS website is \url{www.sdss.org}.
SDSS-IV is managed by the Astrophysical Research Consortium for the Participating Institutions of the SDSS Collaboration including the Brazilian Participation Group, the Carnegie Institution for Science, Carnegie Mellon University, Center for Astrophysics | Harvard \& Smithsonian, the Chilean Participation Group, the French Participation Group, Instituto de Astrof\'isica de Canarias, The Johns Hopkins University, Kavli Institute for the Physics and Mathematics of the Universe (IPMU) / University of Tokyo, the Korean Participation Group, Lawrence Berkeley National Laboratory, Leibniz Institut f\"ur Astrophysik Potsdam (AIP),  Max-Planck-Institut f\"ur Astronomie (MPIA Heidelberg), Max-Planck-Institut f\"ur Astrophysik (MPA Garching), Max-Planck-Institut f\"ur Extraterrestrische Physik (MPE), National Astronomical Observatories of China, New Mexico State University, New York University, University of Notre Dame, Observat\'ario Nacional / MCTI, The Ohio State University, Pennsylvania State University, Shanghai Astronomical Observatory, United Kingdom Participation Group, Universidad Nacional Aut\'onoma de M\'exico, University of Arizona, University of Colorado Boulder, University of Oxford, University of Portsmouth, University of Utah, University of Virginia, University of Washington, University of Wisconsin, Vanderbilt University, and Yale University.
This research has made use of NASA's Astrophysics Data System, TOPCAT\footnote{\url{http://www.star.bris.ac.uk/~mbt/topcat/}} \citep{2005ASPC..347...29T}, JupyterLab\footnote{\url{https://jupyter.org}} \citep{jupyter}, `Aladin sky atlas' \citep[\texttt{v11.0.24};][]{2000A&AS..143...33B} developed at CDS, Strasbourg Observatory, France, and the VizieR catalogue access tool, CDS, Strasbourg, France (DOI: 10.26093/cds/vizier). The original description of the VizieR service was published in \cite{vizier}.
This work made extensive use of the Python packages \texttt{PyCaret}\footnote{\url{https://pycaret.org}} \citep[\texttt{v2.3.10};][]{PyCaret}, \texttt{scikit-learn} \citep[\texttt{v0.23.2};][]{scikit-learn}, \texttt{pandas}\footnote{\url{https://pandas.pydata.org}} \citep[\texttt{v1.4.2};][]{pandas}, \texttt{Astropy}\footnote{\url{https://www.astropy.org}}, a community-developed core Python package for Astronomy \citep[\texttt{v5.0};][]{astropy:2013, astropy:2018, 2022ApJ...935..167A}, \texttt{Matplotlib} \citep[\texttt{v3.5.1};][]{Hunter:2007}, \texttt{betacal}\footnote{\url{https://betacal.github.io}} (\texttt{v1.1.0}), and \texttt{CMasher}\footnote{\url{https://github.com/1313e/CMasher}} \citep[\texttt{v1.6.3};][]{2020JOSS....5.2004V}.
\end{acknowledgements}


\bibliographystyle{bibtex/aa} % style aa.bst
\bibliography{ML_RG_article} % your references Yourfile.bib. bibliog.bib

\begin{appendix} %First appendix

\section{Column names in this study}\label{sec:app_feature_names}

Table~\ref{table:feature_names_in_work} presents the names (and what they represent) of the features, used in throughout this work. This information can be read in combination with the columns presented in Appendix~\ref{sec:app_prediction_results}.

\begin{table}
\setlength{\tabcolsep}{0.45pt}
\caption{Names of columns or features used in the code and what they represent.}             % title of Table
\label{table:feature_names_in_work}      % is used to refer this table in the text
\centering                          % used for centering table
\resizebox{0.99\columnwidth}{!}{
\begin{tabular}{c c c c c c}        % centered columns (6 columns)
\hline\hline                 % inserts double horizontal lines   
\multicolumn{6}{c}{Photometry measurements (magnitudes and fluxes)} \\
Code name       & Feature   & Code name     & Feature   & Code name         & Feature \\
\hline
\texttt{gmag}   & g (PS1)   & \texttt{ymag} & y (PS1)   & \texttt{W1mproPM} & W1 (CW) \\
\texttt{rmag}   & r (PS1)   & \texttt{Jmag} & J (2M)    & \texttt{W1mproPM} & W2 (CW) \\
\texttt{imag}   & i (PS1)   & \texttt{Hmag} & H (2M)    & \texttt{W3mag}    & W3 (AW) \\
\texttt{zmag}   & z (PS1)   & \texttt{Kmag} & Ks (2M)   & \texttt{W4mag}    & W4 (AW) \\[0.25em]
\hline\hline                 % inserts double horizontal lines   
\multicolumn{6}{c}{Colours} \\
\hline
\multicolumn{6}{c}{$66$ colours from all combinations of non-radio magnitudes.}\\
\multicolumn{6}{c}{A sub-sample of them is shown.}\\
\texttt{g\_r}    & g - r (PS1) & \ldots & \ldots & \texttt{W2\_W3} & W2 (CW) - W3 (AW) \\
\texttt{g\_i}    & g - i (PS1) & \ldots & \ldots & \texttt{W2\_W4} & W2 (CW) - W4 (AW) \\
\texttt{g\_z}    & g - z (PS1) & \ldots & \ldots & \texttt{W3\_W4} & W3 - W4 (AW) \\[0.25em]
\hline\hline                 % inserts double horizontal lines   
\multicolumn{6}{c}{Categorical flags} \\
Code name                           & \multicolumn{2}{c}{Feature}           &   &   &   \\
\hline
\multirow{2}{*}{\texttt{band\_num}} & \multicolumn{2}{c}{Number of bands}   &   &   &   \\
                                    & \multicolumn{2}{c}{with measurements} &   &   &   \\[0.25em]
\hline\hline                 % inserts double horizontal lines   
\multicolumn{6}{c}{Boolean flags} \\
Code name       & Feature       & Code name                 & \multicolumn{3}{c}{Feature}                                   \\
\hline
\texttt{class}  & AGN or galaxy & \texttt{radio\_detect}    & \multicolumn{3}{c}{Detection in, at least, one radio band.}   \\[0.25em]
\hline\hline                 % inserts double horizontal lines   
\multicolumn{6}{c}{Redshift} \\
Code name   & \multicolumn{2}{c}{Feature}                   &   &   &   \\
\hline
\texttt{Z}  & \multicolumn{2}{c}{Spectroscopic redshift}    &   &   &   \\[0.25em]
\hline\hline                 % inserts double horizontal lines   
\multicolumn{6}{c}{Outputs of base models} \\
Code name           & Feature       & Code name                     & Feature           & Code name                     & Feature           \\
\hline
\texttt{XGBoost}    & XGBoost       & \texttt{ET}                   & Extra Trees       & \multirow{2}{*}{\texttt{GBR}} & Gradient Boosting \\
\texttt{CatBoost}   & CatBoost      & \multirow{2}{*}{\texttt{GBC}} & Gradient Boosting &                               & Regressor         \\
\texttt{RF}         & Random Forest &                               & Classifier        &                               &                   \\
\hline                                   %inserts single line
\end{tabular}
}
\end{table}

\section{Non-imputed magnitudes}\label{sec:app_nonimputed_mag_dist}

The number of valid measurements in Fig.~\ref{fig:hists_bands_nonimp_HETDEX_S82} for each field and band can be used to determine the relative difference of density of sources between both fields. This density can be obtained by dividing the number of valid measurements over the effective area of each field (Sect.~\ref{sec:data}). Table~\ref{table:magnitude_density} shows these densities.

\begin{table}
\setlength{\tabcolsep}{2.9pt}
\caption{Density of detected sources (in units of sources per square degree) per band and field.}             % title of Table
\label{table:magnitude_density}      % is used to refer this table in the text
\centering                          % used for centering table
\resizebox{0.87\columnwidth}{!}{
\begin{tabular}{c c c c c c c c}        % centered columns (8 columns)
\hline\hline                 % inserts double horizontal lines   
\multicolumn{8}{c}{HETDEX field} \\
Band    & Density         & Band  & Density         & Band  & Density         & Band  & Density       \\
        & (deg$^{-2}$)    &       & (deg$^{-2}$)    &       & (deg$^{-2}$)    &       & (deg$^{-2}$)  \\
\hline
g       &  6\,380.66      & z     & 10\,331.93      & H     &  1\,335.55      & W2    & 35\,700.18    \\
r       &  9\,304.58      & y     &  6\,735.97      & Ks    &  1\,335.55      & W3    & 14\,045.08    \\
i       & 11\,242.35      & J     &  1\,335.55      & W1    & 35\,700.18      & W4    & 14\,044.78    \\[0.25em]
\hline\hline
\multicolumn{8}{c}{S82 field} \\
Band    & Density         & Band  & Density         & Band  & Density         & Band  & Density       \\
        & (deg$^{-2}$)    &       & (deg$^{-2}$)    &       & (deg$^{-2}$)    &       & (deg$^{-2}$)  \\
\hline
g       &  8\,249.04      & z     & 13\,214.70      & H     &  2\,330.92      & W2    & 39\,025.05    \\
r       & 12\,962.35      & y     &  9\,226.45      & Ks    &  2\,330.92      & W3    & 15\,393.12    \\
i       & 14\,507.01      & J     &  2\,330.92      & W1    & 39\,025.01      & W4    & 15\,472.75    \\
\hline
\end{tabular}
}
\end{table}

\section{From scores to probabilities}\label{app:calibration_models}

\begin{figure}
  \centering
  \begin{minipage}{0.24\textwidth}
    \centering
    \includegraphics[width=0.99\textwidth]{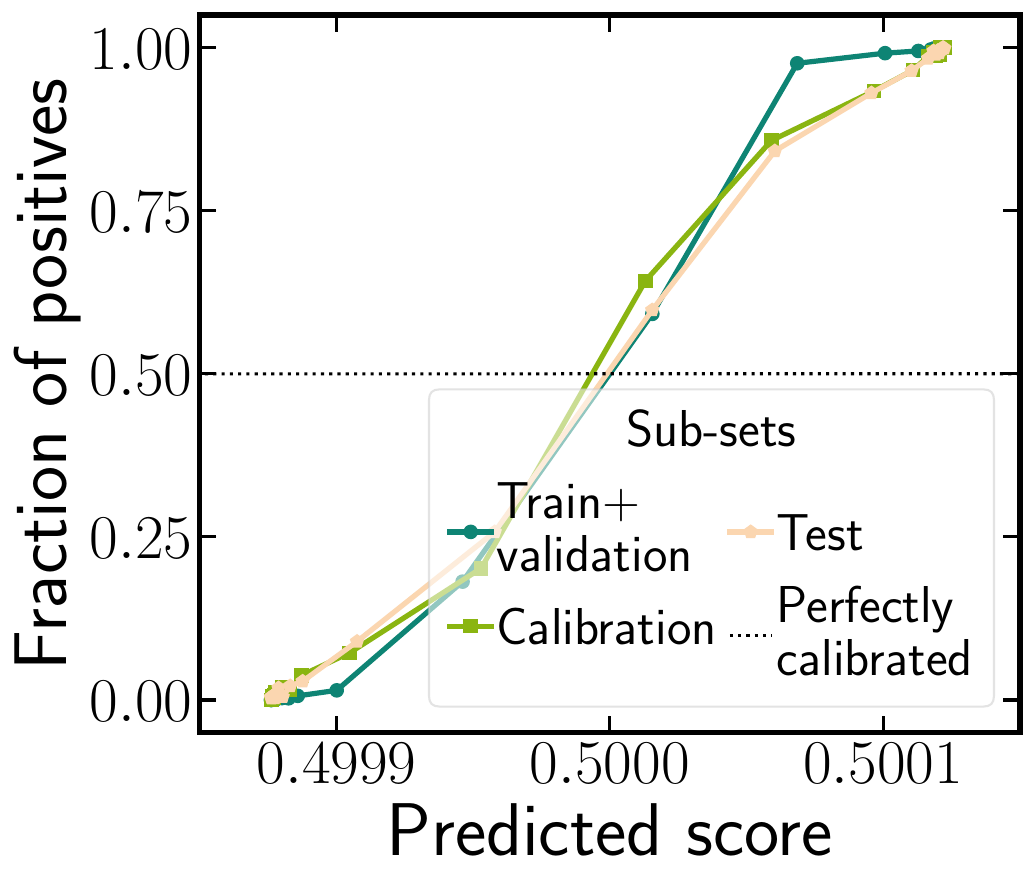}\hfill\break%\linebreak
    {(a) AGN-galaxy}
  \end{minipage}
  \hfill 
  \begin{minipage}{0.24\textwidth}
    \centering
    \includegraphics[width=0.99\textwidth]{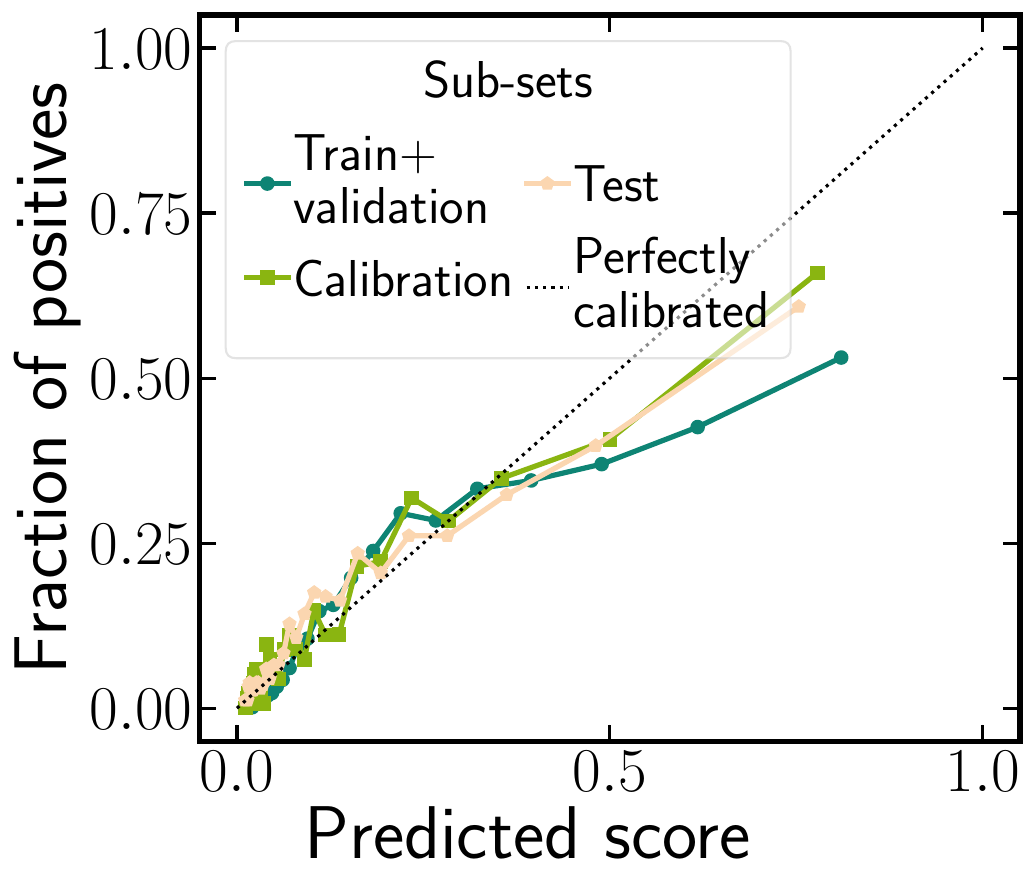}\hfill\break%\linebreak
    {(b) Radio detection}
  \end{minipage}
  \caption{Reliability curves for uncalibrated classifiers. Each line represents the calibration curve for each subset in HETDEX field. Data has been binned and each bin (represented by the points) has the same number of elements per curve. Dashed line represents a perfectly calibrated model. (a) AGN-galaxy classification model. (b) Radio detection model.}
  \label{fig:calibration_curves_classification_pre}
\end{figure}

\begin{figure}
  \centering
  \begin{minipage}{0.24\textwidth}
    \centering
    \includegraphics[width=0.99\textwidth]{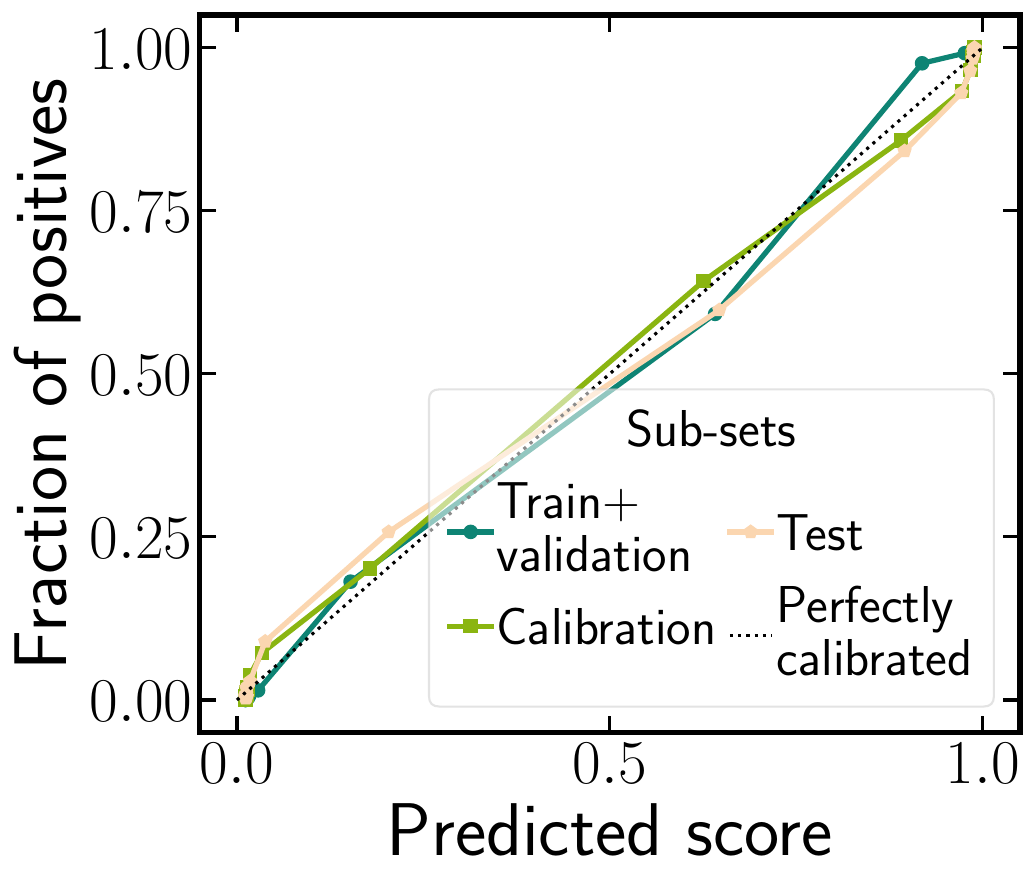}\hfill\break%\linebreak
    {(a) AGN-galaxy}
  \end{minipage}
  \hfill 
  \begin{minipage}{0.24\textwidth}
    \centering
    \includegraphics[width=0.99\textwidth]{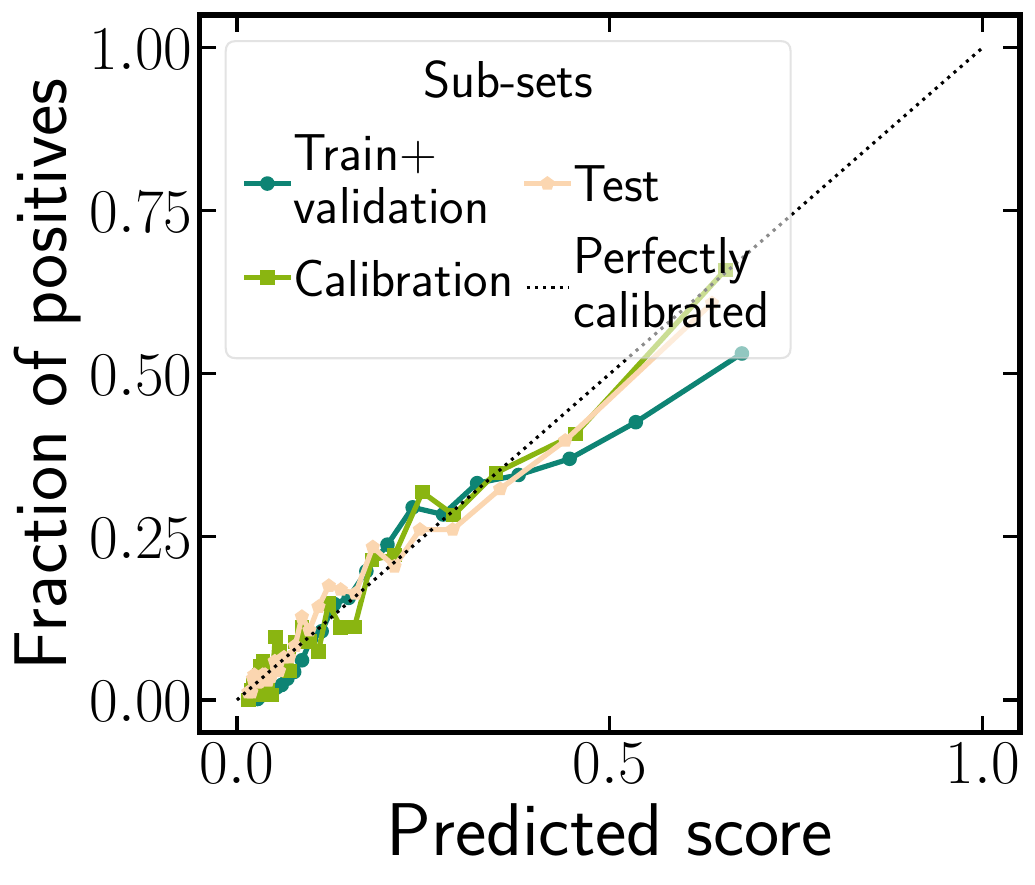}\hfill\break%\linebreak
    {(b) Radio detection}
  \end{minipage}
  \caption{Reliability curves for calibrated classifiers. (a) AGN-galaxy classification model. (b) Radio detection model. Details as in Fig.~\ref{fig:calibration_curves_classification_pre}.}
  \label{fig:calibration_curves_classification_post}
\end{figure}

In general, classifiers deliver scores in the range [$0$, $1$], which could be associated to the probability of a studied source being part of the relevant class (in our work, AGN or radio detectable). The classifier uses a threshold above which, any predicted element would be considered a positive instance. 

With the exception of few algorithms (including the family of logistic regressions), scores from classifiers cannot be directly used as probabilities. As a consequence of this inability, such values cannot be compared from one type of model to some other and can not be combined to obtain a joint score.
Therefore, in order to retrieve joint scores and treat them as probabilities, scores (and, by extension, the classifiers) need to be calibrated. This calibration means that, when taking all predictions with a probability $P$ of being of a class, a fraction $P$ of them really belong to that class \citep[e.g.][]{lichtenstein_1982, SilvaFilho2023}.

Calibration of these scores can be done by applying a transformation to their values. For our work, we will apply a Beta transformation. It allows one to re-distribute the scores of a classifier allowing them to get closer to the definition of probability \citep{10.1214/17-EJS1338SI, pmlr-v54-kull17a}. Calibration steps in our workflow have been applied using the Python package \verb|betacal|. In the case of the radio detection model, the new scores have a wider range than the original, uncalibrated scores.

When obtaining the BSS values for both classification, the AGN-galaxy classifier has a score of ${\mathrm{BSS} = -0.002}$, demonstrating that no major changes were applied to the distribution of scores. For the radio detection classifier, the score is ${\mathrm{BSS} = -0.434}$. Even though the BSS value is slightly negative for the AGN-galaxy classifier, we keep it since its range of values now can be compared and combined with additional probabilities. In the case of the radio detection classifier, the BSS shows a degradation of the calibration, but we will keep the calibrated model given that it provides, overall, better values for the remaining metrics. This effect can be seen, more strongly, with recall.

Calibration (or reliability) plots show how well calibrated the predicted scores of a classifier are by displaying the fraction of sources that are part of a given class as a function of the predicted probability. A perfectly calibrated classifier would have all its prediction lying in the ${x{=}y}$ line. The magnitude of the deviations from that line give information of the miscalibration a model has \citep[see, for instance,][]{ReliabilityofReliabilityDiagrams, VanCalster2019}. In Fig.~\ref{fig:calibration_curves_classification_pre}, we present the reliability curves for the uncalibrated classifiers and, in Fig.~\ref{fig:calibration_curves_classification_post}, for their calibrated versions.

\section{Meta learners hyperparameters}\label{sec:app_hyperpars}

\begin{table}
\setlength{\tabcolsep}{2.9pt}
\caption{Values of Hyperparameters for meta learners after tuning.}             % title of Table
\label{table:hyper_params_meta}      % is used to refer this table in the text
\centering                          % used for centering table
\resizebox{0.85\columnwidth}{!}{
\begin{tabular}{c c c c}        % centered columns (4 columns)
\hline\hline                 % inserts double horizontal lines   
\multicolumn{4}{c}{AGN-galaxy model (\texttt{CatBoost})}                                                \\
Parameter                           & Value             & Parameter                     & Value         \\
\hline
\texttt{learning\_rate}             & 0.0075            & \texttt{random\_strength}     & 0.1           \\
\texttt{depth}                      & 6                 & \texttt{l2\_leaf\_reg}        & 10            \\[0.2em]
\hline\hline
\multicolumn{4}{c}{Radio detection model (\texttt{GradientBoosting})}                                           \\

Parameter                           & Value             & Parameter                     & Value         \\
\hline
\texttt{n\_estimators}              & 187               & \texttt{min\_samples\_leaf}   & 2             \\
\texttt{learning\_rate}             & 0.0560            & \texttt{max\_depth}           & 9             \\
\texttt{subsample}                  & 0.3387            & \texttt{max\_features}        & 0.5248        \\
\texttt{min\_samples\_split}        & 5                 &                               &               \\[0.2em]
\hline\hline
\multicolumn{4}{c}{Redshift prediction model (\texttt{ET})}                                             \\

Parameter                           & Value             & Parameter                     & Value         \\
\hline
\texttt{n\_estimators}              & 100               & \texttt{criterion}            & \texttt{mae}  \\
\texttt{max\_depth}                 & \texttt{None}     & \texttt{min\_samples\_split}  & 2             \\
\texttt{max\_features}              & \texttt{auto}     & \texttt{min\_samples\_leaf}   & 1             \\
\texttt{bootstrap}                  & \texttt{False}    &                               &               \\
\hline
\end{tabular}
}
\tablefoot{This table shows the parameters which were subject to tuning. Remaining hyperparameters used their default values as defined by their developers.}
\end{table}

In Table~\ref{table:hyper_params_meta}, we present the optimised hyperparameters from our meta learners. For all three instances of modelling (AGN-galaxy, radio detection, and redshift), hyperparameters were optimised using the \verb|SkoptSearch| algorithm embedded in the package \verb|tune-sklearn|\footnote{\url{https://github.com/ray-project/tune-sklearn}} \citep[\texttt{v0.4.1};][]{head_tim_2021_5565057}, which implements a Bayesian search in the hyperparameter space.

\section{PR curve threshold optimisation}\label{sec:app_pr_curve}

\begin{figure}
  \centering
  \begin{minipage}{0.24\textwidth}
    \centering
    \includegraphics[width=0.98\textwidth]{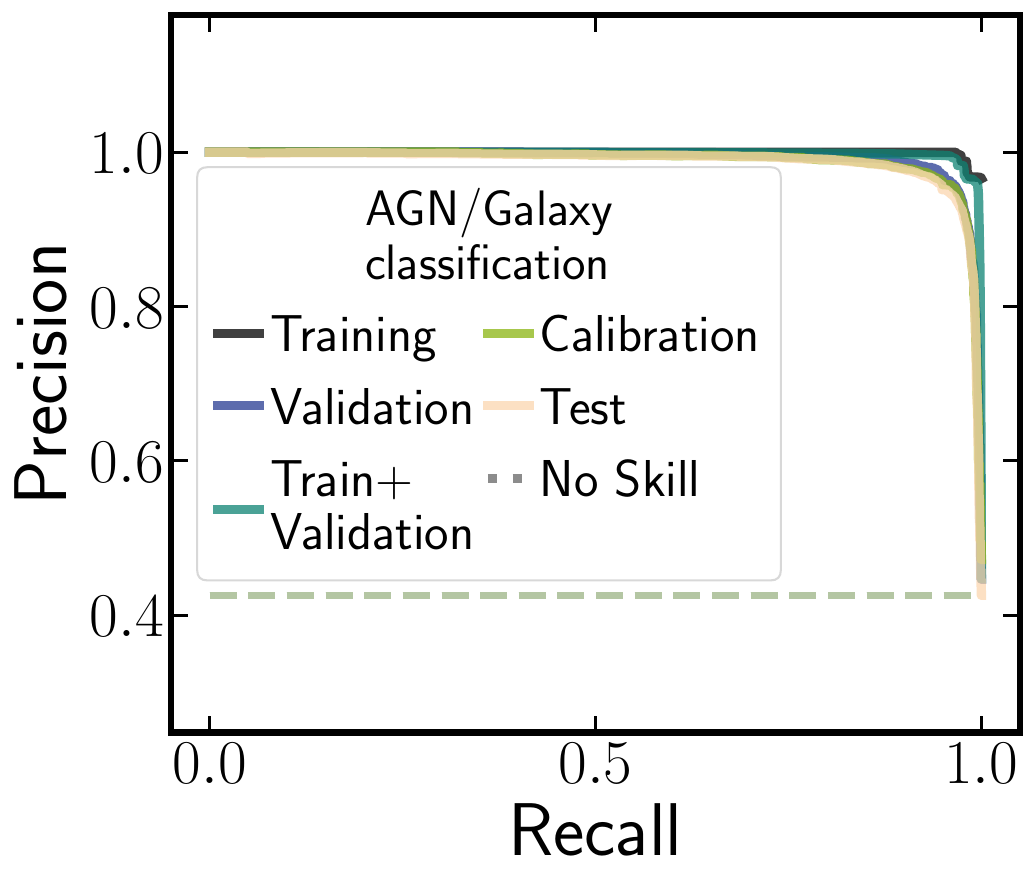}\hfill\break
    {(a) AGN-galaxy} 
  \end{minipage}
  \hfill 
  \begin{minipage}{0.24\textwidth}
    \centering
    \includegraphics[width=0.98\textwidth]{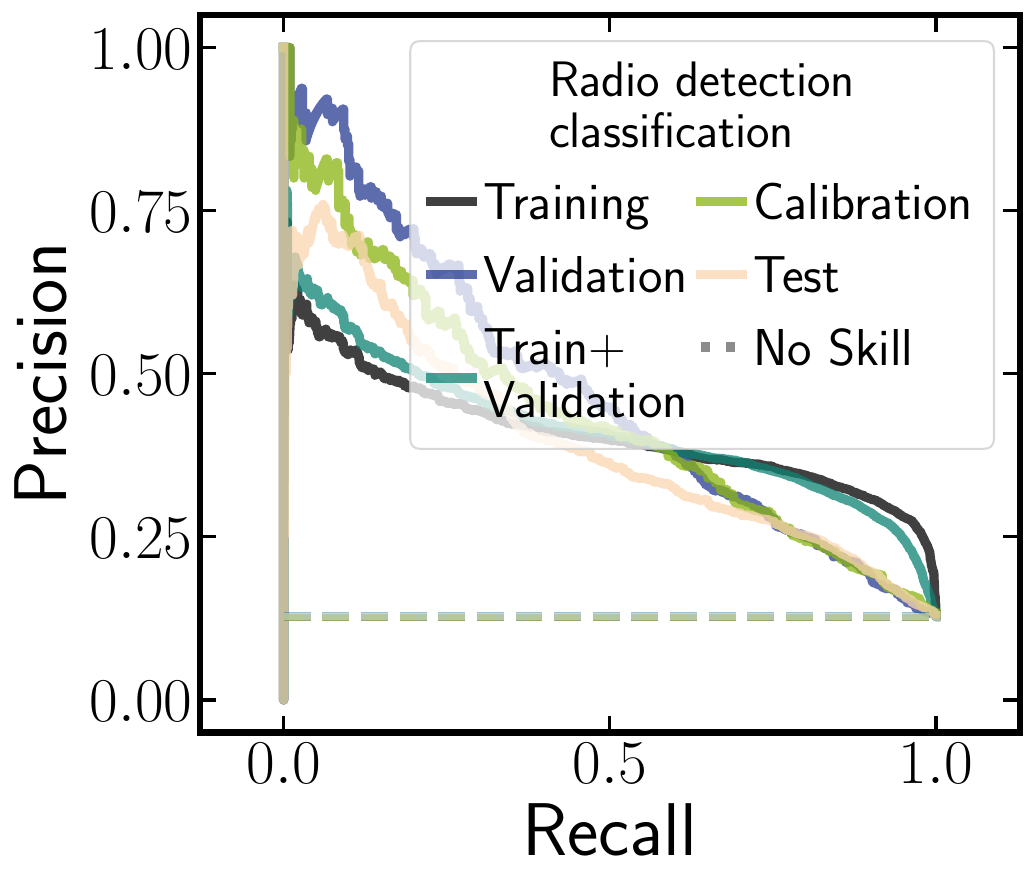}\hfill\break
    {(b) Radio detection}
  \end{minipage}
  \centering
  \caption{Precision-recall curves for the (a) AGN-galaxy and (b) radio detection classification models.}
  \label{fig:PR_curves_classification}
\end{figure}

By maximising the recall  (Eq.~\ref{eq:recall}), we improve the number of recovered elements in each classifier. This can be done by decreasing the threshold by which a source is classified as a positive instances. Setting this threshold to its minimum, $0.0$, would  increase the recall. But every source would be predicted to be an AGN or detected on the radio regardless of their properties.

One statistical tool designed to optimise the classification threshold taking into account the overall model performance is the PR curve. It can help to understand the behaviour of a classifier as a function of its threshold. Both quantities, precision (Eq.~\ref{eq:precision}) and recall, show an inverse correlation, and both depend on the selected threshold. Thus, they can be used to retrieve the score value for which both quantities are balanced. This optimisation is done by finding the threshold that maximises the $\mathrm{F}_{\beta}$ score (Eq.~\ref{eq:f_beta}). This operation is performed over the union of training and validation sets, which have been used to create and train each model. PR curves for all subsets used in our classification models are shown in Fig.~\ref{fig:PR_curves_classification}.

\section{SHAP values for base models}\label{sec:app_shap_base}

\begin{figure}
  \centering
  \begin{minipage}{0.48\columnwidth}
    \centering
    \includegraphics[width=\textwidth]{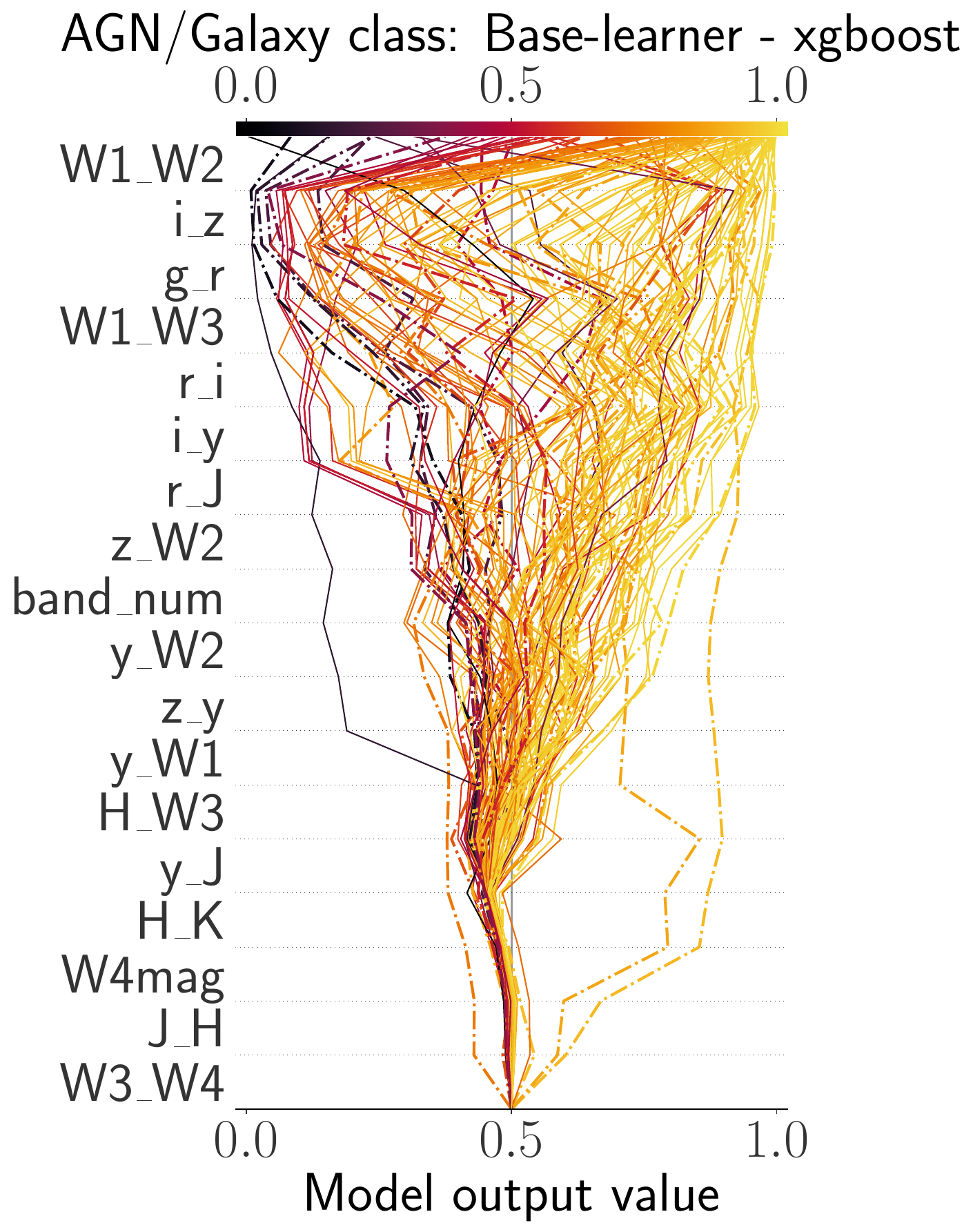}
  \end{minipage}%\\%
  \hfill
  \begin{minipage}{0.48\columnwidth}
    \centering
    \includegraphics[width=\textwidth]{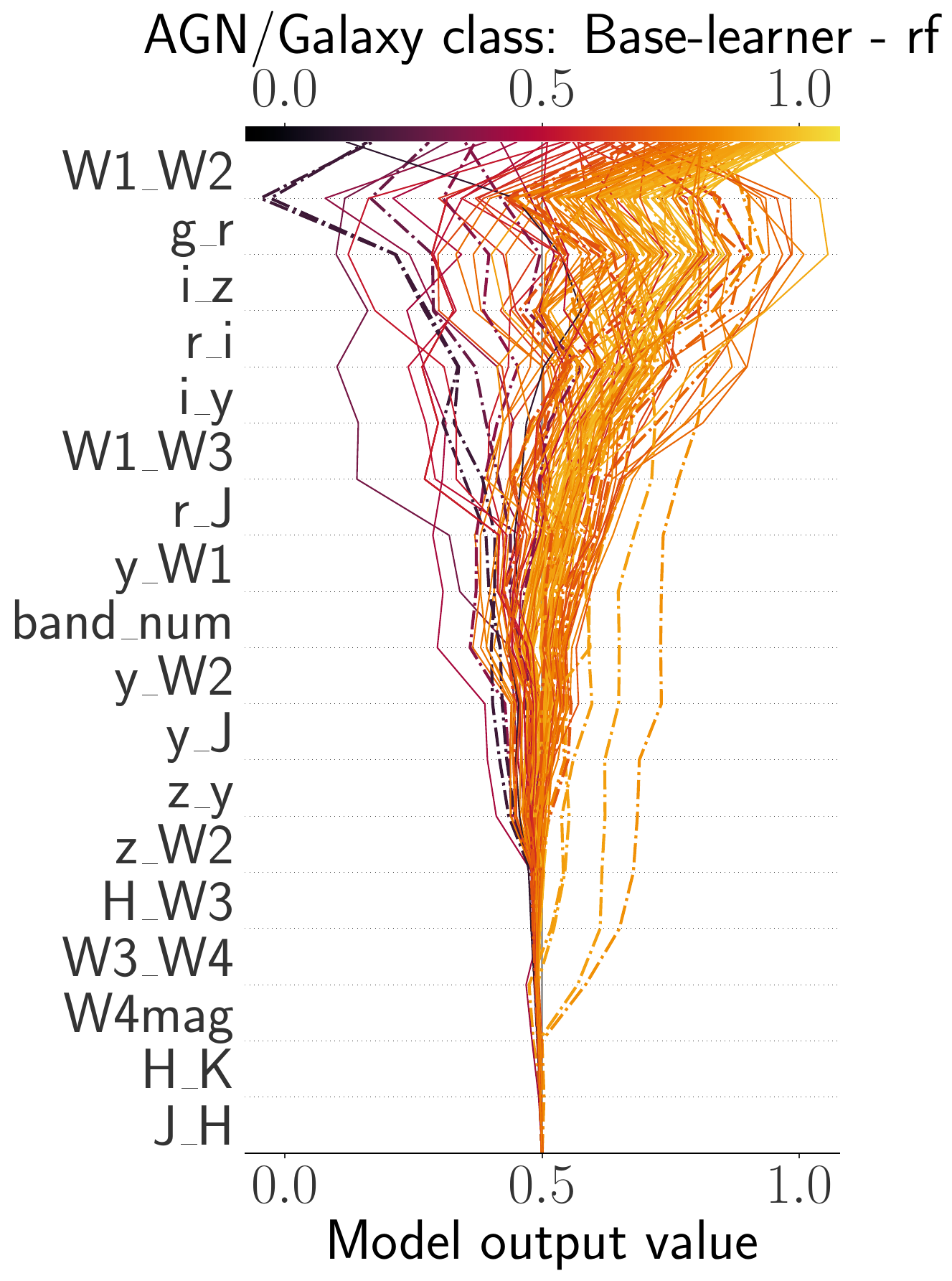}
  \end{minipage}\break%\%
  \begin{minipage}{0.48\columnwidth}
    \centering
    \includegraphics[width=\textwidth]{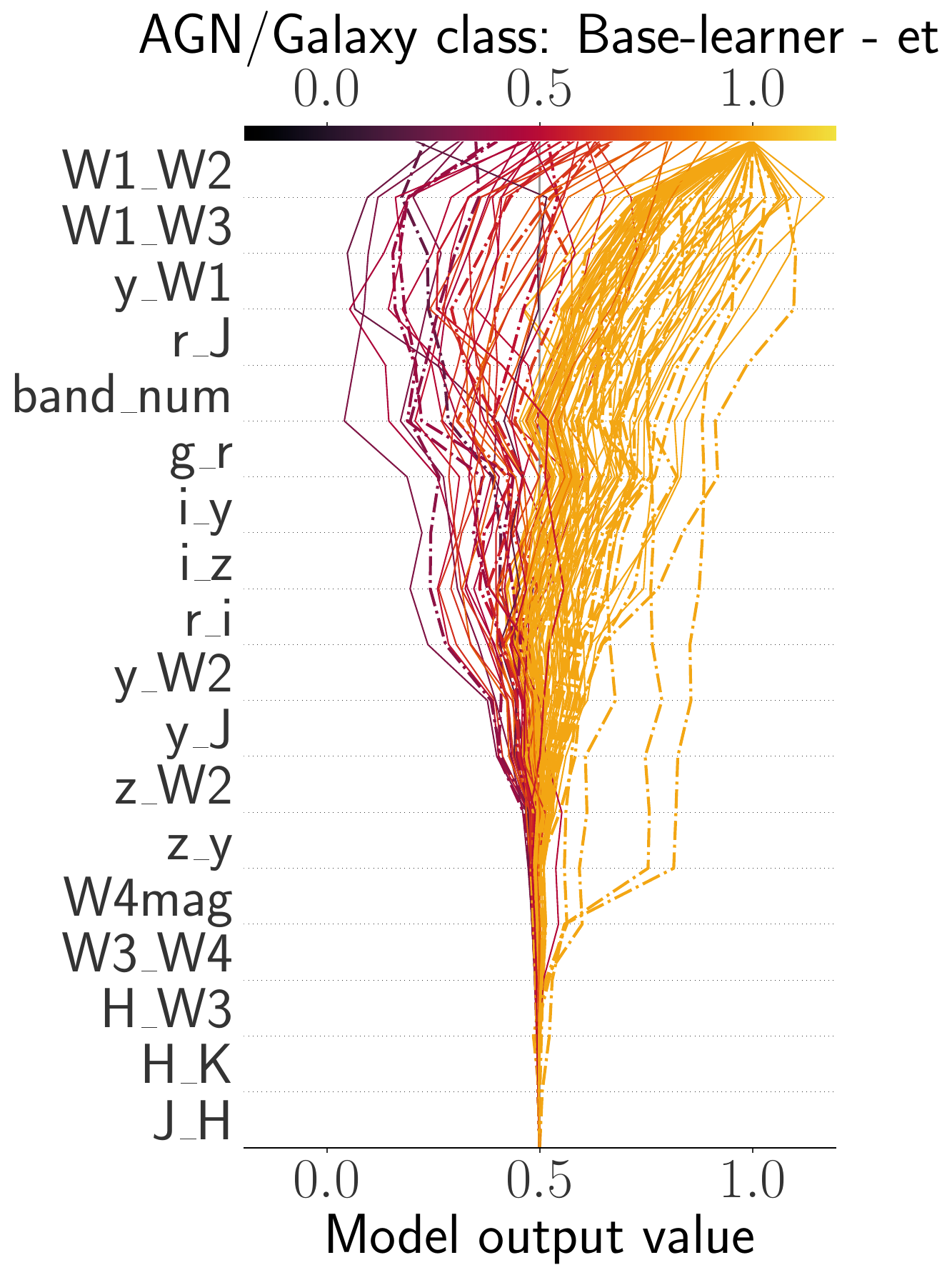}
  \end{minipage}%\\%
  \hfill
  \begin{minipage}{0.48\columnwidth}
    \centering
    \includegraphics[width=\textwidth]{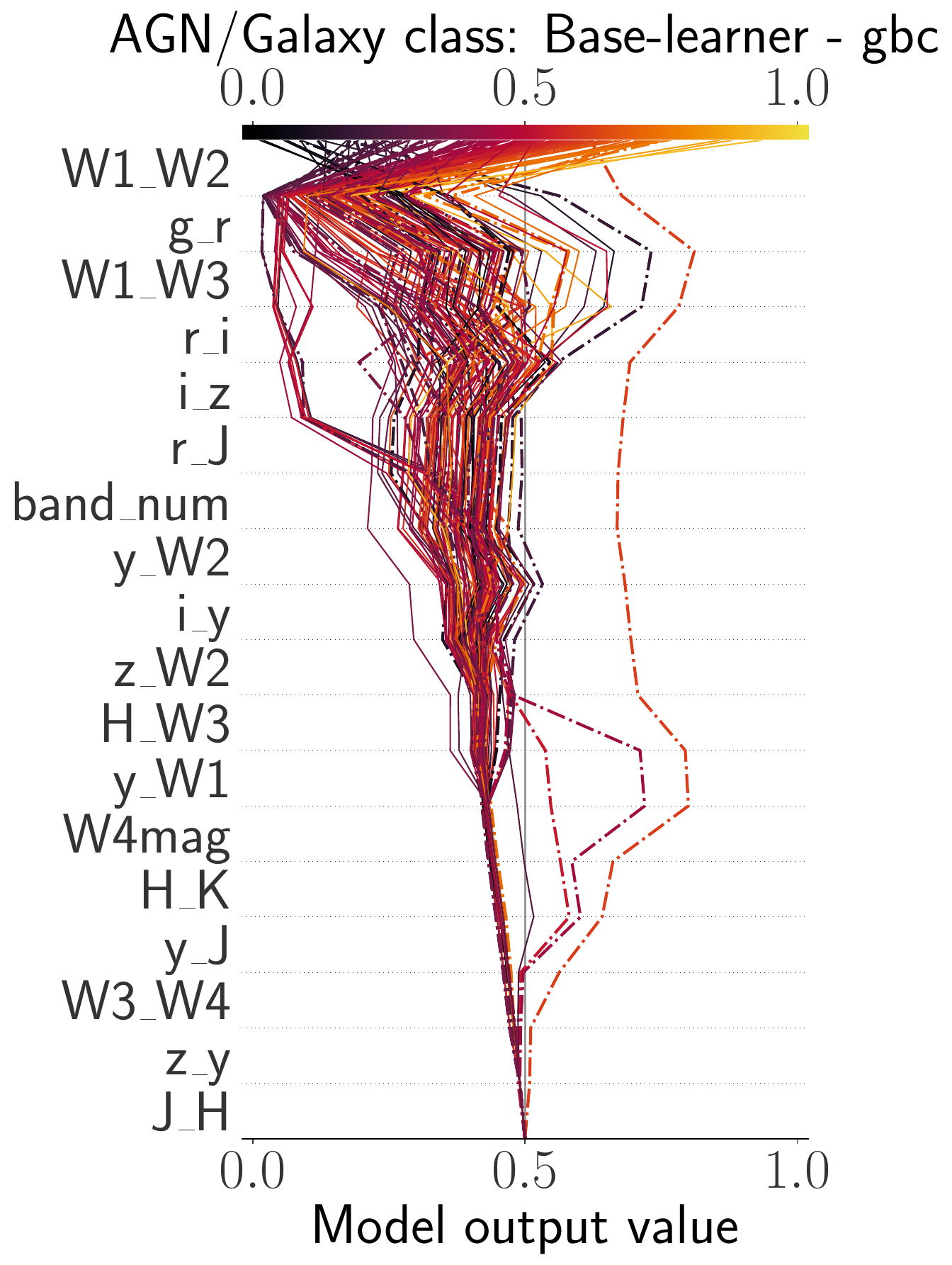}
  \end{minipage}
  \caption{SHAP decision plots for base AGN-galaxy algorithms. Details as described in Figs.~\ref{fig:SHAP_decision_AGN_meta_HETDEX_high_z}. Starting point of predictions is the naive classification threshold. From left to right and from top to bottom, each panel shows the results from \texttt{XGBoost}, \texttt{RF}, \texttt{ET}, and \texttt{GBC}.}
  \label{fig:SHAP_decision_AGN_base_HETDEX_high_z}
\end{figure}

\begin{figure}
  \centering
  \begin{minipage}{0.48\columnwidth}
    \centering
    \includegraphics[width=\textwidth]{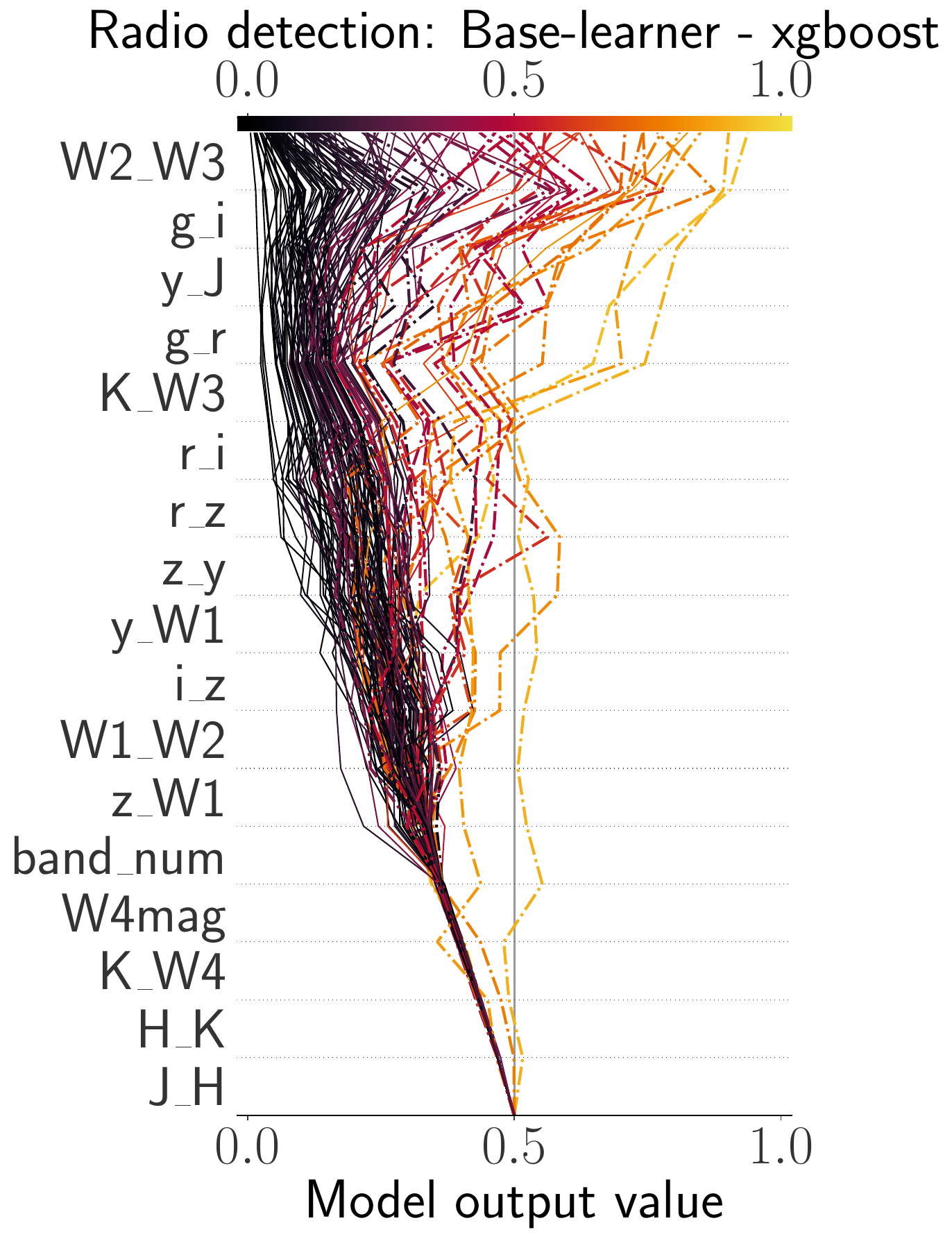}
  \end{minipage}%\\%
  \hfill
  \begin{minipage}{0.48\columnwidth}
    \centering
    \includegraphics[width=\textwidth]{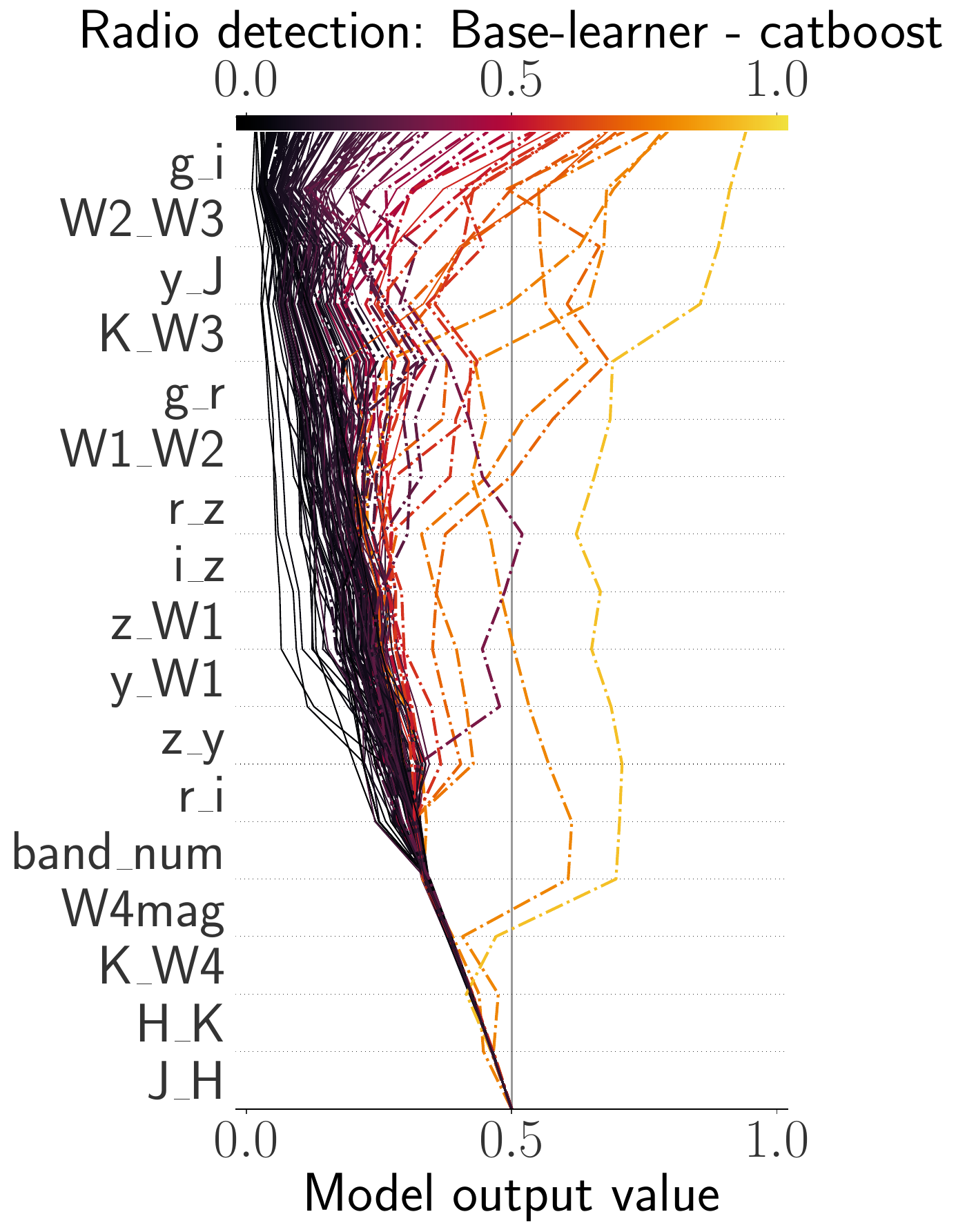}
  \end{minipage}\break%\%
  \begin{minipage}{0.48\columnwidth}
    \centering
    \includegraphics[width=\textwidth]{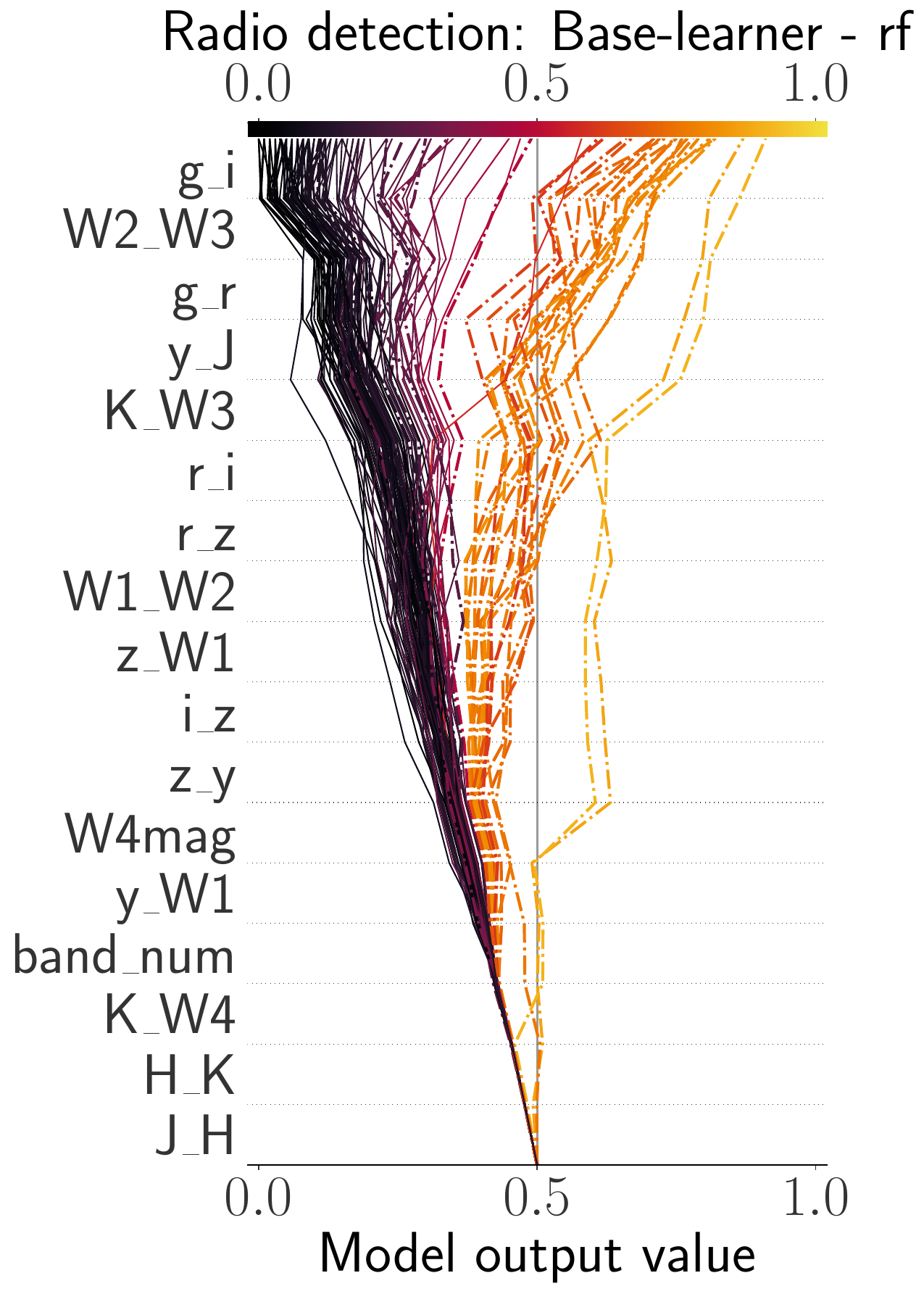}
  \end{minipage}%\\%
  \hfill
  \begin{minipage}{0.48\columnwidth}
    \centering
    \includegraphics[width=\textwidth]{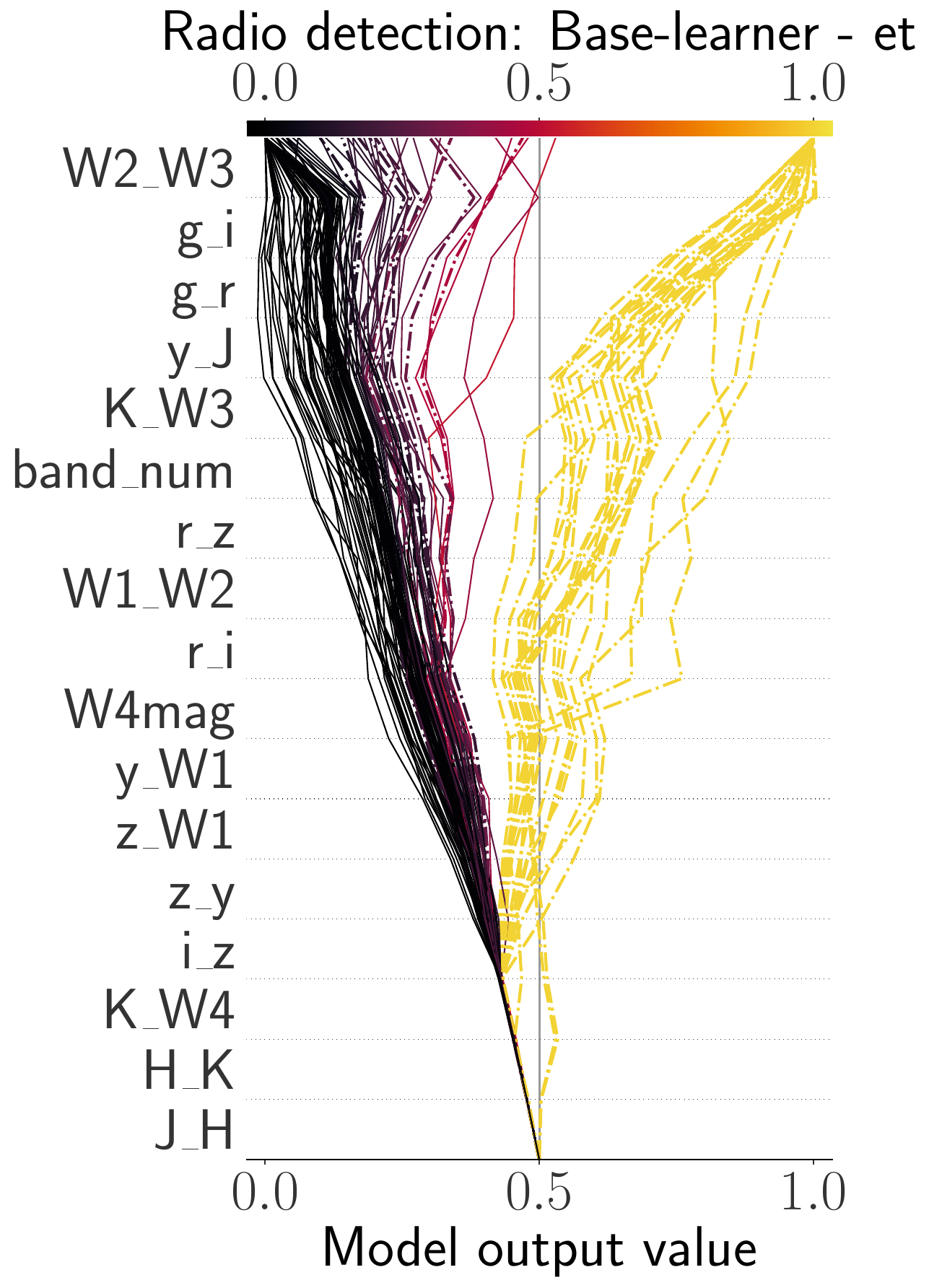}
  \end{minipage}
  \caption{SHAP decision plots from base radio algorithms. Details as Figs.~\ref{fig:SHAP_decision_AGN_meta_HETDEX_high_z} and \ref{fig:SHAP_decision_AGN_base_HETDEX_high_z}. Each panel with results for \texttt{XGBoost}, \texttt{CatBoost}, \texttt{RF}, and \texttt{ET}.}
  \label{fig:SHAP_decision_radio_base_HETDEX_high_z}
\end{figure}

\begin{figure}
  \centering
  \begin{minipage}{0.48\columnwidth}
    \centering
    \includegraphics[width=\textwidth]{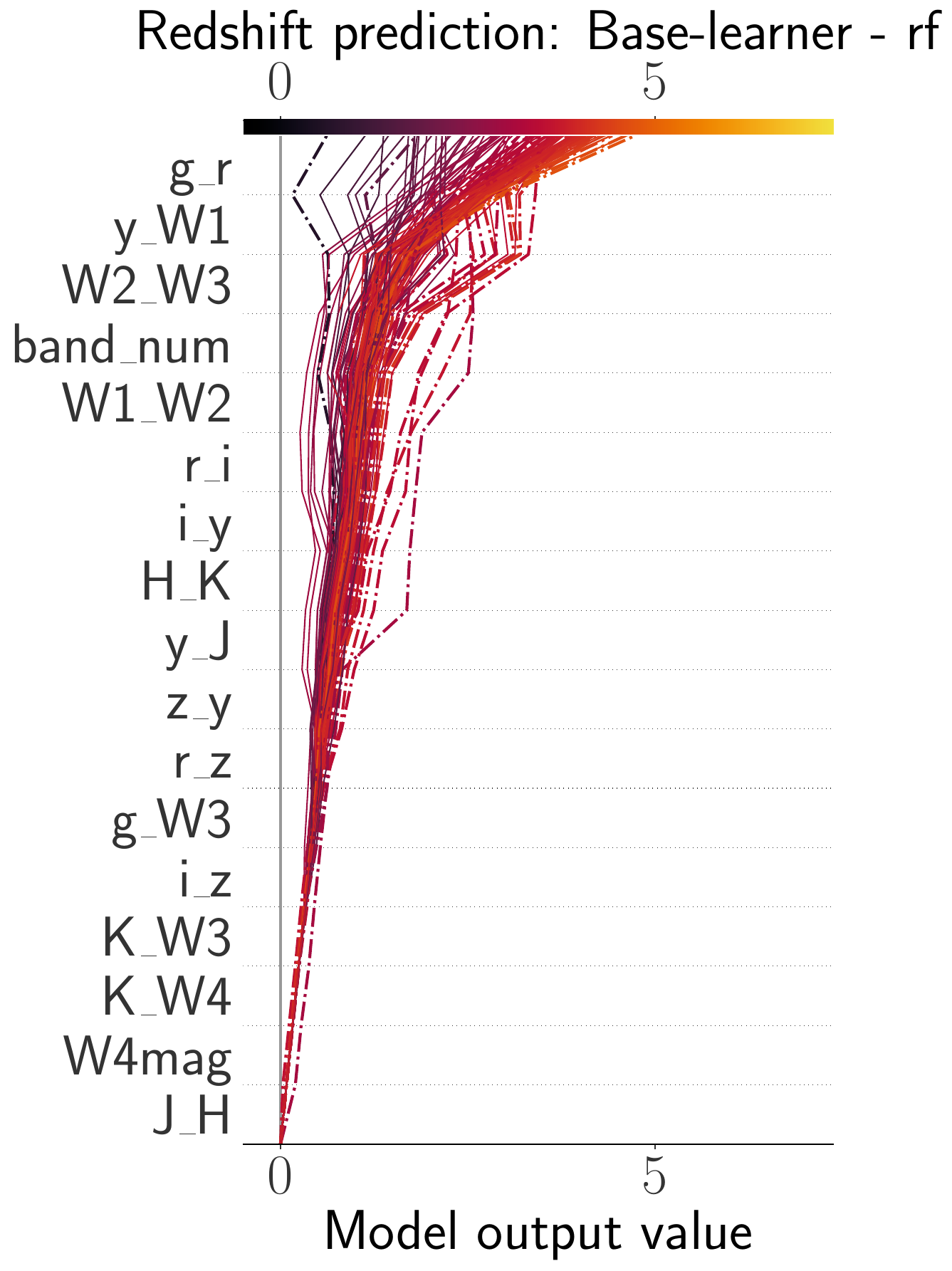}
  \end{minipage}%\\%
  \hfill
  \begin{minipage}{0.48\columnwidth}
    \centering
    \includegraphics[width=\textwidth]{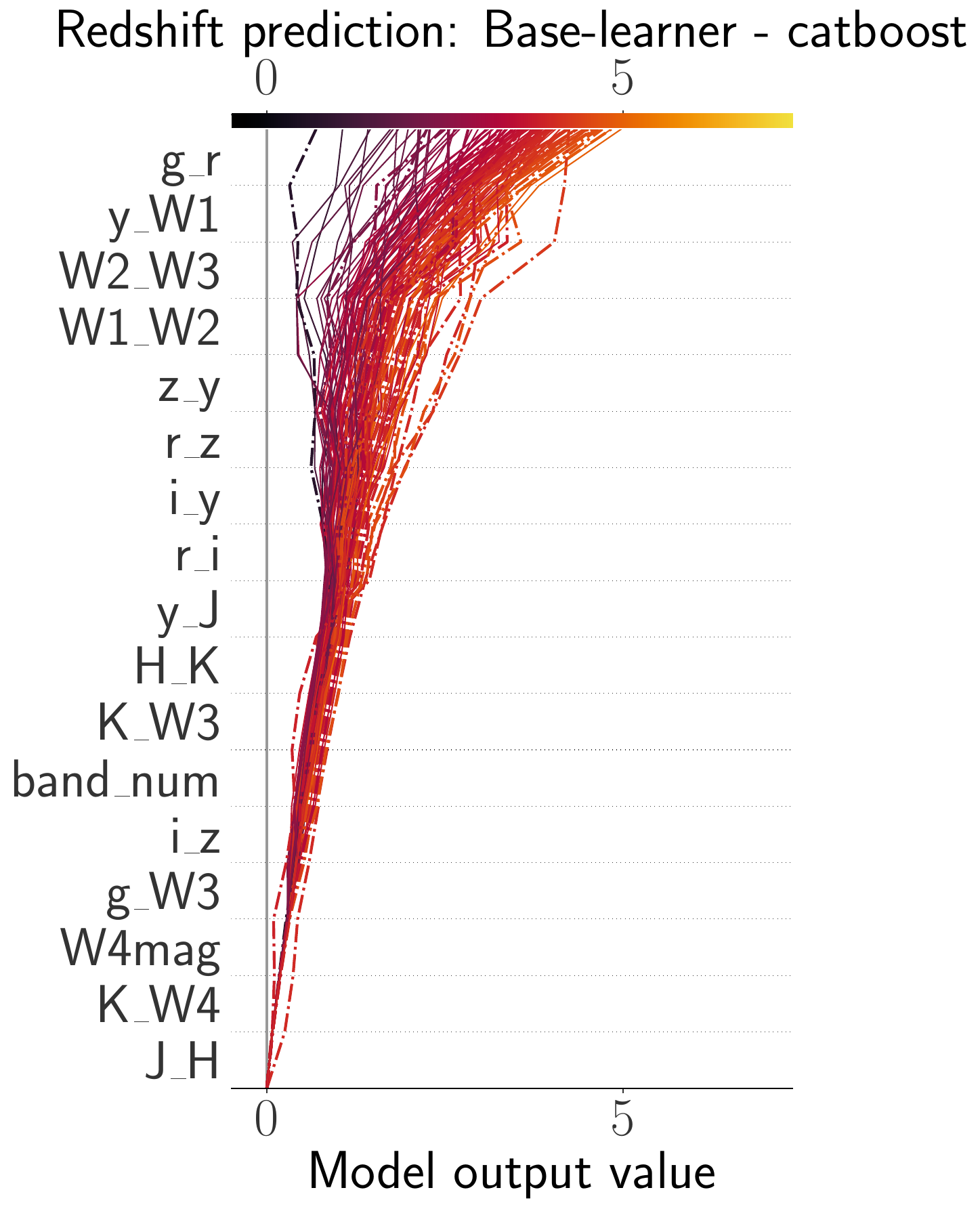}
  \end{minipage}\break%\%
  \begin{minipage}{0.48\columnwidth}
    \centering
    \includegraphics[width=\textwidth]{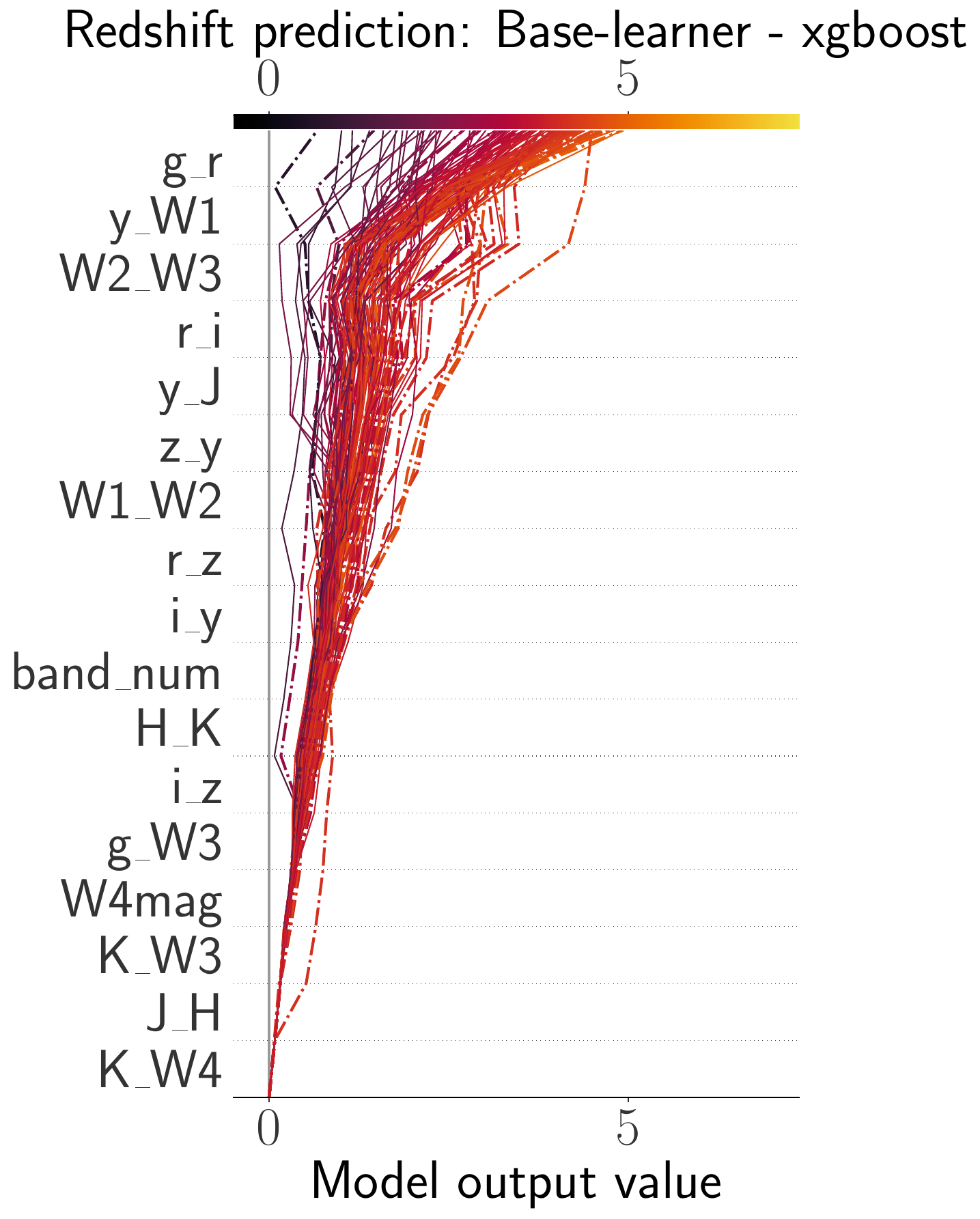}
  \end{minipage}%\\%
  \hfill
  \begin{minipage}{0.48\columnwidth}
    \centering
    \includegraphics[width=\textwidth]{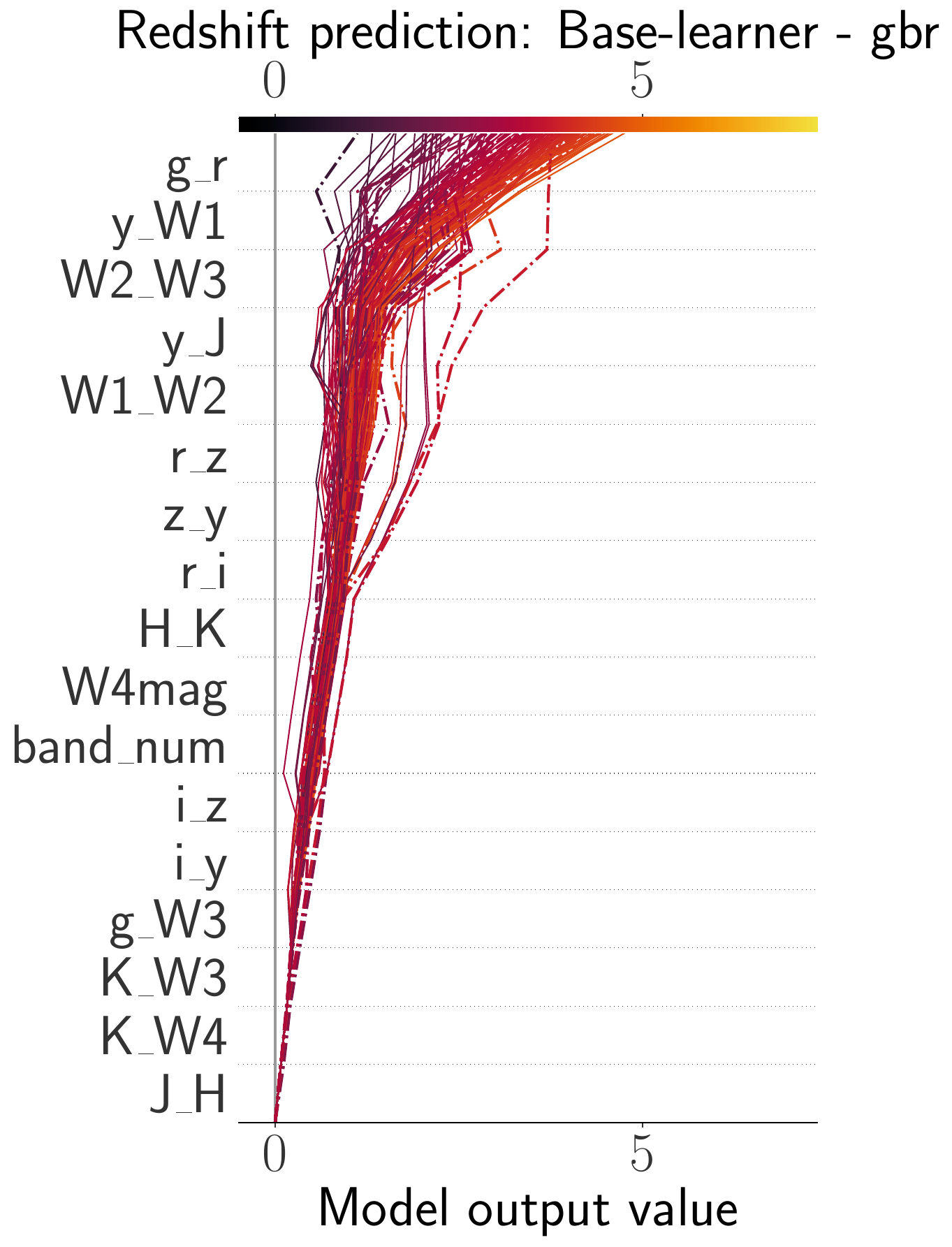}
  \end{minipage}
  \caption{SHAP decision plots from base $z$ algorithms. Details as in Fig~\ref{fig:SHAP_decision_AGN_meta_HETDEX_high_z}. Each panel shows results for \texttt{ET}, \texttt{CatBoost}, \texttt{XGBoost}, and \texttt{GBR}.}
  \label{fig:SHAP_decision_z_base_HETDEX_high_z}
\end{figure}

Figures~\ref{fig:SHAP_decision_AGN_base_HETDEX_high_z}, \ref{fig:SHAP_decision_z_base_HETDEX_high_z}, and \ref{fig:SHAP_decision_radio_base_HETDEX_high_z} show the decision plots for each base learners used in the prediction models of our pipeline (Sect.~\ref{sec:shapley_values}). For the classification algorithms, the starting point of their predictions corresponds to the naive threshold ($0.5$) since no threshold optimisation was applied to them and only the scores are included in the stacking step, not the final probabilities. In the case of the redshift predictors, decision plots start at the value $z {=} 0$, as presented for the meta learner.

\section{Prediction results for radio AGNs}\label{sec:app_prediction_results}

The columns shown in the prediction results for sources in both HETDEX and S82 fields are described as follows. Tables with prediction results and models in pipeline can be retrieved from \url{https://zenodo.org/doi/10.5281/zenodo.10220008}.

\begin{itemize}

\item \texttt{ID}: Internal identification number.
\item \texttt{RA\_ICRS}: Right Ascension (in degrees) of source in CW.
\item \texttt{DE\_ICRS}: Declination (in degrees) of source in CW.
\item \texttt{Name}: Name of the source as it appears in the CW catalogue.
\item \texttt{band\_num}: Number of non-radio bands with a valid measurement per source (cf. Sect.~\ref{sec:feature_creation}).
\item \texttt{class}: \verb|1| if a source is a confirmed AGN by the MQC. \verb|0| if it has been spectroscopically confirmed as a galaxy in SDSS DR16. Sources with no value do not have a spectroscopic classification in this catalogue.
\item \texttt{Sint\_LOFAR} (or \texttt{Fint\_VLAS82}): Imputed integrated flux (in mJy) of source from LOFAR or VLAS82.
\item \texttt{Sint\_LOFAR\_non\_imp} (or \texttt{Fint\_VLAS82\_non\_imp}): Non-imputed integrated flux (in mJy) of source from LOFAR or VLAS82.
\item \texttt{W1mproPM}: Imputed W1 magnitude of source.
\item \texttt{W2mproPM}: Imputed W2 magnitude of source.
\item \texttt{gmag}: Imputed g magnitude of source.
\item \texttt{rmag}: Imputed r magnitude of source.
\item \texttt{imag}: Imputed i magnitude of source.
\item \texttt{zmag}: Imputed z magnitude of source.
\item \texttt{ymag}: Imputed y magnitude of source.
\item \texttt{W3mag}: Imputed W3 magnitude of source.
\item \texttt{W4mag}: Imputed W4 magnitude of source.
\item \texttt{Jmag}: Imputed J magnitude of source.
\item \texttt{Hmag}: Imputed H magnitude of source.
\item \texttt{Kmag}: Imputed Ks magnitude of source.
\item \texttt{Score\_AGN}: Score from the meta AGN-galaxy classifier for a prediction to be an AGN.
\item \texttt{Prob\_AGN}: Probability from the calibrated meta AGN-galaxy classifier for a prediction to be an AGN.
\item \texttt{LOFAR\_detect}: \verb|1| if a source has been detected on the LoTSS survey or in their analogue surveys for fields different to HETDEX (see Sect.~\ref{sec:data_collection}). \verb|0| otherwise.
\item \texttt{Score\_radio\_AGN}: Score from the meta radio detection model for a prediction to be detected in the radio.
\item \texttt{Prob\_radio\_AGN}: Probability, from the calibrated radio detection model for a prediction to be detected in the radio.
\item \texttt{radio\_AGN}: \texttt{class}~$\times$~\texttt{LOFAR\_detect}. \texttt{1} if a source is an AGN and has been detected in the radio. \texttt{0} otherwise.
\item \texttt{Score\_rAGN}: \texttt{Score\_AGN}~$\times$~\texttt{Score\_radio}. Score of a source for it to be an AGN detected in the radio.
\item \texttt{Prob\_rAGN}: \texttt{Prob\_AGN}~$\times$~\texttt{Prob\_radio}. Probability of a source for it to be an AGN detected in the radio.
\item \texttt{Z}: Spectroscopic redshift as listed by the MQC (if available).
\item \texttt{pred\_Z}: Redshift value predicted by our model.

\end{itemize}

\end{appendix}

\end{document}